\documentclass[aps,prb]{revtex4}
\usepackage{graphicx}
\usepackage{bm}
\usepackage{xfrac}
\usepackage{amsmath}
\usepackage{amssymb}
\usepackage{bbold}
\usepackage{caption}
\usepackage{subcaption}
\usepackage{diagbox}
\usepackage{verbatim}

\newcommand{\be}{\begin{equation}}
\newcommand{\ee}{\end{equation}}
\newcommand{\bea}{\begin{eqnarray}}
\newcommand{\eea}{\end{eqnarray}}
\newcommand{\beb}{\begin{eqnarray*}}
\newcommand{\eeb}{\end{eqnarray*}}

\newcommand{\LD}{\langle}
\newcommand{\RD}{\rangle}
\newcommand{\Hh}{\text{H}}

\newcommand{\Tr}{\text{Tr}}

\begin{document}

\title{Phase diagram of a graphene bilayer in the zero-energy Landau level}

\author{Angelika Knothe$^{1,2}$}
\author{Thierry Jolicoeur$^{1}$}
\affiliation{1) Laboratoire de Physique Th\'eorique et Mod\`eles statistiques,
Universit\'e Paris-Sud, 91405 Orsay, France\\
2) Physikalisches Institut, Albert-Ludwigs-Universit\"at Freiburg, Hermann-Herder-Str. 3, D-79104 Freiburg, Germany}

\date{September, 2016}
\begin{abstract}
Bilayer graphene under a magnetic field has an octet of quasidegenerate levels 
due to spin, valley, and orbital degeneracies. This zero-energy Landau level is 
resolved into several incompressible states whose nature is still elusive. We 
use a Hartree-Fock treatment of a realistic tight-binding four-band model to 
understand the quantum ferromagnetism phenomena expected for integer fillings
of the octet levels. We include the exchange interaction with filled Landau 
levels below the octet states. This Lamb-shift-like effect contributes to the 
orbital splitting of the octet. We give phase diagrams as a function of applied 
bias and magnetic field. Some of our findings are in agreement with experiments.
We discuss the possible appearance of phases with orbital coherence.
\end{abstract}
\maketitle

\section{introduction}
The graphene family of new materials has produced novel two-dimensional electron systems. Contrary
to semiconductor devices the reduced dimensionality is due to the atomic structure which is made
of one or few layers. The bilayer graphene (BLG) has been the subject of intense 
scrutiny in the last few years.
Indeed, it has potential electronic instabilities that are different from those of single-layer graphene\cite{Knothe2015}.
There is a flat band contact at the Fermi level and large Berry curvatures. Several instabilities are competing 
even for small  interactions and it is likely that a layer antiferromagnet is the ground state of the neutral BLG system.
This system has very exotic properties under a magnetic field perpendicular to the layers. The Landau levels that appear
have a valley degeneracy and in the case of the central zero-energy level there is also an additional degeneracy
of two orbital states. When considering also the spin degree of freedom this 
means that the zero-energy Landau level
is eight-fold degenerate, i.e., there is an octet of states at zero energy.
Detailed experimental studies\cite{Weitz2010,  Kim2011, Bao2012, 
Jr2012, Maher2013,  VelascoJr2014, Lee2014, Maher2014, Shi2016, Hunt2016} 
of the quantum Hall regime of this octet have revealed that the degeneracy is fully
lifted presumably by an intricate mixture of one-body effects due to the band structure as well as the Coulomb
interactions between electrons.
The quantum Hall regime of the octet of states corresponds to Landau level filling factors
$\nu\in\left[-3,+3\right]$. At these fillings there are incompressible states 
that display various phase transitions
when the bias between layers is varied and/or the magnetic field is varied. Some of the gapped
states do survive the zero-field limit but this is not always the case. With 
increasing quality of samples, the
fractional quantum Hall effect has also been observed.

For integer fillings of the octet levels we expect the appearance of the 
well-studied quantum Hall ferromagnetism
with the added subtlety of orbital/valley degeneracies\cite{Barlas2008,Castro2010,Cote2010}. From a theoretical point of view, it is sensible
to use a Hartree-Fock (HF) approach because in many circumstances the ground state is given by a Slater determinant
provided one neglects Landau level mixing. To confront in some detail the experimental results one has to
first use a tight-binding model that includes all important couplings including some small particle-hole symmetry breaking
terms. When doing a HF calculation it has been pointed out\cite{Shizuya2012} that one has also to include the exchange with the
filled Landau levels that form a ``Dirac sea'' unique to graphene systems. 
Other theoretical approaches that do not use the quantum Hall ferromagnetism 
but a gap equation instead have also been applied to the BLG phase 
diagram\cite{Gorbar2011,Roy2013,Roy2014}.

In this paper we use a refined tight-binding model including all dominant hoppings and we treat the exchange effects
with the Dirac sea. We derive the phase diagram of the BLG octet as a function of applied bias and magnetic field. 
Previous recent HF studies either did not take the Dirac sea into account\cite{Lambert2013} or 
did not consider all relevant tight-binding hoppings\cite{Shizuya2012}.
Here we include the trigonal warping term as well as the next nearest-neighbor 
interlayer hopping which breaks particle-hole symmetry and also lifts the 
degeneracy between $n=0$ and $n=1$ orbitals.
We have only searched for spatially uniform phases in the HF solutions.
The phase diagrams for all fillings as a function bias and magnetic field are 
given in Fig.~(\ref{fig:GSPDs}). There are many phases whose existence is 
limited to a very short range of parameters. They are certainly the phases most 
sensitive to fluctuations beyond HF mean field. So we are more confident 
about the existence of phases which extend in a large domain.
Many of the phases we find can be termed as ``incoherent'', i.e., they are Slater 
determinants of filled levels with eigenstates with well-defined valley 
($\xi$), spin ($\sigma$), and orbital quantum numbers ($n$). ``Coherent'' states 
involve density matrices with some off-diagonal elements 
$
\langle
c^\dagger_{X\sigma\xi n}
c_{X\sigma^\prime \xi^\prime n^\prime} 
\rangle \neq 0
$
for $\sigma,\xi,n\neq\sigma^\prime ,\xi^\prime ,  n^\prime$ 
($X$ being the guiding center coordinate).

For filling factors $\nu=\pm 2$ we find a phase with valley coherence at small 
bias which is quickly destroyed to give way to incoherent phases. For $\nu =0$ 
an incoherent phase is replaced by a phase with spin and valley coherence for 
larger bias which then leads to an incoherent state at even larger bias.
This is exactly what is found in the simplified two-band treatment of Lambert and 
C\^ot\'e\cite{Lambert2013}.
For odd filling factor, the situation is quite different. For $\nu =\pm 1$
a phase with valley coherence is replaced by a phase with orbital coherence 
beyond a critical bias followed then by incoherent phases.
For $\nu =-3$ the situation is similar while for $\nu=+3$ the small bias regime
is now purely an orbital coherent phase and there is no valley coherence.
The phases with orbital coherence at $\nu=\pm 3$ appear for moderate bias and
magnetic field, a regime which is plausibly in the range of current 
experiments.

The paper is organized as follows: in Sec.~\ref{sec:BackgroundMethods}, we define the four-band model we employ to describe 
BLG  and describe how we technically 
proceed to treat the corresponding model Hamiltonian within the HF mean-field 
picture.
In Sec.~\ref{sec:HFRESI}, we give the phase diagram for all filling factors 
of the octet as a function of applied bias and magnetic field.
 We discuss the phase configurations in terms of spin, valley isospin, 
and orbital isospin degrees of freedom.
In Sec.~\ref{sec:HFRESII} the octet 
polarization properties are discussed for different electronic fillings which 
leads us to Hund's rules for the successive occupation of the single particle 
(SP) levels. We further relate these polarization properties to the electronic 
distribution between the two 
layers of the system and comment on the possibility of full layer 
polarization. Finally, we discuss possible extrapolations to stable 
phases at zero magnetic field. 
Section \ref{sec:Relation} contains a comparison of our findings to recent 
experiments as well as to earlier theoretical investigations.  
In Sec.~\ref{sec:DiscConcl} we give some final remarks and present our conclusions.

\section{Background and Methods}
\label{sec:BackgroundMethods}

\subsection{The non-interacting Single Particle Hamiltonian}
\label{ssec:NonIntSP}

\begin{figure}[!htb]
  \centering
\includegraphics[width=0.25\textwidth]{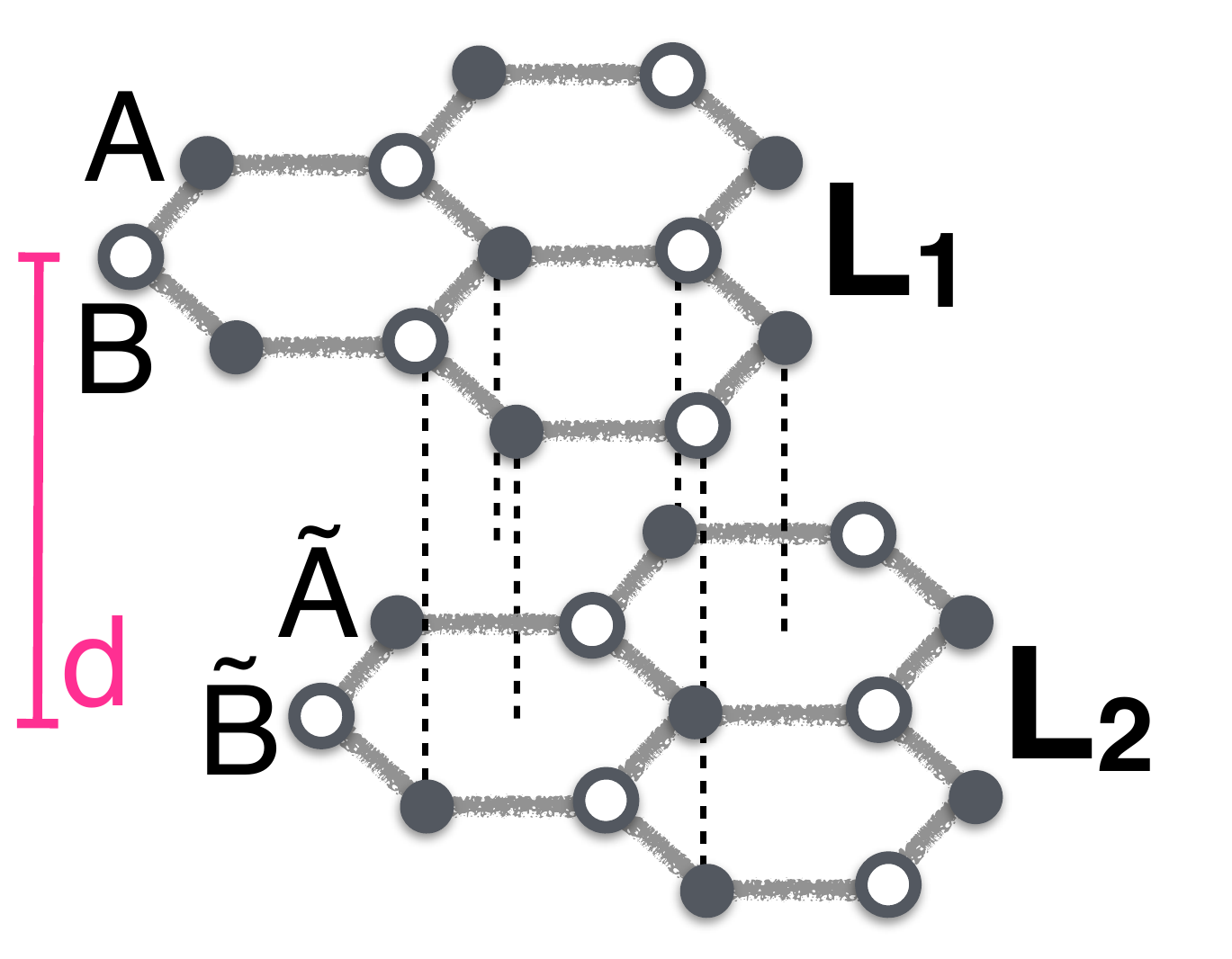}
  \caption{Sketch of the the model for BLG: Two graphene layers $L_1$ and $L_2$ 
are stacked on top of each other at an interlayer distance $d$ in the Bernal 
arrangement. We depict the inequivalent carbon atoms in each monolayer by 
filled and dashed circles, respectively, and label them as $A\;(\tilde{A})$ 
and 
$B\;(\tilde{B})$ on the upper (lower) layer.}
\label{fig:BernalBilGr}
 \end{figure}

A sketch of Bernal stacked BLG is shown in Fig.~\ref{fig:BernalBilGr}. In this 
model describing BLG as two hexagonal lattices on top of each other, we denote 
the constituents as follows: It is composed of an upper layer $L_1$ and a lower 
layer $L_2$  separated by interlayer distance $d$. In each
layer, the hexagonal lattice structure is formed by two trigonal sublattices, in 
which we label the carbon atoms as $A$ and $B$ on the upper layer and 
$\tilde{A}$ and $\tilde{B}$ on the lower layer. This yields a total of four 
atoms per unit cell. We refer to \textit{dimer} sites 
if two atoms lie on top of each other and to \textit{non-dimer} 
sites when this is not the case. 
A tight binding description of the electrons 
on this lattice follows from the so-called Slonczewski-Weiss-McClure model of 
bulk 
graphite\cite{McClure1957, Slonczewski1958}. The
 tight binding hopping parameters are then:
$\gamma_0=\gamma_{A 
\leftrightarrow B}$ describes intralayer coupling, \textit{i.e.}~next neighbors 
in-plane hopping within one graphene layer, 
whereas $\gamma_1=\gamma_{\tilde{A} 
\leftrightarrow B}$ captures interlayer hopping via vertical coupling between 
the pairs of orbitals on the dimer sites. 
For the skew interlayer couplings 
containing both in-plane and vertical components, we write $\gamma_3=\gamma_{A 
\leftrightarrow \tilde{B}}$ for coupling between two non-dimer orbitals and  
$\gamma_4=\gamma_{A \leftrightarrow \tilde{A}}$ for coupling between one dimer 
and one non-dimer orbital. Due to different on-site energies in BLG, 
we also include the splitting $\delta_{A,B}$ for the local 
energies 
between $A$ and $B$ sites on each layer.  

For the hoppings the following relations
hold\cite{Castro2010, McCann2013}: $\gamma_0>\gamma_1>\gamma_3, \gamma_4 \gg 
\delta_{AB}$. Throughout this work, we use the numerical values for the 
parameters listed in Ref.~
\onlinecite{Lambert2013} consistent with previous experimental and 
theoretical investigations\cite{Kuzmenko2009, Yankowitz2014}.
In momentum space, we distinguish between the two inequivalent Dirac points $K$ 
and $K^{\prime}$ and index them by $\zeta=\pm 1$ following the convention 
$\zeta|_{K}=+1$ and $\zeta|_{K^{\prime}}=-1$. After expanding for small momenta 
around these Dirac points the effective Hamiltonian describing the low-energy 
physics can be written as
\begin{equation} \Hh_{\zeta}=\zeta
\begin{bmatrix}
 \frac{1}{2}\Delta_B +\frac{1}{2}  \,\zeta(1+\zeta)\, \delta_{AB} & 
v_3 \mathbf{p}& 
v_4 \mathbf{p}^{\dagger}& 
v_0\mathbf{p}^{\dagger}\\
v_3 \mathbf{p}^{\dagger} &
  -\frac{1}{2}\Delta_B +\frac{1}{2} \,\zeta(1+\zeta)\,  \delta_{AB}& 
v_0\mathbf{p}& 
v_4 \mathbf{p}\\
v_4 \mathbf{p} &
 v_0\mathbf{p}^{\dagger} &
 - \frac{1}{2}  \Delta_B + \frac{1}{2}\,\zeta\,(1-\zeta) \delta_{AB} &
 \gamma_1\\
  v_0\mathbf{p}& 
v_4 \mathbf{p}^{\dagger} &
 \gamma_1 &
 \frac{1}{2}\Delta_B+\frac{1}{2}\,\zeta(1-\zeta)\, \delta_{AB} 
\end{bmatrix},
\label{eqn:H}
\end{equation}

acting respectively on the four-component spinor fields
\begin{equation}\psi_{K}=
\begin{pmatrix} 
\psi_{A}\\
\psi_{\tilde{B}}\\
\psi_{\tilde{A}}\\
\psi_{B}
\end{pmatrix} \; \text{and}\;\;\;
\psi_{K^{\prime}}=
\begin{pmatrix} 
\psi_{\tilde{B}}\\
\psi_{A}\\
\psi_{B}\\
\psi_{\tilde{A}}
\end{pmatrix} .
\label{eqn:Psi}
\end{equation}

In Eq.~\ref{eqn:H}, we use the generalized velocities
$v_i=\frac{\sqrt{3}}{2}\frac{a_L}{\hbar}\gamma_i$ for $\{i=0,1,3,4\}$
written in terms of the lattice constant $a_L$. Besides, $\mathbf{p}=p_x+ip_y$, 
$\mathbf{p}^{\dagger}=p_x-ip_y$ stands for momentum. Additionally, we want to  
capture the effect of an externally applied electric field $E_{\perp}$, 
therefore in Eq.~\ref{eqn:H} we  introduce a bias potential $\Delta_B=e 
d E_{\perp}[\frac{\text{mV}}{nm}]$, with 
$e$  the electric charge.
We  now proceed as follows:  we first neglect 
the smaller parameters $\delta_{AB}, \gamma_3,$ and $\gamma_4$ . In this case 
 exact analytical solutions can be obtained\cite{Mucha-Kruczynski2008, 
Mucha-Kruczynski2009, Mucha-Kruczynski2013, Shizuya2012}. Subsequently, we  
 include the effects of the remaining parameters   in  first order 
perturbation theory.
The following reasoning closely follows the derivation given in Ref.~\onlinecite{Shizuya2012}.
In the $K$ valley, the approximate effective Hamiltonian under study reads:
\begin{equation} \Hh_{K}=
\begin{bmatrix}
 \frac{1}{2}\Delta_B & 0& 0 & v_0 \mathbf{p}^{\dagger}\\
0 &  -\frac{1}{2}\Delta_B & v_0 \mathbf{p}&0\\
0 &  v_0 \mathbf{p}^{\dagger} & - \frac{1}{2}\Delta_B & \gamma_1\\
 v_0 \mathbf{p}& 0 & \gamma_1 & \frac{1}{2}\Delta_B
\end{bmatrix}.
\label{eqn:H_K}
\end{equation}

In the presence of a magnetic field of strength $B$, we replace the canonical 
momentum by the mechanical momentum, 
$\mathbf{p}\rightarrow\pi=\mathbf{p}+e\mathbf{A}$, writing the vector potential 
$\mathbf{A}$ in Landau gauge, $\mathbf{A}=Bx\mathbf{e}_y$. The electronic state 
quantized into the $n$th Landau level is denoted as $|n\RD$, with real space 
representation at guiding center coordinate  $x_p=p\ell_B^2$ given by
\begin{equation}\LD \mathbf{r}|n;p  \RD=\frac{1}{\sqrt{L_y}}
\phi_n(\mathbf{x-x_p}) e^{ipy},
\label{eqn:realspaceWF}
\end{equation}
where $\phi_n$ denotes the n-th harmonic oscillator wave function and $L_y$ 
measures the spatial extension of the system in $y$-direction.
The $\pi$-operators act as ladder operators in the space of Landau functions $\LD \mathbf{r}|n;p  \RD$; the corresponding relations
\begin{align}
&\nonumber\pi^{\dagger} |n\RD  = i\frac{\hbar}{\ell_B}\sqrt{2(n+1)}|n+1\RD,\\
&\pi|n\RD = -i\frac{\hbar}{\ell_B}  \sqrt{2n}|n-1\RD  \text{  for }n>0 \text{   and } \pi|0\RD=0,
\label{eqn:ladder}
\end{align}
so the electronic state of the n-th Landau Level in the valley $K$ is of 
the form (agreeing on $|n\RD\equiv0$ for $n<0$)
\begin{equation}{\psi}^{(n)}_{K}=
\begin{pmatrix} 
b_{(n),1}|n\RD\\
b_{(n),2}|n-2\RD\\
b_{(n),3}|n-1\RD\\
b_{(n),4}|n-1\RD
\end{pmatrix} .
\label{eqn:psi_n}
\end{equation}

It is well-known that the LL spectrum of unbiased BLG shows peculiar behavior 
with respect to the zero energy level\cite{McCann2006, Mucha-Kruczynski2008, 
Mucha-Kruczynski2009}. Indeed, the $n=0$ and the $n=1$ orbitals have zero 
energy. As a consequence, the zero energy state of BLG is eight-fold 
degenerate in the real spin, the valley isospin, and this $n=0,1$ orbital degree 
of freedom. This property has triggered a plethora of studies 
to understand the QH ferromagnetism of the zero energy octet of BLG.
In the case of biased BLG, strictly speaking, this eight-fold degeneracy is no 
longer fully exact but  broken by the presence of a nonzero bias 
potential $\Delta_B$. As long as the bias potential is sufficiently small 
compared to the LL gap it is sensible to focus only on the physics of the octet
and neglect LL mixing.
 We assume all the states of 
energy below the octet states 
$\epsilon_{-n}<\epsilon_{0,1}\approx0$ to be filled. We describe them as a 
manifold of inert levels $n\le -2$ labeled with negative indices 
and 
refer to them as the \textit{Dirac sea}. We discuss in more detail the 
influence of 
the presence of the Dirac sea electrons in the next section \ref{ssec:HF_Int}, 
when interactions between electrons are taken into account.  All the states 
which 
energetically lie above the pseudo zero energy states with $\epsilon_{n}>0$ are 
 empty. The octet has partial fillings $\nu \in [-4,4]$.
For further use we note the explicit forms of the states for the cases $n=0$ and $n=1$:
\begin{equation}\psi^{(0)}_{K}=
\begin{pmatrix} 
|0\RD\\
0\\
0\\
0
\end{pmatrix} ,\;\;\;
\psi^{(1)}_{K}=
\begin{pmatrix} 
b_{(1),1}|1\RD\\
0\\
b_{(1),3}|0\RD\\
b_{(1),4}|0\RD
\end{pmatrix} ,
\label{eqn:H_psi_01}
\end{equation}
with coefficients 
\begin{align}
\nonumber &b_{(1),1}=c_1,\\
\nonumber &b_{(1),3}=-\frac{c_1}{{\gamma}}(1-M^{2}z^2),\\
\nonumber &b_{(1),4}=- c_1 z M,\\
\label{eqn:psi_01_coeff}
\end{align}
in terms of the rescaled parameters ${\gamma}=\frac{\gamma_1}{\hbar\omega_c}$ 
and $2M=\frac{\Delta_B}{\hbar\omega_c}$ and with normalization constant  
$c_1=\frac{1}{\sqrt{1+\frac{1}{{\gamma}^2}(1-M^{2}z^2)^2+z^2M^2}}$ . Here
$\hbar\omega_c\approx 36.3 \, v_0 [10^6 \text{ m/s}]\sqrt{B[\text{T}]} $meV is 
the characteristic cyclotron energy. The parameter z is 
determined as the solution in the range $0\le z\le 1$ to the equation 
$z=\frac{1}{{\gamma}^2}(2-z)(1-M^2 z^2). $
In these expressions, passing from the $K$ to the $K^{\prime}$ valley is 
done by the replacement $M\rightarrow-M$.
For the corresponding lowest energy eigenvalues for the $n=0$ and the $n=1$ 
orbitals we find
\begin{align}
\nonumber &\epsilon_{\xi,n=0}=\xi\frac{1}{2}\Delta_B,\\
&\epsilon_{\xi,n=1}=\xi\frac{1}{2}\Delta_B-\xi\frac{1}{2}z\Delta_B.
\label{eqn:En_01}
\end{align} 
To these solutions of Eq.~\ref{eqn:En_01}, we compute the 
 corrections due to the  parameters $\delta_{AB}, \gamma_3$ and 
$\gamma_4$ as perturbations.  This is done using the states 
of Eq.~\ref{eqn:H_psi_01} for $\Delta_B\equiv0$.
 It turns out that taking into account perturbations up to the first 
order in  $\delta_{AB}, \gamma_3$ and $\gamma_4$ induces a splitting 
between the $n=0$ and the $n=1$ orbitals which reads
\begin{equation}
\Delta_{01}^{\text{pert}}=-\delta_{AB}(1-c^2)-2\frac{\gamma_4}{\gamma_0\gamma_1}c_1^2(\hbar\omega_c)^2.
\label{eqn:D_01_SP}
\end{equation}
In addition there is also the Zeeman effect leading to a gap 
 $\Delta_Z=g \mu_B B$ with $g=2$.
Hence, as the main result of this section \ref{ssec:NonIntSP}, we write the 
effects of Eqs.~\ref{eqn:En_01}, \ref{eqn:D_01_SP}, and the Zeeman splitting  
$\Delta_Z$  into an effective Hamiltonian describing the $n=0,1$ orbitals of 
the 
non-interacting system by writing:
\begin{equation}
\Hh_{0}=  \sum_{p} \sum_{n,\sigma, \xi} \Big[- \frac{\Delta_{B}}{2}\tau_z 
+z\frac{\Delta_{B}}{4}(\tau_z+\lambda_z\tau_z) - 
\frac{\Delta^{\text{pert}}_{01}}{2}\lambda_z - \frac{\Delta_{Z}}{2}\sigma_z 
\Big] c^{\dagger}_{n,\sigma,\xi}(p)\,c_{n,\sigma,\xi}(p),
\label{eqn:H_0}
\end{equation}
where we use the notation 
$\sigma_{\alpha}=\mathbb{1}^{\text{mode}}\otimes\sigma^{\text{spin}}_{\alpha} 
\otimes \mathbb{1}^{\text{valley}}$ , 
$\tau_{\alpha}=\mathbb{1}^{\text{mode}}\otimes\mathbb{1}^{\text{spin}}
\otimes\sigma^{\text{valley}}_{\alpha}$, and 
$\lambda_{\alpha}=\sigma^{\text{mode}}_{\alpha}\otimes\mathbb{1}^{\text{spin}}
\otimes\mathbb{1}^{\text{valley}}$
for the Pauli operators acting in spin, in valley, and in orbital space and 
$\lambda_{\alpha}\tau_{\beta}=\sigma^{\text{mode}}_{\alpha}\otimes\mathbb{1}^{
\text{spin}}\otimes\sigma^{\text{valley}}_{\beta}$.

\begin{table}[!hbt]
   \centering
\begin{tabular}{|c|c||c|c|}
\hline
 \multicolumn{1}{|c}{System Parameters} & \multicolumn{2}{c}{}& \multicolumn{1}{c|}{} \\
\hline
$\gamma_0 $ & 3.1 \text{ eV}  &$\gamma_1$  & 0.39 \text{ eV}  \\
$ \gamma_3 $& 0.1 \text{ eV} & 
$\gamma_4$  & 0.13 \text{ eV} \\
$v_0=\frac{\sqrt{3}}{2}\frac{a_L\gamma_0}{\hbar} $ & $1.1 \times 10^6 \text{ 
m/s}  $& $a_L  $& $0.246 \text{ nm}$ \\
$\hbar\omega_c =\sqrt{2}\frac{\hbar v_F}{\ell_B}$ &$  36.3\; 
v_0 [10^6\frac{m}{s}] \sqrt{B[\text{T}]} \text{ meV} $& d &$ 0.34\text{ nm}$\\
$\ell_B=\sqrt{\frac{\hbar c }{e B}} $& $26\text{nm}\frac{1}{\sqrt{B[\text{T}]}}$ &$ \kappa$ & 5\\
$\delta_{A,B}$&$ 0.016 \text{ eV}$ & &\\
\hline
\multicolumn{1}{|c}{Characteristic energies} & \multicolumn{2}{c}{}& 
\multicolumn{1}{c|}{} \\
\hline
$\Delta_{C}=\sqrt{\frac{\pi}{2}}\alpha=\sqrt{\frac{\pi}{2}}\frac{e^2}{
\kappa\ell_B}  $&$  14.1\sqrt{B[\text{T}]}\text{ meV}  $&$ 
\Delta_{01} $ &$ \frac{1}{8}\Delta_{C}\,c_1^2(4-3c_1^2)  $ \\
$\Delta_{Z}=g\mu_B B$ & $0.11\,B[\text{T}]\text{ meV}  $&$ \Delta_{B}$&$  e d 
E_{\perp}[\frac{\text{mV}}{\text{nm}}] $ \\ 
\hline
  \end{tabular}
  \caption{Numerical values of the system parameters and the energy 
splittings used throughout the analysis.}
\label{tab:SysParam}
\end{table}

\subsection{The Hartree Fock Hamiltonian}
\label{ssec:HF_Int}

We now deal with the Coulomb 
interaction between the electrons:
\begin{equation}
\Hh_C=\frac{1}{2}  
\sum_{n,n^{\prime}}\sum_{\sigma,\sigma^{\prime}}\sum_{\xi,\xi^{\prime}} \iint 
d\mathbf{r} d\mathbf{r}^{\prime} \, \Phi^{\dagger}_{n_1, \sigma, 
\xi}(\mathbf{r}) \Phi^{\dagger}_{n_2, \sigma^{\prime}, 
\xi^{\prime}}(\mathbf{r}^{\prime})\, V^C(\mathbf{r}-\mathbf{r}^{\prime}) \, 
\Phi_{n_3, \sigma^{\prime}, \xi^{\prime}}(\mathbf{r}^{\prime}) \Phi_{n_4, 
\sigma, \xi}(\mathbf{r}),
\label{eqn:H_C}
\end{equation}
written in terms of the field operator $\Phi_{n, \sigma, 
\xi}(\mathbf{r})=\sum_{p}  \LD \mathbf{r}|n,\sigma,\xi;p  \RD 
c_{n,\sigma,\xi}(p)$.
 As a first approximation to the electron-electron 
interaction, the fully symmetric potential is $V^C=\frac{e^2}{\kappa 
|\mathbf{r}-\mathbf{r}^{\prime}|}$ with $\kappa$ the effective dielectric 
constant can be chosen. A 
more realistic approach to the specific geometry of the bilayer system is given 
by a corrected potential which accounts for the finite distance $d$ between the 
upper and the lower graphene layer: $V^C_{\xi,\xi^{\prime}}=\frac{e^2}{\kappa 
|\mathbf{r}-\mathbf{r}^{\prime}+(1-\delta_{\xi,\xi^{\prime}})d\mathbf{e}_z|}$, 
where $\xi,\xi^{\prime}$ is the valley index. Note that within the four-band 
model of BLG, it is not exact to identify the valley index with the 
sublayer index. We discuss the validity of this 
approximation below. To keep calculations as simple 
as possible, we use the corrected Coulomb potential only when it has 
notable effects.
We treat the electron interactions in self-consistent HF theory. 
We decouple the 
interaction operator into a direct Hartree part $\Hh_{C,D}$ and an exchange 
Fock part $\Hh_{C,X}$ in the following way:
\begin{align}
\nonumber \Hh_C  \longrightarrow & \; \Hh_{C,D}+\Hh_{C,X},\\
\nonumber  \LD  c^{\dagger}_{n_1,\sigma,K}&(p_1)  \, 
c^{\dagger}_{n_2,\sigma^{\prime},K^{\prime}}(p_2) \,   
c_{n_3,\sigma,K}(p_3) \,c_{n_4,\sigma^{\prime},K^{\prime}}(p_4)    \RD\\
\nonumber  \longrightarrow & \;\LD  
c^{\dagger}_{n_1,\sigma,K}(p_1)\,c_{n_4,\sigma^{\prime},K^{\prime}}(p_4)  
\RD\,  \LD  
c^{\dagger}_{n_2,\sigma,K}(p_2)\,c_{n_3,\sigma^{\prime},K^{\prime}}(p_3)  \RD\\
 &\;-  \LD  
c^{\dagger}_{n_1,\sigma,K}(p_1)\,c_{n_3,\sigma^{\prime},K^{\prime}}(p_3)  \RD\;
\LD  c^{\dagger}_{n_2,\sigma,K}(p_2)\,c_{n_4,\sigma^{\prime},K^{\prime}}(p_4)  
\RD.
 \label{eqn:HFdecoupling}
\end{align}
The technical details of the HF method employed are given in Sec.~
\ref{subsec:HF}.
 First, we treat the 
interactions of the electrons within the octet sector (n=$0,1$) before 
analyzing the coupling with the electrons filling the Dirac sea ($n\le-2$). 
 Within in $01$-octet, we consider 
interaction between the electrons via the corrected potential 
$V^C_{\xi,\xi^{\prime}}=\frac{e^2}{\kappa 
|\mathbf{r}-\mathbf{r}^{\prime}+(1-\delta_{\xi,\xi^{\prime}})d\mathbf{e}_z|}$. 
When working with an effective two-band model for the electronic 
states of BLG\cite{McCann2006}, within the zero-mode sector there is a direct 
one-to-one correspondence between the valley  degree of freedom and the 
electrons occupation in the the upper or the lower layer,
respectively\cite{Lambert2013}. Within the four-band model applied throughout 
this work, this correspondence valley $\leftrightarrow$ layer within the 
pseudo-zero mode sector is no longer exact. Close investigation of the 
coefficients of Eq.~\ref{eqn:psi_01_coeff} governing the electronic occupation 
of the different atomic sites on the bilayer lattice reveals the 
following. The occupation of the different sublayers which would stay fully 
unoccupied within the two-band model is governed by the coefficient $b_{(1),3}$ 
in $ \psi^{(1)}_{K}$. The four-band model and the two-band model do predict 
different behavior of the layer occupation of BLG. This will be of 
importance  in the subsequent discussion.  
It is thus crucial  to take into account the different behavior of 
the $n=0$ and the $n=1$ modes within the two models. We estimate the error due to the correspondence valley $\leftrightarrow$ 
layer for each valley index: The coefficient $b_{(1),3}$ is largest in magnitude for zero bias - in this case, the relation $b_{(1),3}^2=\frac{b_{(1),1}^2}{\gamma^2}$ holds. Hence, $b_{(1),3}\ll b_{(1),1}$ since $\gamma\gg1$ for the parameters listed in Table \ref{tab:SysParam}. We therefore use the form of 
the corrected Coulomb potential $V^C_{\xi,\xi^{\prime}}$ given above in order 
to include the effect of the anisotropic Coulomb interaction due to the finite 
separation between the layers. 
We perform the HF decoupling of the Coulomb-interaction term 
in the four-band model as calculations within 
an effective two-band model of BLG presented in \onlinecite{Cote2010} and 
\onlinecite{Lambert2013}. 
The contribution from the direct interaction term competes with a positive, 
neutralizing background and yields a capacitive energy \cite{Lambert2013}
\begin{equation}
 \Hh_{D, \text{ Octet}}=\sum_{p} \sum_{n,\sigma, \xi} \alpha 
\frac{d}{\ell_B}\Big(\tilde{v}_{\xi}-\frac{\tilde{v}}{2}\Big)\, 
c^{\dagger}_{n,\sigma,\xi}(p)\,c_{n,\sigma,\xi}(p),
 \label{eqn:H_D}
\end{equation}
where we denote with $\tilde{v}_{\xi}=\sum_p\sum_{n\sigma}\LD 
c^{\dagger}_{n,\sigma,\xi}(p)\,c_{n,\sigma,\xi}(p)  \RD$ the total filling in 
valley $\xi$,  $\tilde{v} = \nu+4$ counts the total number of filled levels in 
the octet, and $\alpha=\frac{e^2}{\kappa\ell_B}$.
From the exchange part of the interaction we obtain the contribution 
\begin{equation}
 \Hh_{X, \text{ 
Octet}}=-\sum_{\substack{p_1,p_2\\p_3,p_4}}\sum_{\substack{n_1,n_3\\n_2,n_4}} 
\sum_{\substack{\sigma, \xi \\\sigma^{\prime}, \xi^{\prime}}} 
\mathfrak{X}^{\xi,\xi^{\prime}}_{\substack{n_1,n_3\\n_2,n_4}}(0) \,\LD 
c^{\dagger}_{n_1,\sigma,\xi}(p_1)\,c_{n_3,\sigma^{\prime},\xi^{\prime}}(p_3)  
\RD\,  
c^{\dagger}_{n_2,\sigma^{\prime},\xi^{\prime}}(p_2)\,c_{n_4,\sigma,\xi}(p_4),
 \label{eqn:H_Octet,X}
\end{equation}
where, following previous definitions, we find the exchange matrix elements 
\begin{equation}
\mathfrak{X}^{\xi,\xi^{\prime}}_{\substack{n_1,n_2\\n_3,n_4}}(\mathbf{q})=\alpha 
\int 
\frac{d\mathbf{p}\ell_B^2}{2\pi}\frac{1}{p\ell_B}e^{-pd(1-\delta_{\xi,\xi^{
\prime}})}\mathfrak{K}_{n_1,n_4}(p)\mathfrak{K}_{n_3,n_2}(-p)e^{i\mathbf{p}
\times\mathbf{q}\ell_B^2},
\label{eqn:X}
\end{equation}
with 
\begin{align}
\nonumber &\mathfrak{K}_{0,0}(\mathbf{p}) =e^{-\frac{\ell^2_B p^2}{4}}\\
\nonumber &\mathfrak{K}_{0,1}(\mathbf{p}) =e^{-\frac{\ell^2_B p^2}{4}}\frac{c_1 \ell_B}{\sqrt{2}}(ip_x+p_y)\\
\nonumber &\mathfrak{K}_{1,0}(\mathbf{p}) =e^{-\frac{\ell^2_B p^2}{4}}\frac{c_1 \ell_B}{\sqrt{2}}(ip_x-p_y)\\
&\mathfrak{K}_{1,1}(\mathbf{p})=e^{-\frac{\ell^2_B p^2}{4}}(1-c_1\frac{\ell^2_B p^2}{2}).
  \label{eqn:K01}
\end{align}
For future use we introduce the notation $\Delta_{n_1 n_2 n_3 n_4}:= 
\mathfrak{X}^{\xi,\xi}_{\substack{n_1,n_2\\n_3,n_4}}$ for the terms conserving 
the valley index and $X_{n_1 n_2 n_3 n_4}:= 
\mathfrak{X}^{\xi,\xi^{\prime}}_{\substack{n_1,n_2\\n_3,n_4}}$ in the case $\xi 
\neq \xi^{\prime}$ for the valley index non-conserving terms.

 In Ref.~
\onlinecite{Shizuya2012}, Shizuya has shown that exchange
interactions between the electrons in the Dirac sea within the four-band model 
of BLG leads to a splitting $\Delta_{01}^{\text{int}}$ between the 
 $n=0$ and $n=1$ orbitals. 
This exchange phenomenon analogous to the Lamb shift of atomic energy levels
leads to a term of the form (where the LL index only runs over $n=0,1$):
\begin{equation}
 \Hh_{X, \text{ Dirac}}=  \sum_{p} \sum_{n,\sigma, \xi} 
\frac{\Delta_{01}^{\text{int}}}{2}\lambda_z\, 
c^{\dagger}_{n,\sigma,\xi}(p)\,c_{n,\sigma,\xi}(p),
\label{eqn:H_01}
\end{equation}

with 
$\lambda_{\alpha}=\sigma^{\text{mode}}_{\alpha}\otimes\mathbb{1}^{\text{spin}}
\otimes\mathbb{1}^{\text{valley}}$
for the Pauli operators acting in $01$-orbital space and $ 
\Delta_{01}^{\text{int}} 
 = \frac{1}{8}\Delta_{C}\,c_1^2(4-3c_1^2) $ is the splitting induced by the 
presence of the Dirac sea, where we defined 
$\Delta_C=\sqrt{\frac{\pi}{2}}\alpha=\sqrt{\frac{\pi}{2}}\frac{e^2}{\kappa\ell_B
}$.

Considering 
the anisotropic interlayer Coulomb interaction merely entails a simple 
rescaling (at first order in $d/\ell_B$)
$\Delta_B\rightarrow\Delta_{B,eff}=(1-16\frac{W}{\hbar\omega_c})\Delta_B$.
 Assembling all terms from above discussion, we arrive at the HF Hamiltonian 
\begin{equation}
\Hh_{HF}= \Hh_0+ \Hh_{X, \text{ Octet}}+  \Hh_{D, \text{ Octet}} + \Hh_{X, \text{ Dirac}} .
  \label{eqn:H_HFfull}
\end{equation}

Hence, in terms of the order parameter $P_{\substack{n^{\prime},n\\ 
\sigma^{\prime},\sigma; \xi^{\prime},\xi}}(p):=\LD 
c^{\dagger}_{n,\sigma,\xi}(p)\,c_{n^{\prime},\sigma^{\prime},\xi^{\prime}}(p) 
\RD $, and within a local approximation $P({p})\approx P({p^{\prime}})$ for a 
state uniform or varying sufficiently slowly  in space, we  obtain for the HF 
energy functional (suppressing summation over $p$):
\begin{align}
\nonumber E_{HF}&=
-\frac{1}{2} \sum_{\substack{n_1,n_3\\n_2,n_4}} 
\sum_{\substack{\sigma, \xi \\\sigma^{\prime}, \xi^{\prime}}} 
\mathfrak{X}^{\xi,\xi^{\prime}}_{\substack{n_1,n_3\\n_2,n_4}}(\mathbf{0})P_{
\substack { n_3 ,
n_1\\ \sigma^{\prime},\sigma\\ \xi^{\prime},\xi}}P_{\substack{n_4,n_2\\ 
\sigma,\sigma^{\prime}\\ \xi,\xi^{\prime}}}
+ \frac{\alpha}{4}\frac{d}{\ell_B} 
(\tilde{\nu}_K-\tilde{\nu}_{K^{\prime}})^2    \\
&+ \frac{\Delta_{01}}{2}\Tr[\lambda_z P]+ \frac{\Delta_{Z}}{2}\Tr[\sigma_z P]+ 
\frac{\Delta_{B,eff}}{2}\Tr[\tau_z 
P]+z\frac{\Delta_{B,eff}}{4}\Tr[(\tau_z+\lambda_z\tau_z) P],
  \label{eqn:E_HFfull}
\end{align}

where we summarized $\Delta_{01}= \Delta_{01}^{\text{pert}} + 
\Delta_{01}^{\text{int}} $ as the total splitting in orbital space 
induced by the different effects discussed above. We search only uniform HF 
solutions.

\subsection{HF method}
\label{subsec:HF}

In this work, we study the model Hamiltonian $\Hh_{HF}$ given in 
Eq.~\ref{eqn:H_HFfull} within HF theory. In Eq.~\ref{eqn:H_HFfull}, the 
Hamiltonian 
$\Hh_{HF}$ depends on $P$ which in turn itself is determined by the 
lowest-energy solution to the corresponding eigenvalue problem. The numerical 
procedure leading to its solution must thus be carried out self-consistently. We 
briefly sketch the algorithm used in our analysis. 
We fix the total number of electrons in the octet. Here, 
the \textit{filling factor} $\nu$ of the octet is defined with respect to the 
half-filled, charge neutral case: We write $\nu=-3 \; (-2,-1,0,1,2,3)$ for 1 
(2,3,4,5,6,7) out of the eight available zero-energy levels being occupied. In 
Secs.~\ref{sec:HFRESI} and \ref{sec:HFRESII}, we present investigations and 
discussions of all the different possible fillings factors $\nu\in [-3,3]$.
The density matrix is assumed to be independent of the guiding center coordinate
so we are looking only for spatially uniform solutions.
For a given filling factor $\nu$ implying n occupied levels, we start by 
initializing n SP vectors $|i\RD$: The eight entries each are taken  
from a random uniform distribution, thereby respecting normalization. The 
density matrix 
$P^{\text{int}}=\sum_i^n |i\RD\LD i|$ built from these vectors serves as a 
starting point for the self-consistent HF minimization procedure. 

Iteration schemes similar to the one used here and equally based on the 
so-called Roothaan algorithm for self-consistent HF iteration\cite{Roothaan1951} 
have been applied earlier in HF studies of QH systems\cite{Sohrmann2007, 
Romer2008}. 
%

A check for proper convergence to a true 
solution of the HF equations is performed by always requiring the SP energy 
eigenvalues to reproduce the energy yielded by the HF energy functional of 
Eq.~\ref{eqn:E_HFfull} up to a precision better than $10^{-5}$.

From the final converged density matrix $P$ we 
calculate the components of the spin $\mathbf{S}$, the valley isospin 
$\mathbf{T}$, and the orbital isospin degree of freedom $\mathbf{L}$ according 
to 

\begin{equation}
S_{\alpha}=\frac{1}{2}\Tr[\sigma_{\alpha}P], \hskip10pt 
T_{\alpha}=\frac{1}{2}\Tr[\tau_{\alpha}P], \hskip10pt 
L_{\alpha}=\frac{1}{2}\Tr[\lambda_{\alpha}P],
\end{equation}

for $\alpha\in\{x,y,z\}$. We identify different  phases by different 
configurations of the spin and isospin degrees of freedom. By tracing their 
evolution as functions of the external parameters, i.e., the bias 
potential $\Delta_B$ and the magnetic field $B$, we determine the  phase 
diagrams in the $\{\Delta_B$-$B\}$-plane. From this numerical HF procedure we 
furthermore gain information about the HF SP eigenstates and eigenvalues for 
each value of  $\Delta_B$ and $B$. Hence we can infer the structure of the 
occupied and unoccupied SP states for each phase within this HF MF picture. This 
knowledge about the GS structure allows us to proceed further by analytical 
means: Using a particular structure of the GS eigenvectors to construct the 
corresponding density matrix $P$ and minimizing the HF energy functional given 
in Eq.~\ref{eqn:E_HFfull} for this $P$, allows us to compute analytically 
properties of the various phases such as canting angles of the energetically 
favorable spin and isospin orientation or phase boundaries between different GS 
phases. 

We have not tried to search for spatially non-uniform HF solutions. There is no 
clear experimental evidence for such states so far. The HF investigations of 
Lambert and C\^ot\'e have found such solutions only at very large bias.

\section{HF  Phase Diagram}
\label{sec:HFRESI}

\begin{figure}[!htb]
  \centering
  \begin{subfigure}[!htb]{0.33\textwidth}
\includegraphics[width=\textwidth]{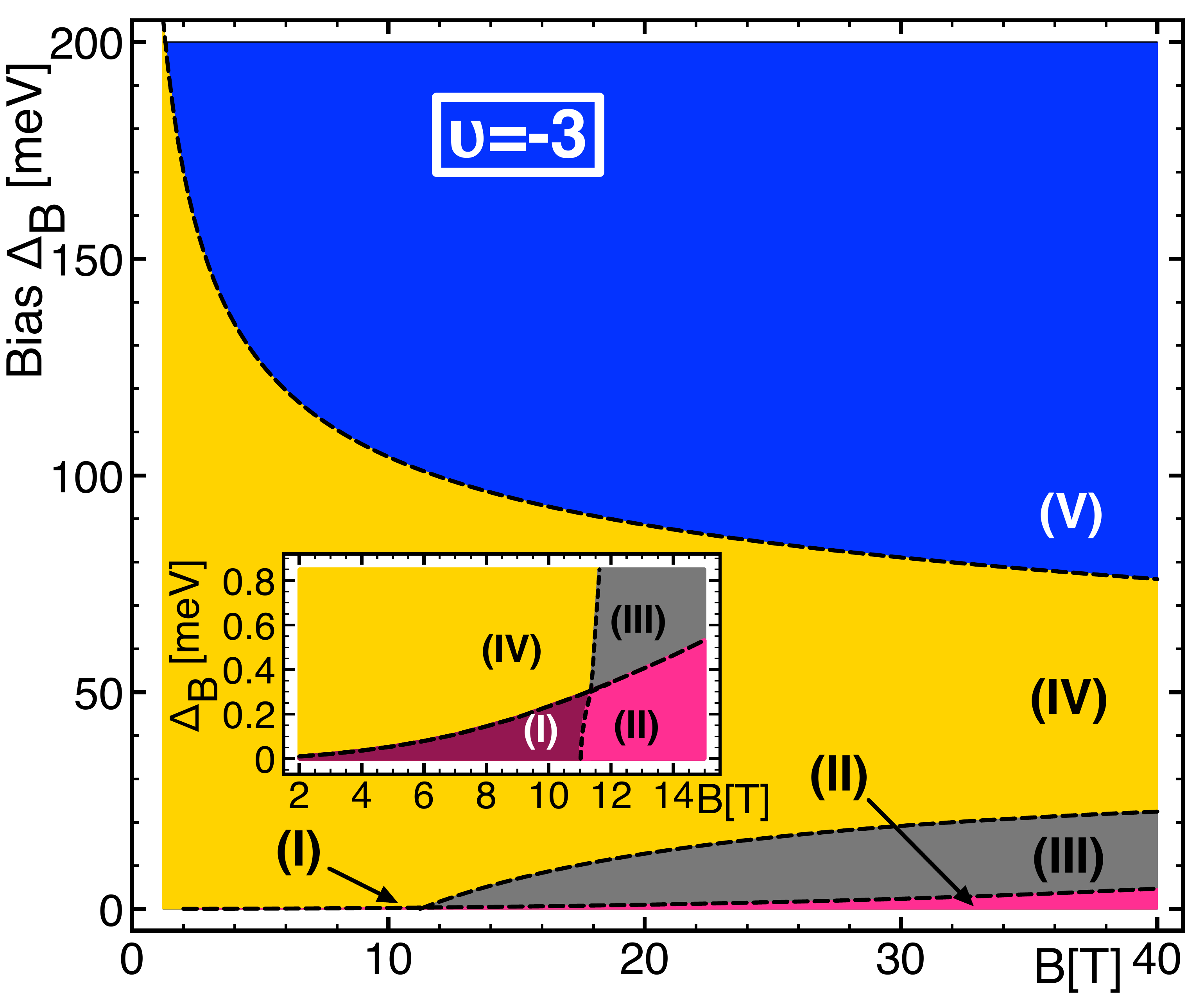}
\label{fig:b1iofM}
 \end{subfigure}\,
  \begin{subfigure}[!htb]{0.33\textwidth}
\includegraphics[width=\textwidth]{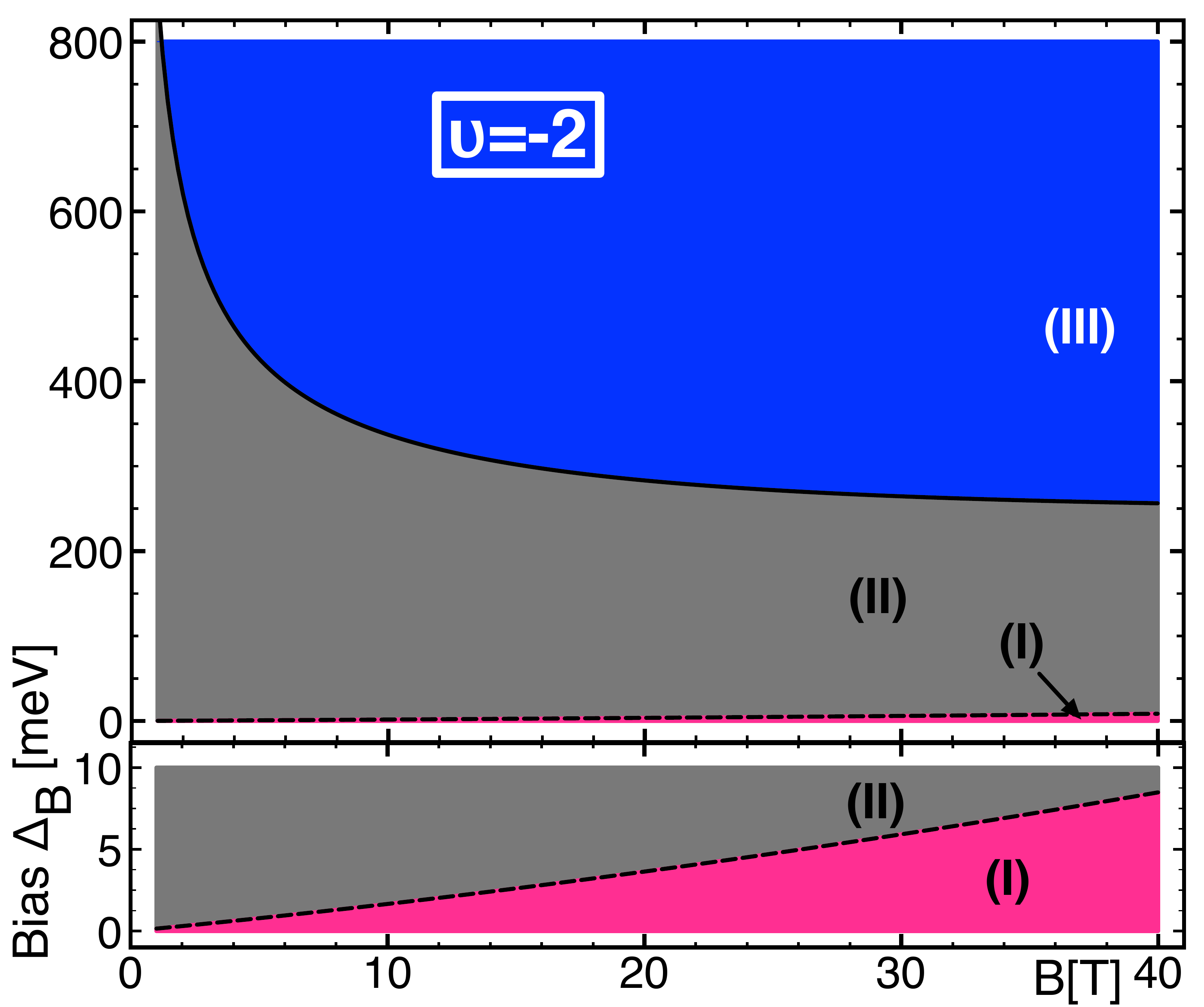}
\label{fig:b1iofM}
 \end{subfigure}\,
  \begin{subfigure}[!htb]{0.33\textwidth}
\includegraphics[width=\textwidth]{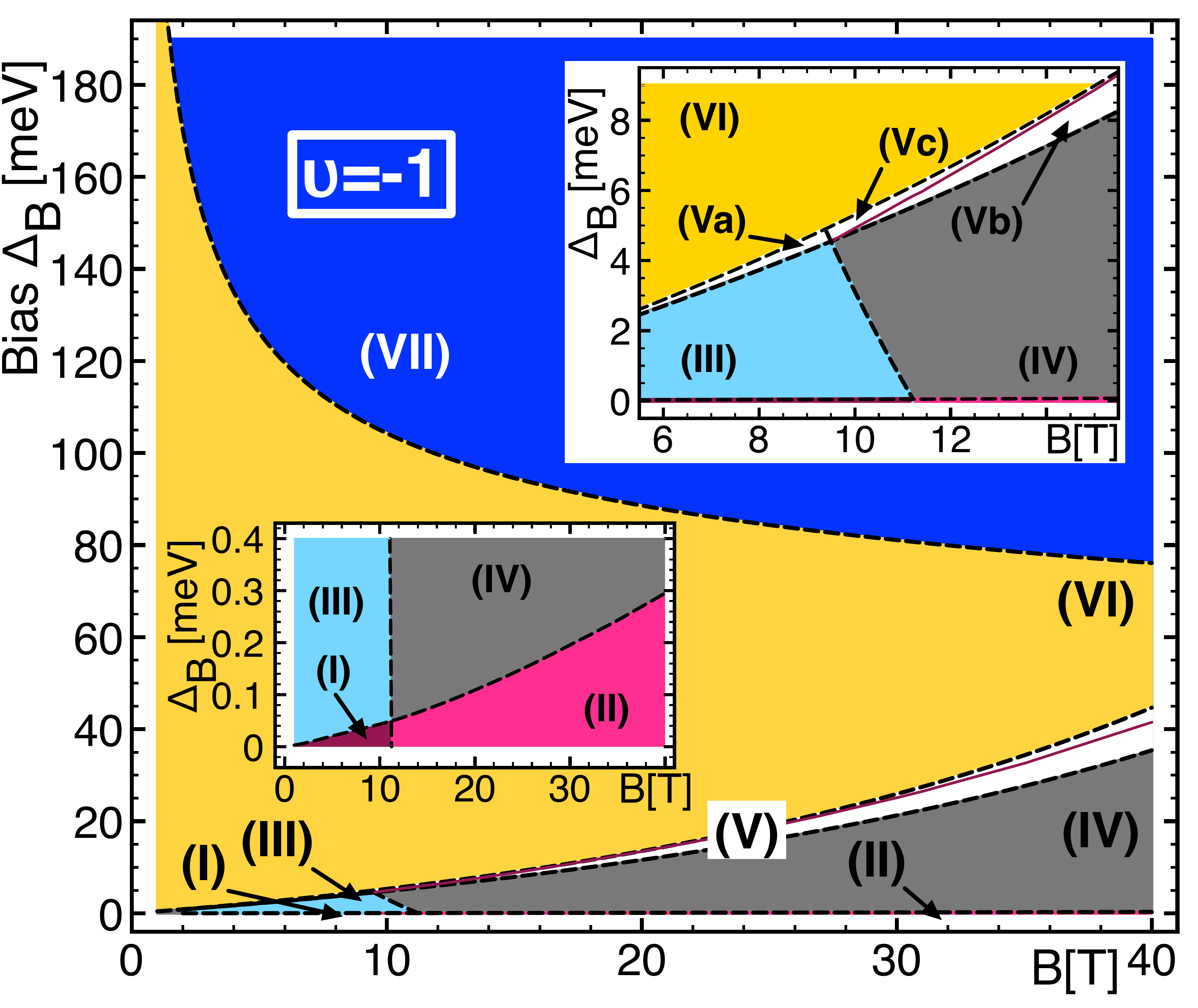}
\label{fig:b1iofM}
 \end{subfigure}
   \begin{subfigure}[!htb]{0.33\textwidth}
\includegraphics[width=\textwidth]{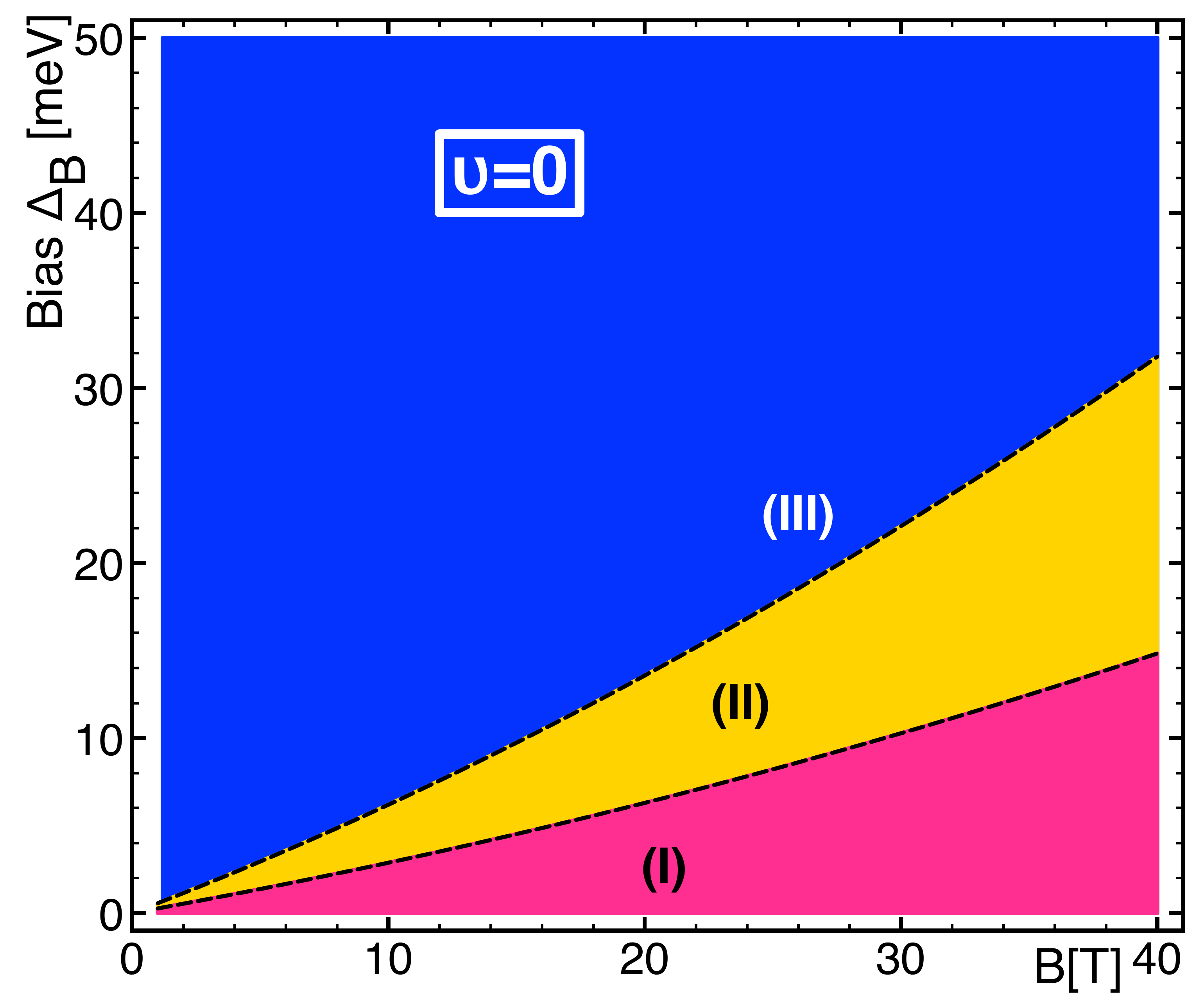}
\label{fig:b1iofM}
 \end{subfigure}
   \begin{subfigure}[!htb]{0.33\textwidth}
\includegraphics[width=\textwidth]{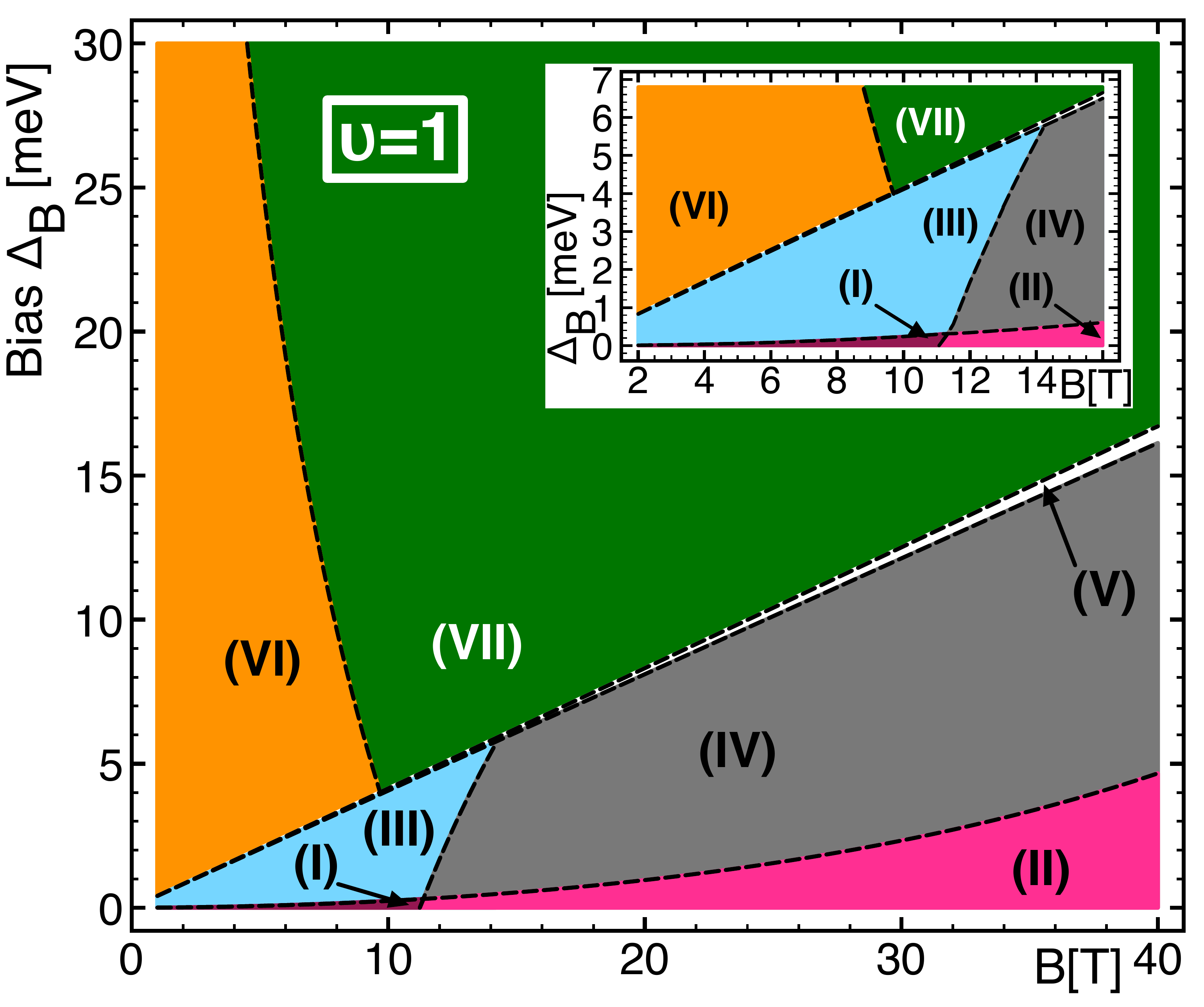}
\label{fig:b1iofM}
 \end{subfigure}
   \begin{subfigure}[!htb]{0.33\textwidth}
\includegraphics[width=\textwidth]{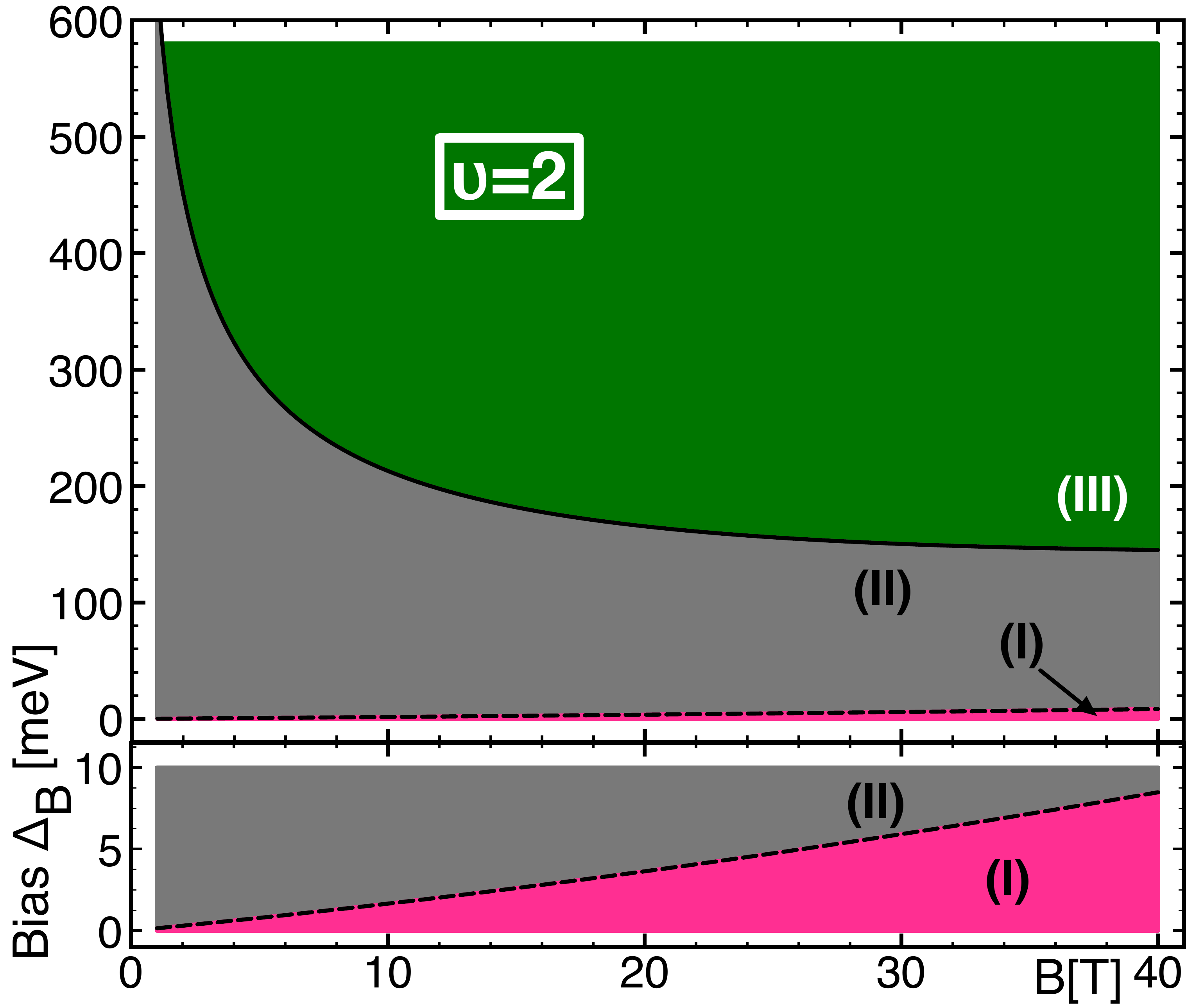}
\label{fig:b1iofM}
 \end{subfigure}
   \begin{subfigure}[!htb]{0.33\textwidth}
\includegraphics[width=\textwidth]{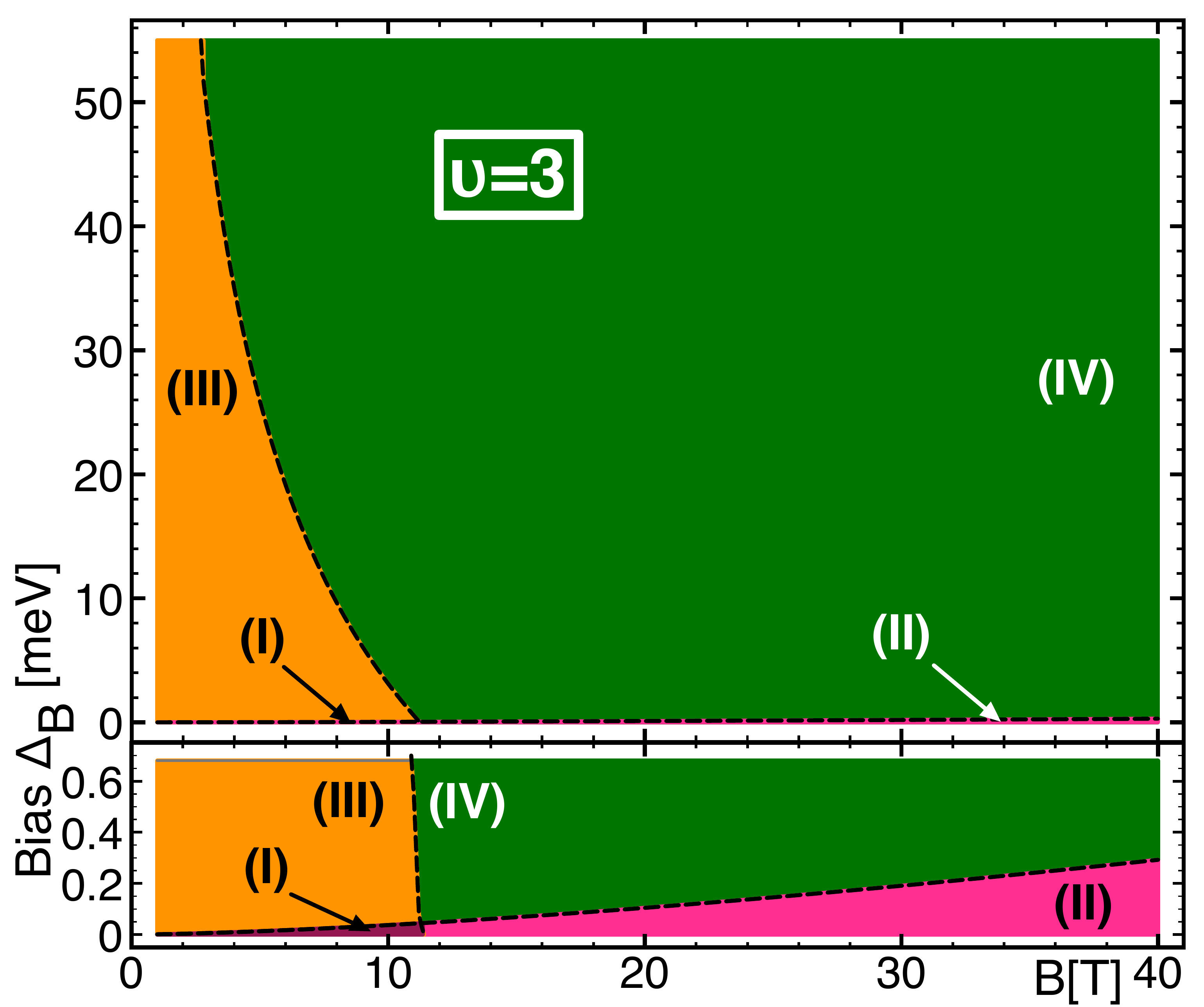}
\label{fig:b1iofM}
 \end{subfigure}
   \caption{Phase diagrams obtained for BLG at the different 
filling factors $\nu\in[-3,3]$. The GS behavior of the Hamiltonian $\Hh_{HF}$ of 
Eq.~\ref{eqn:H_HFfull} is studied with the HF methods described in Sec.~
\ref{subsec:HF}. We employ the following color code for the phases: bordeaux/magenta: $\mathbf{S} \propto \mathbf{e}_z$ and $\mathbf{T}$ in a canted state $\rightarrow$ valley coherence;  yellow/orange: $\mathbf{S} \propto \mathbf{e}_z$ and $\mathbf{L}$ in a canted state $\rightarrow$ orbital coherence; gray/blue/green:  $ \mathbf{S},\mathbf{T} \propto \mathbf{e}_z$ for $\mathbf{L} \equiv 0$ or $\mathbf{L}  \propto \pm \mathbf{e}_z$$\rightarrow$ partial polarization.  }
 \label{fig:GSPDs}
\end{figure}

We present the phase diagrams of BLG obtained for different filling factors 
$\nu$ using the HF procedure described in the previous section \ref{subsec:HF}. 
 In Fig.~\ref{fig:GSPDs}, we plot the phase diagrams for the different $\nu$ in the plane 
spanned by the bias $\Delta_B$ and the magnetic field $B$. From these phase 
diagrams, we identify a total of 32 different phases of the BLG system at 
different 
filling factors. The explicit form of the respective phases are listed in  
Tables \ref{tab:GS_M3}-\ref{tab:GS_3}. Their spin and isospin polarization 
properties are given in Tables \ref{tab:PolProp_negMu} and 
\ref{tab:PolProp_posMu} 
for negative and positive filling factors, respectively. In the first section of 
the text, we successively discuss the respective cases for each different 
filling factor.  We next 
summarize these results and compare our findings for different $\nu$ among 
each other.


Figure \ref{fig:GSPDs} shows the collection of  phase diagrams obtained for the 
different filling factors $\nu \in [-3,3]$.

\subsection{One electron: $\nu=-3$}
With one electron per orbital in the system, we find the following different 
phases:

Unbiased case ($\Delta_B\equiv0$, evolution as a function of $B$): the GS is 
polarized along the $z$-axis in the spin degree of freedom, but not in the 
valley degree of freedom, where the corresponding isospin vector lies in 
the $\{x$-$y\}$-plane. The isospin corresponding to the orbital mode is in a 
canted configuration, thus we find a phase with orbital coherence. The optimal 
canting angle in orbital space $\theta_0$, plotted in 
Fig.~\ref{fig:AngleB20}, varies as a function of $B$ between 
$\theta_0\rightarrow\frac{\pi}{4}$ at vanishing magnetic field 
and $\theta_0\equiv0$ at sufficiently high magnetic field strengths above a 
certain critical value $B_{crit}\approx11$ T. It is given by the relation
\begin{equation}
\cos\theta_0=\frac{\sqrt{- \Delta_{0011}+2 
\Delta_{01}-\Delta_{1001}+\Delta_{0000}-X_{0011}-X_{1001}+X_{0000}}}{\sqrt{
\Delta_{0000}-2 \Delta_{0011} -2\Delta_{1001}+\Delta_{1111}+ X_{0000}-2 
X_{0011}-2  X_{1001}+X_{1111}}}.
\label{eqn:NuM3_t0}
\end{equation}

Along the line of zero bias, the GS hence undergoes a transition from a canted 
to a fully polarized state in orbital isospin with increasing magnetic field 
strength $B$.

Phase (I) occurs at small but nonzero bias $\Delta_B>0 $ and below a 
critical magnetic field strength $B<B_{crit}$: the GS is spin polarized and 
canted both in valley  and orbital degrees of freedom. In phase (I), cuts 
along lines of increasing bias $\Delta_B$, for any strength of the magnetic 
field $B$, correspond to a rotation of the valley-isospin vector from a 
configuration in the $\{x$-$y\}$-plane to a state fully polarized along the 
$z$-axis. At the same time, there is orbital coherence, the orbital isospin
components taking non-trivial values $0 < \,L_{x},\, L_z \,<1$.

Phase (II): still in the regime of very small values of the bias  $\Delta_B$, 
but for larger magnetic fields  $B>B_{crit} $, the system is fully polarized in 
the spin and orbital isospin degree of freedom. The valley isospin, however, is 
in a canted configuration,
where the optimal angle is determined by 
\begin{equation}
\cos2\theta_{II} =\frac{\Delta_{B,eff}\; {\ell_B} (z-1)}{\alpha  d+{\ell_B} 
(\Delta_{1111}-X_{1111})}.
\end{equation}

Hence, in phase (II), along any cut at a fixed $B>B_{crit}$, as  the bias 
$\Delta_B$ increases, the state undergoes a rotation of the valley-isospin from 
$\mathbf{T}$ lying in the $\{x$-$y\}$-plane at $\Delta_B=0$ to a fully valley 
polarized state at sufficiently large $\Delta_B$.

Phase (III) emerges as an intermediate phase at sufficiently large values of 
the magnetic field $B>B_{crit}\approx11$ T when the bias potential is raised 
beyond 
the regime of phase (II): over a certain parameter range of bias and magnetic 
field strength, the system becomes a fully polarized ferromagnet in all spins 
and isospins.

Phase (IV) dominates the intermediate part of the $\nu=-3$ phase diagram over 
the whole parameter range of bias $\Delta_B$ and magnetic field strength $B$. It 
is characterized by full ferromagnetic polarization of the spin and valley 
isospin, but canting of the orbital isospin resulting in an orbital coherent 
phase. 
For the optimal canting angle in orbital space we find the expression
\begin{equation}
\cos2\theta_{IV}=\frac{-\Delta_{0000}-2 \Delta_{01}+\Delta_{1111}+
z\Delta_{B,eff}}{\Delta_{0000}-2 \Delta_{0011}-2 \Delta_{1001}+\Delta _{1111}}.
\label{eqn:NuM3_tIV}
\end{equation}

In phase (IV), cuts as a function of increasing bias $\Delta_B$ at any value of 
the magnetic field hence trace the rotation of the orbital isospin vector to the 
fully antiferromagnetically-polarized state in orbital space.

Phase (V): for sufficiently large values of the bias, we find the limiting case 
for the GS to be fully polarized in spin and valley isospin, but 
antiferromagnetically-polarized 
in the orbital degree of freedom.
At $\nu=-3$, all phases transform into one another via smooth rotations of the 
respective isospin degrees of freedom. All transitions between different 
phases therefore are of second order in this case. These phase transitions occur 
at the following critical values of the bias, respectively:

(II) $\rightarrow$  (III):
\begin{equation}
\Delta_{B,eff}^{crit}=\frac{-\alpha  d-\Delta_{1111} {\ell_B}+{\ell_B} 
X_{1111}}{{\ell} (z-1)},
\end{equation}

(III) $\rightarrow$ (IV):
\begin{equation}
\Delta_{B,eff}^{crit}=\frac{2}{z} (\Delta_{0011}+\Delta_{01}+\Delta_{1001}-\Delta_{1111}),
\end{equation}

(IV) $\rightarrow$ (V):
\begin{equation}
\Delta_{B,eff}^{crit}=\frac{2}{z} (\Delta_{0000}-\Delta_{0011}+\Delta_{01}-\Delta_{1001}).
\end{equation}

\begin{table}
\centering
\begin{tabular}{ |l | c |}
\hline
   $\Delta_B\equiv0$ & $ 
|v_1\RD=\frac{1}{\sqrt{2}}\cos\theta\Big[\,|1,\uparrow,+\RD+|1,\uparrow,-\RD\,
\Big]+\frac{1}{\sqrt{2}}\sin\theta\Big[\,|0,\uparrow,+\RD+|0,\uparrow,-\RD\,\Big
] $ \\
\hline  Phase (I) &$ |v_1\RD=a_1 |1,\uparrow,+\RD+ a_2  |1,\uparrow,-\RD + a_5|0,\uparrow,+\RD+ a_6|0,\uparrow,-\RD$ \\
\hline  Phase (II) & $|v_1\RD=\sin\theta|1,\uparrow,+\RD + \cos\theta|1,\uparrow,-\RD$  \\
\hline    Phase (III) & $|v_1\RD=|1,\uparrow,+\RD $  \\
\hline      Phase (IV) & $|v_1\RD=\sin\theta|1,\uparrow,+\RD + \cos\theta|0,\uparrow,+\RD$  \\
\hline        Phase (V) & $|v_1\RD=|0,\uparrow,+\RD$  \\
\hline
\end{tabular}
\caption{GS configurations for the different phases for filling factor $\nu=-3$.}
\label{tab:GS_M3}
\end{table}

\subsection{Two electrons: $\nu=-2$}

When there are two electrons per state within the octet, 
the GS structure of the system is the following:

Phase (I): for small values of the bias $\Delta_B<\Delta_B^{crit}$, the GS is 
partially polarized in the spin, whereas the valley isospin is canted and the 
orbital isospin is ordered in an antiferromagnetic way. The optimal valley 
canting angle is determined by 
\begin{equation}
\cos2\theta_{I}=\frac{\Delta_{B,eff} {\ell_B}( z-2) }{-4 \alpha  
d-\Delta_{0000} 
{\ell_B}-2 \Delta_{0011} {\ell_B}-\Delta_{1111} {\ell_B}+{\ell_B} X_{0000}+2 
{\ell_B} 
X_{0011}+{\ell_B} X_{1111}}.
\end{equation}

Hence, in this phase, cuts along lines of increasing bias $\Delta_B$, for any 
strength of the magnetic field $B$, correspond to a rotation of the 
valley-isospin vector from a configuration in the $x$-$y$-plane to fully aligned 
along the $z$-axis.

 Phase (II): within an intermediate range of the bias $\Delta_B$, the GS 
is a fully polarized ferromagnet in spin and valley isospin. The orbital 
isospin degree of freedom, however, is in an antiferromagnetic configuration 
yielding zero overall orbital polarization.

 Phase (III): in the limit of a sufficiently large bias $\Delta_B$, we find the 
GS to be an antiferromagnet in spin. The valley isospin is fully polarized, 
whereas the orbital isospin turns out to be fully 
antiferromagnetically polarized.

\begin{table}
\centering
\begin{tabular}{| l | c |}
\hline
  $\Delta_B\equiv0$&$  |v_1\RD= \frac{1}{\sqrt{2}} \Big[\, |1,\uparrow,+\RD+   |1,\uparrow,-\RD\, \Big],
 |v_2\RD= \frac{1}{\sqrt{2}} \Big[\,  |0,\uparrow,+\RD+   |0,\uparrow,-\RD\, \Big], $ \\
\hline
  Phase (I) &$  |v_1\RD=\cos\theta |1,\uparrow,+\RD+ \sin\theta  |1,\uparrow,-\RD,
 |v_2\RD=\cos\theta |0,\uparrow,+\RD+ \sin\theta  |0,\uparrow,-\RD, $ \\
 \hline
  Phase (II) &$ \nonumber|v_1\RD=|1,\uparrow,+\RD , 
|v_2\RD=|0,\uparrow,+\RD. $  \\
    \hline
    Phase (III) &$|v_1\RD=|0,\uparrow,+\RD ,
|v_2\RD=|0,\downarrow,+\RD,$ \\
\hline
\end{tabular}\caption{The different GS configurations which occur at filling $\nu=-2$.}\label{tab:GS_M2}\end{table}

At filling factor $\nu=-2$, we observe two different types of phase transitions: 
going from phase (I) to phase (II) is achieved by a smooth rotation of the 
valley isospin. This is a  second order transition. From phase (II) to 
phase (III), however, the system undergoes jumps in spin and orbital isospin 
degree of freedom, which characterizes a discontinuous first order phase 
transition. The critical values of the bias for these transitions read, 
respectively:

(I) $\rightarrow$ (II) :
\begin{equation}
\Delta^{crit}_{B,eff}=\frac{-4 \alpha  d-\Delta_{0000} {\ell_B}-2 \Delta_{0011} 
{\ell_B}- \Delta_{1111} {\ell_B}+{\ell_B}  X_{0000}+2 {\ell_B}  
X_{0011}+{\ell_B} 
X_{1111}}{{\ell_B} (z-2)}.
\end{equation}

(II) $\rightarrow$ (III):
The GS of phase (II) is lower in energy than the GS of phase (III) up to a critical bias 
\begin{equation}
\Delta^{crit}_{B,eff}=\frac{1}{z}(\Delta_{0000}-2 \Delta_{0011}+2 \Delta_{01}-\Delta_{1111}+2  \Delta_{Z}).
\end{equation}

\subsection{Three electrons: $\nu=-1$}

When there are three electrons in the system, we find the following GS
structure:

Unbiased case ($\Delta_{B}\equiv0$, evolution as a function of $B$): the GS is 
a 
fully polarized spin ferromagnet, while its valley isospin lies in the 
$\{x$-$y\}$-plane,  and the orbital isospin is canted in an orbital coherent 
phase. The optimal canting angle in orbital space $\theta_0$ as shown in 
Fig.~\ref{fig:AngleB20} varies as a function of $B$ between 
$\theta_0\rightarrow\frac{\pi}{4}$ at vanishing magnetic field $B\rightarrow0$ 
and $\theta_0=\frac{\pi}{2}$ at  magnetic field strengths above 
 $B_{crit}\approx11.3$ T. It fulfills the relation
\begin{equation}
\cos 2\theta_0= \frac{-3  \Delta_{0000}-4  \Delta_{01}+3  
\Delta_{1111}+X_{0000}-X_{1111}}{\Delta_{0000}-2 ( 
\Delta_{0011}+\Delta_{1001})+\Delta_{1111}+ X_{0000}-2 ( X_{0011}+ 
X_{1001})+X_{1111}}.
\label{eqn:NuM1_t0}
\end{equation}

Along the line of zero bias, as a function of increasing magnetic field strength 
$B$, the GS hence undergoes a transition from a canted state in the orbital 
isospin to a partially polarized state.

Phases (I) and (II): in the regime of very small bias $\Delta_B$, we find a 
rotation of the valley-isospin at either canted or partially aligned orbital 
isospin, respectively. In both phases (I) and (II) the GS is a fully 
polarized spin ferromagnet. The valley isospin assumes non-trivial 
configurations $0\le T_x, T_z\le1$, $T_y\equiv0$. Phase (I) occurs for 
sufficiently small values of the magnetic field, $B<B_{crit}$; the corresponding 
GS is given by an involved superposition of different SP states (cf.~ table 
\ref{tab:GS_M1}) which leads to a non-trivial isospin configuration. In phase 
(II), however, \textit{i.e.}~at field values above the critical magnetic field, 
we can describe the valley isospin in simple terms with the valley canting angle 
$\theta$ as the only parameter, where the optimal angle turns out to be
\begin{equation}
\cos 2\theta_{(II)}= \frac{ \Delta_{B,eff} {\ell_B}}{\alpha  d+{\ell_B} ( 
\Delta_{0000}-X_{0000})}.
\end{equation}

The orbital isospin in phase (I) is in a canted 
configuration, $0\le L_x, L_z\le1$, $L_y\equiv0$, whereas phase (II) is 
partially polarized in orbital space.
Hence, in phase (I) and (II), cuts along lines of increasing bias $\Delta_B$ for 
any strength of the magnetic field $B$ correspond to a rotation of the 
valley-isospin vector from a configuration in the $\{x$-$y\}$-plane to a state 
fully aligned along the $z$-axis. At the same time, increasing $B$ at a fixed 
value of the bias $\Delta_B$ corresponds to rotating the LL-isospin from a 
canted configuration in phase (I) to a partly polarized configuration in phase 
(II).

 Phase (III) and (IV): At larger values of the bias $\Delta_B$ we find pendants 
of phase (I) and (II), now at polarized configurations of the valley isospin: 
here, the GS is a fully polarized spin ferromagnet and a partially 
polarized valley isospin, while the orbital isospin degree of freedom again 
varies as function of the bias  $\Delta_B$ and the magnetic field strength $B$: 
It is canted for small values of the magnetic field in phase (III) with optimal 
canting angle
\begin{equation}
\cos2\theta_{\text{(III)}} =\frac{ \Delta_{0000}+2  \Delta_{01}- 
\Delta_{1111}+z\Delta_{B,eff} }{ \Delta_{0000}-2 ( \Delta_{0011}+ 
\Delta_{1001})+ \Delta_{1111}},
\end{equation}
 
 which evolves into the partially polarized phase (IV)  above a critical 
value of the field $B_{crit}$. Hence, increasing the magnetic field strength $B$ 
corresponds to rotating the orbital isospin from a canted configuration in phase 
(III) to a partially polarized state in phase (IV). Both the spin and the 
valley isospin vectors remain constant in these phases for all values of 
$\Delta_B$ and $B$.

 Phase (V): A narrow transition regime is established with complex behavior of 
the GS configuration. All spin and isospin degrees of freedom take nontrivial 
values and  evolve as functions of $\Delta_B$ and $B$. Exploiting the 
notation of the states as given in Table \ref{tab:GS_M1}, we write  
$S_z=1+\frac{1}{2}(c_{1}^2-c_{2}^2+c_{3}^2-c_{4}^2),\; S_x\equiv  S_y\equiv0$, 
$T_z=1-\frac{1}{2}(c_{1}^2-c_{2}^2+c_{3}^2-c_{4}^2),\; T_x\equiv   T_y\equiv0$, 
and $L_z=\frac{1}{2}(c_{1}^2+c_{2}^2-c_{3}^2-c_{4}^2)  ,\; L_x=-(c_{1} 
c_{3}+c_{2} c_{4} ) ,\;  L_y\equiv0.$ Within the parameter range of phase (V) 
one can distinguish between the following regimes:
Phase (Va): For $B<B_{crit}$,  all four entries $c_i\neq0$ evolve smoothly as 
functions of the bias $\Delta_B$ and the magnetic field strength $B$. For 
increasing  $\Delta_B$ across phase (Va), this leads to smooth evolution of the 
spin and valley isospins from $S_z=\frac{3}{2}$ to $S_z=\frac{1}{2}$ and  from 
$T_z=\frac{1}{2}$ to $T_z=\frac{3}{2}$, respectively, accompanied by kinks in 
the orbital isospin components which are nonzero within this range: 
$0<L_z<\frac{1}{2}$ and $0<L_x<\frac{1}{2}$. 
Phases (Vb) and (Vc): For $B>B_{crit}$, two competing transitions occur within  
the parameter range of  phase (V): there is a smooth evolution of  $S_z$ and 
$T_z$ as in the former case. It is governed by smoothly evolving entries $c_1$ 
and $c_2$ while $c_3\equiv c_4\equiv0$ (so that 
$L_x\equiv0$ and $L_z\equiv\frac{1}{2}$ fixed by normalization). At a 
sufficiently high value of $\Delta_B$, eventually, $c_4$ jumps to a nonzero 
value, thereby inducing nonzero values of $L_z$ and $L_x$ and nontrivial 
evolution of all spin and isospin degrees of freedom. 
Phase (VI) occupies a wide parameter range including all magnetic field 
strengths and intermediate values of the bias $\Delta_B$. While the spin is 
partially polarized and the valley isospin is fully polarized, the orbital 
isospin is in a canted configuration, assuming the optimal canting angle
\begin{equation}
\cos2\theta_{\text{(VI)}} =  \frac{ \Delta _{0000}+2 \Delta _{01}- \Delta 
_{1111}-z \Delta _{Beff} }{ \Delta _{0000}-2 (\Delta _{0011}+ \Delta 
_{1001})+\Delta _{1111}}.
\label{eqn:NuM1_tVI}
\end{equation}
Hence, for any value of the magnetic field $B$, with rising bias $\Delta_B$ the 
orbital isospin performs a rotation to a partially antiferromagnetically 
polarized configuration: 
$\mathbf{L}\rightarrow-\frac{1}{2}\mathbf{e}_z$.

 Phase (VII): For sufficiently large values of the bias, the GS phase eventually 
reaches a configuration which is partially polarized in spin, fully polarized in 
the valley isospin, and partially antiferromagnetically-polarized in the orbital 
isospin degree 
of freedom.
 Except for the transition regime of phase (V) described above, all phase 
transitions of the $\nu=-1$ phase diagram go along with smooth rotations of the 
respective isospin vectors and therefore are of second order. The most prominent transitions 
occur at the following critical values of the bias:
 
  (II) $\rightarrow$ (IV):
\begin{equation}
\Delta_{B,eff}^{crit}= \frac{\alpha  d+ \Delta_{0000} {\ell_B}-{\ell_B}  
X_{0000}}{{\ell_B}},
\end{equation}

(III) $\rightarrow$ (IV):
\begin{equation}
\Delta_{B,eff}^{crit}= -\frac{2}{z} ( \Delta_{0011}+ \Delta_{01}+ \Delta_{1001}- \Delta_{1111}),
\end{equation}

(IV) $\rightarrow$ (Vb):
\begin{equation}
\Delta_{B,eff}^{crit}=\frac{-\alpha  d+\Delta_{1111} {\ell_B}- \Delta_{Z} 
{\ell_B}-{\ell_B} X_{1111}}{{\ell_B} (z-1)},
\end{equation}

 (VI) $\rightarrow$ (VII):
\begin{equation}
\Delta_{B,eff}^{crit}= \frac{2}{z} ( \Delta_{0000}- \Delta _{0011}+ \Delta _{01}- \Delta _{1001}).
\end{equation}

\begin{table}
\centering
\begin{tabular}{ |l | c |}
\hline
  $\Delta_B=0$ &$    |v_1\RD= -\frac{1}{\sqrt{2}} \Big[\,|1,\uparrow,+\RD+|1,\uparrow,-\RD\,\Big], 
  |v_2\RD= \frac{1}{\sqrt{2}} \Big[\,|0,\uparrow,+\RD+|0,\uparrow,-\RD\,\Big]$,\\
  
&$|v_3\RD=-\frac{1}{\sqrt{2}}\sin\theta\Big[\,|1,\uparrow,+\RD-|1,\uparrow,-\RD\
,\Big]-\frac{1}{\sqrt{2}}\cos\theta\Big[\,|0,\uparrow,+\RD-|0,\uparrow,-\RD\,
\Big],$\\
\hline
  Phase (I) & $ |v_1\RD= -a_1|1,\uparrow,+\RD- a_2|1,\uparrow,-\RD+ b_1|0,\uparrow,+\RD+b_2|0,\uparrow,-\RD, $\\
 & $ |v_2\RD= b_1|1,\uparrow,+\RD+b_2 |1,\uparrow,-\RD+ a_1|0,\uparrow,+\RD+a_2|0,\uparrow,-\RD $,\\
 & $|v_3\RD=c_1|1,\uparrow,+\RD- c_2|1,\uparrow,-\RD- c_3|0,\uparrow,+\RD+c_4|0,\uparrow,-\RD,$\\
\hline
  Phase (II) & $ |v_1\RD= -\cos\theta|1,\uparrow,+\RD- \sin\theta|1,\uparrow,-\RD , 
  |v_2\RD=   \cos\theta|0,\uparrow,+\RD+\sin\theta|0,\uparrow,-\RD ,$\\
& $|v_3\RD=\sin\theta|1,\uparrow,+\RD- \cos\theta|1,\uparrow,-\RD,   $\\
    \hline
    Phase (III) & $   |v_1\RD= | 1,\uparrow,+\RD  ,  |v_2\RD=   |0,\uparrow,+\RD , |v_3\RD=\cos\theta|1,\uparrow,-\RD+\sin\theta|0,\uparrow,-\RD$\\
\hline
 Phase (IV) & $  |v_1\RD=  1,\uparrow,+\RD  , 
  |v_2\RD=   |0,\uparrow,+\RD ,
 |v_3\RD= |1,\uparrow,-\RD$\\
 \hline
 Phase (V) & $|v_1\RD=  |1,\uparrow,+\RD,   |v_2\RD=  |0,\uparrow,+\RD ,$\\
&$|v_3\RD=c_1|1,\uparrow,-\RD+ c_2|1,\downarrow,+\RD-c_3|0,\uparrow,-\RD-c_4|0,\downarrow,+\RD$\\
 \hline
 Phase (VI) & $ |v_1\RD=  |1,\uparrow,+\RD  , |v_2\RD=  |0,\uparrow,+\RD , 
|v_3\RD=  \cos\theta|1,\downarrow,+\RD +\sin\theta|0,\downarrow,+\RD$\\
  \hline
 Phase (VII) & $ |v_1\RD=  |1,\uparrow,+\RD  , |v_2\RD=  |0,\uparrow,+\RD , |v_3\RD= |0,\downarrow,+\RD$\\
\hline
\end{tabular}\caption{The possible GS configurations identified for the phase 
diagram at filling factor $\nu=-1$.}\label{tab:GS_M1}\end{table}

\subsection{Four electrons: $\nu=0$}

The bilayer system is charge neutral when there are four electrons per 
state occupying 
exactly half of the states within the octet. For this configuration of half 
filling 
we find the following different GS phases:

Phase (I): in the unbiased configuration as well as for sufficiently small 
values of the bias $\Delta_B$, the GS is  a fully polarized spin ferromagnet, 
while it is an antiferromagnet both in valley and in orbital space, leading to 
vanishing overall valley and orbital polarization.
 Phase (II): for all magnetic field strengths and in an intermediate regime of 
the bias, the spin and the valley isospin undergo evolution as functions of 
$\Delta_B$ and $B$ as a function of one angle $\theta$, which minimizes the 
energy for
\begin{equation}
\cos2\theta_{(II)}=\frac{4 \alpha  d+2 \Delta _{Z} {\ell_B}+ \Delta _{Beff} 
{\ell_B} 
(z-2)}{4 \alpha  d+{\ell_B} ( \Delta _{0000}+2 \Delta_{0011}+ \Delta_{1111}- 
X_{0000}-2 X_{0011}- X_{1111})}.
\end{equation}

In orbital space, the state is an antiferromagnet, with zero orbital 
polarization.
Hence, for a given value of the magnetic field $B$, upon increasing the bias 
$\Delta_B$ over the parameter range of phase (II), the total spin evolves from a 
fully aligned state to a state with zero total spin, while contrarily the total 
valley isospin evolves from zero to a fully polarized valley ferromagnet state: 
$\mathbf{S}=2\mathbf{e}_z \longrightarrow \mathbf{S}\equiv0$, 
$\;\mathbf{T}\equiv0 \longrightarrow \mathbf{T}=2\mathbf{e}_z$.

 Phase (III): For sufficiently large values of the bias $\Delta_B$ the GS
assumes antiferromagnetic order in both spin space and in the space of the 
orbital isospin, while being a fully polarized ferromagnet in valley space. 
 The transitions between the different GS phases of $\nu=0$ are all 
characterized by smooth rotations of the isospin degrees of freedom indicating 
continuous second order transitions. We give the critical values of the bias at 
which these phase transitions occur:
 
 (I) $\rightarrow$ (II):
\begin{equation}
\Delta^{crit}_{B,eff}=  \frac{\Delta_{0000}+2 \Delta_{0011}+\Delta_{1111}-2 \Delta_{Z}-X_{0000}-2 X_{0011}-X_{1111}}{z-2}  ,
\end{equation}

(II) $\rightarrow$ (III) :
\begin{equation}
\Delta^{crit}_{B,eff}=  \frac{{\ell_B} (- \Delta_{0000}-2  \Delta_{0011}- 
\Delta_{1111}-2  \Delta_{Z}+ X_{0000}+2  X_{0011}+ X_{1111})-8 \alpha  
d}{{\ell_B} 
(z-2)}.
\end{equation}
 
\begin{table}
\centering
\begin{tabular}{| l | c |}
\hline
  Phase (I) &$ |v_1\RD=  |1,\uparrow,+\RD ,
 |v_2\RD=  |1,\uparrow,-\RD ,
 |v_3\RD=  |0,\uparrow,+\RD ,
  |v_4\RD=  |0,\uparrow,-\RD $\\
 \hline
  Phase (II) &  $|v_1\RD=|1,\uparrow,+\RD ,
 |v_2\RD= \cos\theta |1,\uparrow,-\RD + \sin\theta  |1,\downarrow,+\RD $,\\
&  $ |v_3\RD=|0,\uparrow,+\RD,
 |v_4\RD= \cos\theta |0,\uparrow,-\RD + \sin\theta  |0,\downarrow,+\RD$ \\
    \hline
    Phase (III) & $|v_1\RD=  |1,\uparrow,+\RD ,
 |v_2\RD=  |1,\downarrow,+\RD ,
 |v_3\RD=  |0,\uparrow,+\RD ,
 |v_4\RD=  |0,\downarrow,+\RD $  \\
\hline
\end{tabular}\caption{The three different GS we identified for the phase diagram of $\nu=0$.}\label{tab:GS_0}\end{table}

\subsection{Five electrons: $\nu=1$}

For the case of five  electrons  we identify the following GS 
structure:

Unbiased case ($\Delta_{B}\equiv0$, evolution as a function of $B$): At zero 
bias, we find a GS configuration in which the spin is partially polarized, while 
the valley isospin vector lies in the $\{xy\}$-plane. The orbital isospin 
assumes a canted configuration, thus exhibiting non-trivial orbital coherence. 
The optimal canting angle in orbital space $\theta_0$, as shown in 
Fig.~\ref{fig:AngleB20}, varies as a function of $B$ between 
$\theta_0\rightarrow\frac{\pi}{4}$ at vanishing magnetic field $B\rightarrow0$ 
and $\theta_0=0$  above a  
critical value $B_{crit}\approx 11$ T. This angle fulfills the relation
\begin{equation}
\cos \theta_0= \frac{\sqrt{\Delta_{0000}- \Delta_{0011}+2  \Delta _{01}- \Delta 
_{1001}+ X_{0000}- X_{0011}- X_{1001}}}{\sqrt{ \Delta_{0000}-2  \Delta _{0011}-2 
 \Delta _{1001}+ \Delta _{1111}+ X_{0000}-2  X_{0011}-2  X_{1001}+ X_{1111}}}.
\label{eqn:Nu1_t0}
\end{equation}

Along the line of zero bias, as a function of increasing magnetic field strength 
$B$, the GS hence undergoes a transition from a canted state in the orbital 
isospin to a partially polarized state.


Phases (I) and (II): at small values of the bias $\Delta_B$, these phases are
in a partially polarized spin state, while the valley isospin takes 
non-trivial values $0\le T_x, T_z \le 1$. Meanwhile, the orbital isospin is 
either in canted configuration with $0\le L_x, L_z \le 1$ (phase I, for 
sufficiently small values of the magnetic field) or is partially polarized  
(phase II, above some critical magnetic field strength). The former case being 
more involved, in the latter phase (II) we find a single parameter dependence of 
the states' configuration with one optimal angle $\theta$ determined by
\begin{equation}
\cos 2\theta_{(II)}= \frac{\Delta_{B,eff} {\ell_B} (z-1)}{{\ell_B} (X_{1111}- 
\Delta _{1111})-\alpha  d},
\end{equation}

governing the canting in valley space. Hence, in phase (I) and (II), cuts along 
lines of increasing bias $\Delta_B$ for any strength of the magnetic field $B$ 
correspond to a rotation of the valley-isospin vector from a configuration in 
the $\{x$-$y\}$-plane to a state aligned along the $z$-axis: 
$\mathbf{T}=\frac{1}{2}\mathbf{e}_x\; \longrightarrow 
\;\mathbf{T}=\frac{1}{2}\mathbf{e}_z$. Meanwhile, increasing $B$ at fixed value 
of the bias $\Delta_B$ corresponds to rotating the orbital isospin from a canted 
configuration in phase (I) to a partially polarized configuration in phase 
(II).

 Phase (III) and (IV): at larger values of the bias $\Delta_B$, similar 
behavior 
as in phases (I) and (II) translates into valley polarized phases: we find the 
GS to be partially polarized both in spin space and in the space of the valley 
isospin, while the orbital isospin is either canted (below a critical magnetic 
field in phase III) or partially polarized (for sufficiently large magnetic 
field values in phase IV). The optimal canted angle of the orbital isospin is 
determined by 
\begin{equation}
\cos2\theta_{\text{(III)}} = \frac{ \Delta_{0000}+2 \Delta_{01}- \Delta _{1111}- 
z\Delta_{B,eff} }{ \Delta_{0000}-2 ( \Delta_{0011}+ \Delta_{1001})+ 
\Delta_{1111}}.
\end{equation}

This angle varies as function of the bias  $\Delta_B$ and the magnetic field 
strength $B$. At any value of  $\Delta_B$, when $B$ increases, the angle 
rotates 
until it reaches zero, leading to the partially polarized orbital state. Hence, 
increasing the magnetic field strength $B$ corresponds to rotating the orbital 
isospin from a canted configuration in phase (III) to a partially polarized 
state in phase (IV). Both the spin and the valley isospin vectors remain 
constant in these phases for all values of $\Delta_B$ and $B$.

Phase (V): within a narrow range of the bias $\Delta_B$, there is an 
intermediate transition regime: We find a complex GS structure in which all the 
spin and isospin degrees of freedom take nontrivial values and evolve as 
functions of  $\Delta_B$ and the magnetic field strength $B$. With the notation 
of the states used in Table \ref{tab:GS_1}, the spin and isospin configurations 
read $ S_z=1+\frac{1}{2}(  
{a_1}^2-{a_2}^2+{b_1}^2-{b_2}^2+{c_1}^2-{c_2}^2+{c_3}^2-{c_4}^2  ),\; S_x\equiv  
S_y\equiv0$,  
$T_z=1-\frac{1}{2}({a_1}^2-{a_2}^2+{b_1}^2-{b_2}^2+{c_1}^2-{c_2}^2+{c_3}^2-{c_4}
^2),\; T_x\equiv   T_y\equiv0$, and
$L_z=\frac{1}{2}( 
{a_1}^2+{a_2}^2-{b_1}^2-{b_2}^2+{c_1}^2+{c_2}^2-{c_3}^2-{c_4}^2  )  ,\; 
L_x=c_{1} c_{3}+c_{2} c_{4} ,\;  L_y\equiv0$.

Cuts of increasing bias $\Delta_B$ at fixed $B$ may exhibit  three 
different types of behavior depending on the value of $B$:

 (Va):
For small $B$, the phase (Va) is located between the phases (III) and (VI): in 
both these phases, the orbital isospin is neither zero nor fully polarized,  
exhibiting non-zero value $L_x\neq0$.  Phase (Va) now smoothly connects between 
these two phases with all four entries $a_i\neq0,$ $b_i\neq0$, and $c_i\neq0$ 
evolving smoothly as functions of the bias $\Delta_B$ and the magnetic field 
strength $B$. For increasing  $\Delta_B$ across phase (Va), this leads to smooth 
evolution of the spin and valley isospins from $S_z=\frac{3}{2}$ to 
$S_z=\frac{1}{2}$ and  from $T_z=\frac{1}{2}$ to $T_z=\frac{3}{2}$, 
respectively, accompanied by kinks in the orbital isospin components being 
nonzero $0<L_z<\frac{1}{2}$ and $0<L_x<\frac{1}{2}$. 

(Vb):
For an intermediate value of $B$, the phase (Vb) emerges between the phases
(III) and (VII). In the former, the orbital isospin is not fully polarized with 
$L_x\neq0$, whereas in the latter only the x-component is nonzero: 
$L_z=\frac{1}{2}$ and $L_x\equiv L_y\equiv0$. This transition is accomplished 
within phase (Vb) by first a sudden jump of the orbital isospin $L_x \rightarrow 
0$, $L_z\rightarrow\frac{1}{2}$ as the coefficients $c_3$ and $c_4$ suddenly 
jump to zero; Subsequently, the smooth rotations of the spin and the valley 
isospin degrees of freedom are governed by the remaining coefficients $a_i$ and 
$b_i$ evolving smoothly. 

(Vc):
For large enough values of the magnetic field, the phase (Vc) is located between 
the phases (IV) and (VII)  - both these phases exhibit the same configuration of 
the orbital isospin degree of freedom: $\mathbf{L}=\frac{1}{2}\mathbf{e}_z$. 
Here, the coefficients $c_i$ are zero throughout the phase (Vc): $c_1\equiv 
c_2\equiv0$. The orbital isospin therefore remains constant within this regime. 
The remaining coefficients $a_i$ and $b_i$ evolve smoothly as functions of 
$\Delta_B$ and $B$, smooth rotations $S_z=\frac{3}{2}\rightarrow 
S_z=\frac{1}{2}$ $T_z=\frac{1}{2} \rightarrow T_z=\frac{3}{2}$ across phase 
(Vc).

Phase (VI) and (VII): For sufficiently large values of the bias $\Delta_B$, we 
observe GS structures akin to those of phases (III) and (IV), but here at full 
valley polarization: At partially polarized spin and fully valley polarized 
isospin, the orbital isospin rotates from a canted position we call phase (VI) 
at sufficiently small magnetic fields to a partially polarized state, i.e., phase (VII), above a certain critical magnetic field strength. The 
optimal orbital canting angle hereby is determined by
\begin{equation}
\cos2\theta_{\text{(VI)}} = \frac{ \Delta_{0000}+2 \Delta _{01}- \Delta 
_{1111}+z\Delta _{Beff} }{ \Delta _{0000}-2 ( \Delta _{0011}+ \Delta _{1001})+ 
\Delta _{1111}} .
\label{eqn:Nu1_tVI}
\end{equation}

Hence, for any value of the magnetic field $B$, with rising bias $\Delta_B$ the 
orbital isospin performs a rotation to a configuration aligned along the 
$z$-axis: $\mathbf{L}\rightarrow\frac{1}{2}\mathbf{e}_z$. At sufficiently large 
values of the bias, phase (VII) eventually is fully 
polarized along the $z$-axis in valley isospin, and partially polarized in the 
spin and orbital isospin degree of freedom.

Except for the more complicated transition regime within phase (V), all phase 
transitions observed for the $\nu=1$ state are of continuous second order 
nature. We compute the critical values of the bias for these transitions, 
respectively:

(II) $\rightarrow$ (IV):
\begin{equation}
\Delta_{B,eff}^{crit}=  \frac{-\alpha  d- \Delta_{1111} {\ell_B}+{\ell_B} 
X_{1111}}{{\ell_B} (z-1)},
\end{equation}

(III) $\rightarrow$ (IV):
\begin{equation}
\Delta_{B,eff}^{crit}=  \frac{2 (\Delta_{0011}+\Delta_{01}+\Delta_{1001}- \Delta_{1111})}{z},
\end{equation}

(IV) $\rightarrow$ (Vc):
\begin{equation}
\Delta_{B,eff}^{crit}=  \frac{\alpha  d- \Delta_{0000} {\ell_B}+ \Delta_{Z} 
{\ell_B}+{\ell_B} X_{0000}}{{\ell_B}},
\end{equation}

(Vc) $\rightarrow$ (VII):
\begin{equation}
\Delta_{B,eff}^{crit}= \frac{3 \alpha  d+\Delta_{0000} {\ell_B}+\Delta _{Z} 
{\ell_B}-{\ell_B} X_{0000}}{{\ell_B}},
\end{equation}

 (VI) $\rightarrow$ (VII):
\begin{equation}
\Delta_{B,eff}^{crit}=  -\frac{2 (\Delta_{0011}+\Delta _{01}+\Delta_{1001}-\Delta_{1111})}{z}.
\end{equation}

\begin{table}
\centering
\begin{tabular}{ |l | c |}
\hline
  $\Delta_B=0$ & $ |v_1\RD=  |1,\uparrow,+\RD , 
  |v_2\RD=  |1,\uparrow,-\RD , 
  |v_3\RD= |0,\uparrow,+\RD,
  |v_4\RD= |0,\uparrow,-\RD$ ,\\
&$|v_5\RD=\frac{1}{\sqrt{2}}\cos\theta\Big[\,|1,\downarrow,+\RD+|1,\downarrow,
-\RD\,\Big]-\frac{1}{\sqrt{2}}\sin\theta\Big[\,|0,\downarrow,+\RD+|0,\downarrow,
-\RD\,\Big]$ \\
\hline
  Phase (I) & $ |v_1\RD=  |1,\uparrow,+\RD , 
  |v_2\RD=  |1,\uparrow,-\RD , 
  |v_3\RD= |0,\uparrow,+\RD,
  |v_4\RD= |0,\uparrow,-\RD$ ,\\
&$|v_5\RD=a_1|1,\downarrow,+\RD+a_2|1,\downarrow,-\RD- a_3|0,\downarrow,+\RD-a_4|0,\downarrow,-\RD$  \\
\hline
  Phase (II) & $ |v_1\RD=  |1,\uparrow,+\RD , 
  |v_2\RD=  |1,\uparrow,-\RD , 
  |v_3\RD= |0,\uparrow,+\RD,
  |v_4\RD= |0,\uparrow,-\RD ,|v_5\RD=\cos\theta|1,\downarrow,+\RD+\sin\theta|1,\downarrow,-\RD$ \\
    \hline
    Phase (III) &$ |v_1\RD=  |1,\uparrow,+\RD , 
  |v_2\RD=  |1,\uparrow,-\RD , 
  |v_3\RD= |0,\uparrow,+\RD,
  |v_4\RD= |0,\uparrow,-\RD , |v_5\RD=\cos\theta|1,\downarrow,+\RD+\sin\theta|0,\downarrow,+\RD$  \\
\hline
 Phase (IV) & $  |v_1\RD=  |1,\uparrow,+\RD , 
  |v_2\RD=  |1,\uparrow,-\RD , 
  |v_3\RD= |0,\uparrow,+\RD,
  |v_4\RD= |0,\uparrow,-\RD ,
|v_5\RD= |1,\downarrow,+\RD $\\
 \hline
 Phase (V) & $ |v_1\RD=  |1,\uparrow,+\RD , 
  |v_2\RD=  a_1|1,\uparrow,-\RD+a_2 |1,\downarrow,+\RD , 
  |v_3\RD= |0,\uparrow,+\RD,
  |v_4\RD= b_1|0,\uparrow,-\RD+b_2 |0,\downarrow,+\RD $ ,\\
& $|v_5\RD=c_1|1,\uparrow,-\RD+ c_2|1,\downarrow,+\RD+c_3|0,\uparrow,-\RD+c_4|0,\downarrow,+\RD$ \\
 \hline
 Phase (VI) & $ |v_1\RD=  |1,\uparrow,+\RD , 
  |v_2\RD=  |1,\downarrow,+\RD , 
  |v_3\RD= |0,\uparrow,+\RD,
  |v_4\RD= |0,\downarrow,+\RD, |v_5\RD=  \cos\theta|1,\uparrow,-\RD 
+\sin\theta|0,\uparrow,-\RD$  \\
\hline
 Phase (VII) & $ |v_1\RD=  |1,\uparrow,+\RD , 
\nonumber  |v_2\RD=  |1,\downarrow,+\RD , 
\nonumber  |v_3\RD= |0,\uparrow,+\RD,
\nonumber  |v_4\RD= |0,\downarrow,+\RD ,
|v_5\RD=  |1,\uparrow,-\RD $\\
\hline
\end{tabular}
\caption{GS configurations identified in the phase diagram for filling factor $\nu=1.$}\label{tab:GS_1}\end{table}

\subsection{Six electrons: $\nu=2$}

If there are six electrons occupying octet states, the structure we find for the GS is the following:

Phase (I): In the regime of sufficiently small bias, we observe a GS which is 
partially polarized in spin and exhibits valley coherence in a valley canted 
phase where the optimal valley canting angle is given by
\begin{equation}
\cos 2\theta_{(I)}=  -\frac{ \Delta_{B,eff} {\ell_B} (z-2)}{{\ell_B} (- 
\Delta_{0000}-2  \Delta _{0011}- \Delta _{1111}+X_{0000}+2 X_{0011}+X_{1111})-4 
\alpha  d}.
\end{equation}

The orbital order is antiferromagnetic, which leads to vanishing overall orbital 
polarization. Hence, in phase (I), cuts along lines of increasing bias 
$\Delta_B$ for any strength of the magnetic field $B$ correspond to a rotation 
of the valley-isospin vector from a configuration in the $\{x$-$y\}$-plane to a 
state partially polarized along the $z$-axis: $\mathbf{T}= \mathbf{e}_x\; 
\longrightarrow \;\mathbf{T}= \mathbf{e}_z$. 

Phase (II):  For larger values of the bias $\Delta_B$, we observe an 
intermediate regime in which the GS exhibits partial polarization both in the 
spin and the valley isospin degree of freedom and antiferromagnetic ordering in 
the orbital isospin. Hence, the GS is a partially aligned spin and 
valley ferromagnet over a broad parameter range. 

Phase (III): When the system is biased sufficiently strongly, we find the GS to 
be a polarized state both for the valley and the orbital isospin. 
Meanwhile, due to antiferromagnetic ordering of the spin degree of freedom, the 
overall spin polarization vanishes. This phase  for large values of the bias 
$\Delta_B$ is established for all magnetic field strengths. 
For the system at filling $\nu=2$ we identify two different types of phase 
transitions as functions of $\Delta_B$ and $B$: The small bias transition (I) to 
(II) comes with a smooth rotation of the valley isospin and therefore is of 
continuous second order. For larger bias, however, the system jumps from phase 
(II) to phase (III) in a discontinuous fashion characterizing a first order 
transition. We give the values of the critical bias at which these phase 
transitions occur:

 (I) $\rightarrow$ (II):
\begin{equation}
\Delta_{B,eff}^{crit}=   \frac{{\ell_B} (X_{0000}+2 X_{0011}+X_{1111}- \Delta_{0000}-2  \Delta _{0011}- 
\Delta _{1111})-4 \alpha  d}{{\ell_B} (z-2)},
\end{equation}

(II) $\rightarrow$ (III): Phase (II) is energetically favorable over phase (III) 
up to a the critical bias
\begin{equation}
\Delta_{B,eff}^{crit}=\frac{- \Delta_{0000}-2 \Delta_{0011}-2 \Delta_{01}+\Delta_{1111}+2 \Delta_{Z}}{z}.
\end{equation} 

\begin{table}
\centering
\begin{tabular}{| l | c |}
\hline
 $\Delta_B\equiv0$& $ |v_1\RD=  |1,\uparrow,+\RD , 
  |v_2\RD=  |1,\uparrow,-\RD , 
  |v_3\RD= |0,\uparrow,+\RD,
  |v_4\RD= |0,\uparrow,-\RD$ ,\\
 &$|v_5\RD= \frac{1}{\sqrt{2}}\Big[\, |1,\downarrow,+\RD+ |1,\downarrow,-\RD\, \Big] ,
|v_6\RD=  \frac{1}{\sqrt{2}}\Big[\, |0,\downarrow,+\RD+ |0,\downarrow,-\RD\,\Big]$\\

\hline
  Phase (I) & $ |v_1\RD=  |1,\uparrow,+\RD , 
  |v_2\RD=  |1,\uparrow,-\RD , 
  |v_3\RD= |0,\uparrow,+\RD,
  |v_4\RD= |0,\uparrow,-\RD$ ,\\
 &$|v_5\RD=\sin\theta|1,\downarrow,+\RD+ \cos\theta|1,\downarrow,-\RD ,
|v_6\RD=\sin\theta|0,\downarrow,+\RD+ \cos\theta|0,\downarrow,-\RD$\\
 \hline
  Phase (II) &$  |v_1\RD=  |1,\uparrow,+\RD , 
  |v_2\RD=  |1,\uparrow,-\RD , 
  |v_3\RD= |0,\uparrow,+\RD,
  |v_4\RD= |0,\uparrow,-\RD ,
 |v_5\RD= |1,\downarrow,+\RD,\
|v_6\RD=|0,\downarrow,+\RD $ \\
    \hline
    Phase (III) &  $|v_1\RD=  |1,\uparrow,+\RD , 
  |v_2\RD=  |1,\uparrow,-\RD , 
  |v_3\RD= |0,\uparrow,+\RD,
 |v_4\RD= |1,\downarrow,-\RD ,
 |v_5\RD= |1,\downarrow,+\RD,
|v_6\RD=|0,\downarrow,+\RD  $ \\
\hline
\end{tabular}\caption{The different possible GS configurations we observed at filling factor $\nu=2$.}\label{tab:GS_2}\end{table}

\subsection{Seven electrons: $\nu=3$}

With only one hole in the octet, the GS exhibit the following 
structure: in the
unbiased case ($\Delta_{B}\equiv0$, evolution as a function of $B$), we find a 
partially spin polarized GS, while the valley isospin is aligned along the 
$x$-axis. The state exhibits orbital coherence as the orbital isospin is in a 
canted configuration, where the optimal angle $\theta_0$ varies as a function of 
$B$ between $\theta_0\rightarrow\frac{\pi}{4}$ at vanishing magnetic field 
$B\rightarrow0$ and $\theta_0=\frac{\pi}{2}$ at sufficiently high magnetic field 
strengths above a certain critical value $B_{crit}\approx 11.3$ T. It fulfills 
the relation
\begin{equation}
\cos 2\theta_0=  \frac{-3 \Delta_{0000}-4  \Delta_{01}+3  \Delta 
_{1111}+X_{0000}-X_{1111}}{\Delta_{0000}-2 \Delta_{0011}-2 \Delta _{1001}+\Delta 
_{1111}+X_{0000}-2 X_{0011}-2 X_{1001}+X_{1111}}.
\label{eqn:Nu3_t0}
\end{equation}

Along the line of zero bias, as a function of increasing magnetic field strength 
$B$ the GS hence undergoes a transition from a canted state in the orbital 
isospin to a partially aligned state.

Phases (I) and (II): in this regime of sufficiently weak bias $\Delta_B$, the 
GS 
has partial polarization in spin space, while the valley isospin 
undergoes a rotation and therefore takes nontrivial values $0\le T_x,T_z \le \frac{1}{2}$. 
Meanwhile, the orbital isospin is either canted with  $0\le L_x,L_z \le \frac{1}{2}$ for 
small magnetic field strengths in phase (I), or partially polarized in phase 
(II) at sufficiently large magnetic fields. While the dependencies of 
the isospins in phase (I) being more involved, we can express the valley isospin 
in phase (II) in terms of one valley tilting angle for which the optimal 
configuration is determined by
\begin{equation}
\cos 2\theta_{(II)}= \frac{\Delta _{B,eff} {\ell_B}}{\alpha  d+{\ell_B} 
(\Delta_{0000}-X_{0000})}.
\end{equation}

Hence, in phase (I) and (II), cuts along lines of increasing bias $\Delta_B$ for 
any strength of the magnetic field $B$ correspond to a rotation of the valley 
isospin vector from a configuration in the $\{x$-$y\}$-plane to a state aligned 
along the $z$-axis: $\mathbf{T}=\frac{1}{2}\mathbf{e}_x\; \longrightarrow 
\;\mathbf{T}=\frac{1}{2}\mathbf{e}_z$. At the same time, increasing $B$ at a 
fixed value of the bias $\Delta_B$ corresponds to rotating the orbital isospin 
from a canted configuration in phase (I) to an aligned configuration, 
$\mathbf{L}=\frac{1}{2}\mathbf{e}_z$ in phase (II).

Phase (III): at small magnetic fields and large values of the bias, the spin 
and 
the valley isospin degree of freedom are equally partially polarized, while the 
orbital isospin undergoes a rotation through a canted state, thereby exhibiting 
nontrivial orbital coherence. The optimal angle in orbital space is determined 
by 
\begin{equation}
\cos2\theta_{\text{(III)}} =  \frac{\Delta_{0000}+2  \Delta_{01}- \Delta_{1111}+ z\Delta_{B,eff} }{\Delta_{0000}-2 ( \Delta_{0011}+ \Delta_{1001})+\Delta_{1111}},
\label{eqn:Nu3_tIII}
\end{equation}

varying as function of the bias  $\Delta_B$ and the magnetic field strength $B$.

Phase (IV): when both magnetic field strength $B$ and  bias $\Delta_B$ 
are sufficiently large, the GS adopts a configuration in which all spin and 
isospin degrees of freedom are equally partially polarized.

All the phase transitions for filling factor $\nu=3$ are continuous second order 
transitions, which occur via smooth rotations of the respective isospin degrees 
of freedom. The critical values of the bias for these transitions are given by

 (II) $\rightarrow$ (IV):
\begin{equation}
\Delta_{B,eff}^{crit}=   \frac{\alpha  d}{{\ell_B}}+\Delta_{0000}-X_{0000},
\end{equation}

(III) $\rightarrow$ (IV):
\begin{equation}
\Delta_{B,eff}^{crit}= -\frac{2 ( \Delta_{0011}+\Delta_{01}+\Delta_{1001}-\Delta_{1111})}{z}.
\end{equation}
 
\begin{table}
\centering
\begin{tabular}{ |l | c |}
\hline
  $\Delta_B=0$ & $|v_1\RD=  |1,\uparrow,+\RD , 
 |v_2\RD=  |1,\uparrow,-\RD , 
 |v_3\RD= |0,\uparrow,+\RD,
 |v_4\RD= |0,\uparrow,-\RD$ ,\\
&$|v_5\RD=\frac{1}{\sqrt{2}}\cos\theta\Big[\,|1,\downarrow,+\RD+|1,\downarrow,
-\RD\,\Big]+\frac{1}{\sqrt{2}}\sin\theta\Big[\,|0,\downarrow,+\RD+|0,\downarrow,
-\RD\,\Big],$ \\
&$|v_6\RD=\frac{1}{\sqrt{2}}\sin\theta\Big[\,|1,\downarrow,+\RD+|1,\downarrow,-\RD\,\Big]-\frac{1}{\sqrt{2}}\cos\theta\Big[\,|0,\downarrow,+\RD+|0,\downarrow,-\RD\,\Big]$,\\
&$|v_7\RD=\frac{1}{\sqrt{2}}\sin\theta\Big[\,|1,\downarrow,+\RD-|1,\downarrow,-\RD\,\Big]+\frac{1}{\sqrt{2}}\cos\theta\Big[\,|0,\downarrow,+\RD-|0,\downarrow,-\RD\,\Big]  $\\
\hline
  Phase (I) & $ |v_1\RD=  |1,\uparrow,+\RD, 
 |v_2\RD=  |1,\uparrow,-\RD, 
 |v_3\RD= |0,\uparrow,+\RD,
 |v_4\RD= |0,\uparrow,-\RD$,\\
& $|v_5\RD=-a_1|1,\downarrow,+\RD+a_2|1,\downarrow,-\RD- a_3|0,\downarrow,+\RD+a_4|0,\downarrow,-\RD$,\\
 & $|v_6\RD=a_3|1,\downarrow,+\RD-a_4|1,\downarrow,-\RD- a_1|0,\downarrow,+\RD+a_2|0,\downarrow,-\RD$,\\
 & $|v_7\RD=a_4|1,\downarrow,+\RD+a_3|1,\downarrow,-\RD+ a_2|0,\downarrow,+\RD+a_1|0,\downarrow,-\RD$  \\
\hline
  Phase (II) & $ |v_1\RD=  |1,\uparrow,+\RD, 
 |v_2\RD=  |1,\uparrow,-\RD , 
 |v_3\RD= |0,\uparrow,+\RD,
 |v_4\RD= |0,\uparrow,-\RD$ ,\\
& $|v_5\RD=- \cos\theta|0,\downarrow,+\RD+\sin\theta|0,\downarrow,-\RD, |v_6\RD= \cos\theta|1,\downarrow,+\RD-\sin\theta|1,\downarrow,-\RD$,\\
 & $|v_7\RD=\sin\theta|1,\downarrow,+\RD+\cos\theta|1,\downarrow,-\RD$  \\
    \hline
    Phase (III) &  $|v_1\RD=  |1,\uparrow,+\RD, 
  |v_2\RD=  |1,\uparrow,-\RD, 
 |v_3\RD= |0,\uparrow,+\RD,
 |v_4\RD= |0,\uparrow,-\RD$,\\
  & $|v_5\RD= |1,\downarrow,+\RD,
 |v_6\RD= |0,\downarrow,+\RD,
|v_7\RD=\cos\theta|1,\downarrow,-\RD+\sin\theta|0,\downarrow,-\RD$ \\
\hline
 Phase (IV) &  $ |v_1\RD=  |1,\uparrow,+\RD , 
|v_2\RD=  |1,\uparrow,-\RD, 
 |v_3\RD= |0,\uparrow,+\RD,
 |v_4\RD= |0,\uparrow,-\RD$,\\
  &$ |v_5\RD= |1,\downarrow,+\RD ,
 |v_6\RD= |0,\downarrow,+\RD,
|v_7\RD= |1,\downarrow,-\RD $\\
 \hline
\end{tabular}\caption{The different possible configurations which occur in the 
$\nu=3$ phase diagram.}\label{tab:GS_3}\end{table}

\section{Comparative description}
\label{comparo}

\begin{table}
\centering
\begin{tabular}{ |l | c |c| c| c|c| c| c|c| c|}
\hline
\diagbox{Phase}{$\nu$}&-3 & -2 & -1 & 0  \\
\hline
$\Delta_B\equiv0$  & $\mathbf{S}=\frac{1}{2}\mathbf{e}_z$  & $\mathbf{S}= \mathbf{e}_z$  &  $\mathbf{S}=\frac{3}{2}\mathbf{e}_z$ & $\mathbf{S}= {2}\mathbf{e}_z$ \\
 & $\mathbf{T}=\frac{1}{2}\mathbf{e}_x$  & $\mathbf{T}= \mathbf{e}_x$ &  $\mathbf{T}=\frac{1}{2}\mathbf{e}_z$ & $\mathbf{T}\equiv0$\\
  & $\mathbf{L}=\frac{1}{2}\sin2\theta\mathbf{e}_x+\frac{1}{2}\cos2\theta\mathbf{e}_z$  & $\mathbf{L}\equiv0$  & $\mathbf{L}=\frac{1}{2}\sin2\theta\mathbf{e}_x-\frac{1}{2}\cos2\theta\mathbf{e}_z$  & $\mathbf{L}\equiv0$\\
\hline
I  &  $\mathbf{S}=\frac{1}{2}\mathbf{e}_z$  & $\mathbf{S}= \mathbf{e}_z$ & $\mathbf{S}=\frac{3}{2}\mathbf{e}_z$& $\mathbf{S}= {2}\mathbf{e}_z$  \\
& $0\le T_z, T_x\le\frac{1}{2}$ , $T_y\equiv0$& $\mathbf{T}=\sin2\theta\mathbf{e}_x+\cos2\theta\mathbf{e}_z$  & $0\le T_z, T_x\le\frac{1}{2}$, $T_y\equiv0$ & $\mathbf{T}\equiv0$  \\
& $0\le L_z, L_x\le\frac{1}{2}$, $L_y\equiv0$   & $\mathbf{L}\equiv0$& $0\le L_z, L_x\le\frac{1}{2}$, $L_y\equiv0$  & $\mathbf{L}\equiv0$  \\
\hline
II & $\mathbf{S}=\frac{1}{2}\mathbf{e}_z$   & $\mathbf{S}=\mathbf{e}_z$  &  $\mathbf{S}=\frac{3}{2}\mathbf{e}_z$& $\mathbf{S}= {2}\cos^2\theta\;\mathbf{e}_z$  \\
  & $\mathbf{T}=\frac{1}{2}\sin2\theta\mathbf{e}_x-\frac{1}{2}\cos2\theta\mathbf{e}_z$ & $\mathbf{T}=\mathbf{e}_z$ & $\mathbf{T}=\frac{1}{2}\sin2\theta\mathbf{e}_x+\frac{1}{2}\cos2\theta\mathbf{e}_z$ & $\mathbf{T}= {2}\sin^2\theta\;\mathbf{e}_z$ \\
   &  $\mathbf{L}=\frac{1}{2}\mathbf{e}_z$& $\mathbf{L}\equiv0$  &  $\mathbf{L}=\frac{1}{2}\mathbf{e}_z$& $\mathbf{L}\equiv0$  \\
    \hline
 III & $\mathbf{S}=\frac{1}{2}\mathbf{e}_z$& $\mathbf{S}\equiv0$ &  $\mathbf{S}=\frac{3}{2}\mathbf{e}_z$ & $\mathbf{S}\equiv0$  \\
 & $\mathbf{T}=\frac{1}{2}\mathbf{e}_z$ & $\mathbf{T}=\mathbf{e}_z$&  $\mathbf{T}=\frac{1}{2}\mathbf{e}_z$ & $\mathbf{T}= {2}\mathbf{e}_z$ \\    
 & $\mathbf{L}=\frac{1}{2}\mathbf{e}_z$ & $\mathbf{L}=-\mathbf{e}_z$ & $\mathbf{L}=\frac{1}{2}\sin2\theta\mathbf{e}_x+\frac{1}{2}\cos2\theta\mathbf{e}_z$ & $\mathbf{L}\equiv0$  \\   
\hline
IV&    $\mathbf{S}=\frac{1}{2}\mathbf{e}_z$& &  $\mathbf{S}=\frac{3}{2}\mathbf{e}_z$ & \\
 & $\mathbf{T}=\frac{1}{2}\mathbf{e}_z$ & - &  $\mathbf{T}=\frac{1}{2}\mathbf{e}_z$&- \\ 
 & $\mathbf{L}=\frac{1}{2}\sin2\theta\mathbf{e}_x-\frac{1}{2}\cos2\theta\mathbf{e}_z$ & &  $\mathbf{L}=\frac{1}{2}\mathbf{e}_z$& \\
 \hline
 V & $\mathbf{S}=\frac{1}{2}\mathbf{e}_z$& & $\frac{1}{2}\le S_z\le\frac{3}{2}$, $S_x\equiv S_y\equiv 0$&    \\
 & $\mathbf{T}=\frac{1}{2}\mathbf{e}_z$ &-  & $\frac{1}{2}\le T_z \le\frac{3}{2}$, $T_x\equiv T_y\equiv 0$ & -  \\    
 & $\mathbf{L}=-\frac{1}{2}\mathbf{e}_z$ &  & $0\le L_z, L_x\le\frac{1}{2}$, $L_y\equiv0$ &  \\  
    \hline
 VI &    & & $\mathbf{S}=\frac{1}{2}\mathbf{e}_z$& \\
 &   - &- & $\mathbf{T}=\frac{3}{2}\mathbf{e}_z$& -\\
 &     & & $\mathbf{L}=\frac{1}{2}\sin2\theta\mathbf{e}_x+\frac{1}{2}\cos2\theta\mathbf{e}_z$& \\
\hline
 VII &    &  & $\mathbf{S}=\frac{1}{2}\mathbf{e}_z$& \\
 &   - &- & $\mathbf{T}=\frac{3}{2}\mathbf{e}_z$& -\\
 &    & & $\mathbf{L}=-\frac{1}{2}\mathbf{e}_z$& \\
\hline
\end{tabular}\caption{Spin and isospin properties of the different phases 
observed for negative filling factors.}\label{tab:PolProp_negMu}\end{table}

\begin{table}
\centering
\begin{tabular}{ |l | c |c| c| c|c| c| c|c| c|}
\hline
\diagbox{Phase}{$\nu$}&  1 & 2 & 3  \\
\hline
$\Delta_B\equiv0$  & $\mathbf{S}=\frac{3}{2}\mathbf{e}_z$  & $\mathbf{S}= \mathbf{e}_z$ &  $\mathbf{S}=\frac{1}{2}\mathbf{e}_z$ \\
 & $\mathbf{T}=\frac{1}{2}\mathbf{e}_x$  & $\mathbf{T}= \mathbf{e}_x$ & $\mathbf{T}=\frac{1}{2}\mathbf{e}_x$ \\
 &  $\mathbf{L}=-\frac{1}{2}\sin2\theta\mathbf{e}_x+\frac{1}{2}\cos2\theta\mathbf{e}_z$ & $\mathbf{L}\equiv0$ &  $\mathbf{L}=\frac{1}{2}\sin2\theta\mathbf{e}_x-\frac{1}{2}\cos2\theta\mathbf{e}_z$  \\
\hline
I  & $\mathbf{S}=\frac{3}{2}\mathbf{e}_z$ & $\mathbf{S}=\mathbf{e}_z$  &  $\mathbf{S}=\frac{1}{2}\mathbf{e}_z$ \\ 
& $0\le T_z, T_x\le\frac{1}{2}$, $T_y\equiv0$ & $\mathbf{T}=\sin2\theta\mathbf{e}_x-\cos2\theta\mathbf{e}_z$ & $0\le T_z, T_x\le\frac{1}{2}$, $T_y\equiv0$  \\
& $0\le L_z, L_x\le\frac{1}{2}$, $L_y\equiv0$ &$\mathbf{L}\equiv0$ &    $0\le L_z, L_x\le\frac{1}{2}$, $L_y\equiv0$\\
\hline
II &  $\mathbf{S}=\frac{3}{2}\mathbf{e}_z$  &  $\mathbf{S}= \mathbf{e}_z$  &  $\mathbf{S}=\frac{1}{2}\mathbf{e}_z$  \\
 & $\mathbf{T}=\frac{1}{2}\sin2\theta\mathbf{e}_x+\frac{1}{2}\cos2\theta\mathbf{e}_z$ &  $\mathbf{T}= \mathbf{e}_z$  & $\mathbf{T}=-\frac{1}{2}\sin2\theta\mathbf{e}_x+\frac{1}{2}\cos2\theta\mathbf{e}_z$  \\
  & $\mathbf{L}=\frac{1}{2}\mathbf{e}_z$ & $\mathbf{L}\equiv0$ & $\mathbf{L}=\frac{1}{2} \mathbf{e}_z$  \\
\hline
 III &  $\mathbf{S}=\frac{3}{2}\mathbf{e}_z$   & $\mathbf{S}\equiv0$ &  $\mathbf{S}=\frac{1}{2}\mathbf{e}_z$  \\   
  & $\mathbf{T}=\frac{1}{2}\mathbf{e}_z$ &  $\mathbf{T}= \mathbf{e}_z$ &  $\mathbf{T}=\frac{1}{2}\mathbf{e}_z$  \\
&   $\mathbf{L}=\frac{1}{2}\sin2\theta\mathbf{e}_x+\frac{1}{2}\cos2\theta\mathbf{e}_z$ &  $\mathbf{L}= \mathbf{e}_z$ & $\mathbf{L}=\frac{1}{2}\sin2\theta\mathbf{e}_x+\frac{1}{2}\cos2\theta\mathbf{e}_z$ \\
\hline
IV &  $\mathbf{S}=\frac{3}{2}\mathbf{e}_z$   & &  $\mathbf{S}=\frac{1}{2}\mathbf{e}_z$    \\   
& $\mathbf{T}=\frac{1}{2}\mathbf{e}_z$ & - &  $\mathbf{T}=\frac{1}{2}\mathbf{e}_z$   \\
  & $\mathbf{L}=\frac{1}{2}\mathbf{e}_z$ &  &  $\mathbf{L}=\frac{1}{2}\mathbf{e}_z$   \\
 \hline
 V &  $\frac{1}{2}\le S_z \le\frac{3}{2}$,$S_x\equiv S_y\equiv 0$  & &\\  
 & $\frac{1}{2}\le T_z \le\frac{3}{2}$, $T_x\equiv T_y\equiv 0$ & - &-\\
 & $0\le L_z, L_x\le\frac{1}{2}$, $L_y\equiv0$  & &\\
    \hline
 VI &   $\mathbf{S}=\frac{1}{2}\mathbf{e}_z$ &  &\\
 & $\mathbf{T}=\frac{3}{2}\mathbf{e}_z$ & - & - \\
  &  $\mathbf{L}=\frac{1}{2}\sin2\theta\mathbf{e}_x+\frac{1}{2}\cos2\theta\mathbf{e}_z$ & &\\
\hline
 VII &   $\mathbf{S}=\frac{1}{2}\mathbf{e}_z$  & &\\
  & $\mathbf{T}=\frac{3}{2}\mathbf{e}_z$ & - & - \\
   &   $\mathbf{L}=\frac{1}{2}\mathbf{e}_z$  & & \\
\hline
\end{tabular}\caption{ Spin and isospin properties of the different  phases  
for the bilayer system at positive fillings}\label{tab:PolProp_posMu}\end{table}

\subsection{General features}

The phase diagrams of Fig.~\ref{fig:GSPDs} displaying the  
 spin and isospin configurations as functions of $\Delta_B$  and $B$ share 
some common features for all filling factors $\nu\in[-3,3]$. 
In general, we observe many different spin and isospin structures: 
Among these, the valley and the orbital isospin can be in canted configurations, 
thus exhibiting non-trivial coherence. In 
Fig.~\ref{fig:GSPDs}, the regions where such phases occur are drawn in bordeaux, 
pink, turquoise, yellow, or orange, respectively.  The spin and 
isospin configurations for all the different possible phases are summarized 
in Tables \ref{tab:PolProp_negMu} and \ref{tab:PolProp_posMu}.

The unbiased system  $\Delta_B\equiv0$ is spin polarized for all values of 
the filling factor. This also remains true for sufficiently small values of the 
bias in every case (in Fig.~\ref{fig:GSPDs}, all phases except the blue or green 
ones at even filling factors). In the opposite limit of large bias, valley 
polarization emerges for all $\nu$ (blue or green regions in 
Fig.~\ref{fig:GSPDs}). Qualitatively, this is in accordance with 
experimental\cite{Kim2011} as well as previous theoretical\cite{Lambert2013} 
investigations which suggest an evolution towards a valley polarized state with 
increasing bias. The values of the critical bias and the critical magnetic 
field strength below (above) which the system is spin (valley) polarized, 
however, differ for different values of $\nu$. Furthermore, it depends on the 
filling factor whether the respective polarized phase formed in these two 
limits is partially polarized or fully polarized 
 in spin or valley space. 

\subsection{Odd versus even filling factors}

We now compare the behavior of the orbital degree of freedom for the GS phase 
diagrams obtained at odd filling factors $\nu=-3,-1,1,3$. For the unbiased 
system,i.e., along the line of zero bias $\Delta_B\equiv0$, all the 
systems with $\nu$ odd undergo a similar evolution of the orbital isospin: 
at 
small $B$, we find a canted configuration, then with rising  $B$ the orbital 
isospin
rotates smoothly until it reaches a polarized state above some critical magnetic 
field strength $B_{crit}$. For non-zero values of the bias this 
transition in the orbital configuration is translated in the upper half of 
each 
phase diagram: for every odd filling factor, we find a large phase exhibiting 
orbital coherence at any $\Delta_B>0$ (yellow or orange regions in Fig.~\ref{fig:GSPDs}). 
These orbitally coherent phases then respectively evolve into orbitally 
polarized 
configurations (blue or green phases in Fig.~\ref{fig:GSPDs}, respectively) by smooth rotations of the 
orbital isospin when $B$ is increased for any $\Delta_B$ held fix.

For even filling factors $\nu=-2,0,2$, however, we do not observe any phase
 with orbital coherence. There is no phase transition 
as a function of $B$ along the line of zero bias, but the GS is in a spin 
polarized configuration with vanishing orbital isospin stable for all $B$. Some 
of the  phases at $\nu$ even carry orbital polarization, \textit{i.e.}~the 
total orbital isospin is of the form $\mathbf{L}\propto\mathbf{e}_z$ (blue phase 
at $\nu=-2$ and green phase at $\nu=2$ in Fig.~\ref{fig:GSPDs}). The remaining 
 phases at even fillings show antiferromagnetic orbital order, i.e., the overall orbital polarization vanishes and we find 
$\mathbf{L}\equiv0$. 

\subsection{Negative against positive filling factors}

The most striking feature when comparing negative  
$\nu=-3,-2,-1,0$ (Table \ref{tab:PolProp_negMu}) to positive filling factors 
$\nu=1,2,3$  (Table \ref{tab:PolProp_posMu}) has to do with the
orbital  polarization. In the limit of large values of the bias 
$\Delta_B$, the system  exhibits orbital polarization for all values of the 
filling factor (in fig.~\ref{fig:GSPDs}, these phases are drawn in blue or in 
green). For negative filling factors, however, this polarization is negative, 
$\mathbf{L}\propto-\mathbf{e}_z$ (blue phases in Fig.~\ref{fig:GSPDs}), whereas 
the GS for positive filling factors for sufficiently large bias turns out to be 
positively polarized, $\mathbf{L}\propto+\mathbf{e}_z$ (green phases in 
Fig.~\ref{fig:GSPDs}). Physically, this indicates that at negative filling 
factors, it is energetically favorable to have predominantly the $n=0$ orbitals 
populated, while at higher filling factors the systems prefer to successively 
populate $n=1$ orbitals. 

\section{HF Results II: Physical Properties of the States}
\label{sec:HFRESII}

\subsection{Octet Polarization, Hund's Rules, Layer Distribution}

\begin{figure}[!htb]
  \centering
\includegraphics[width=0.55\textwidth]{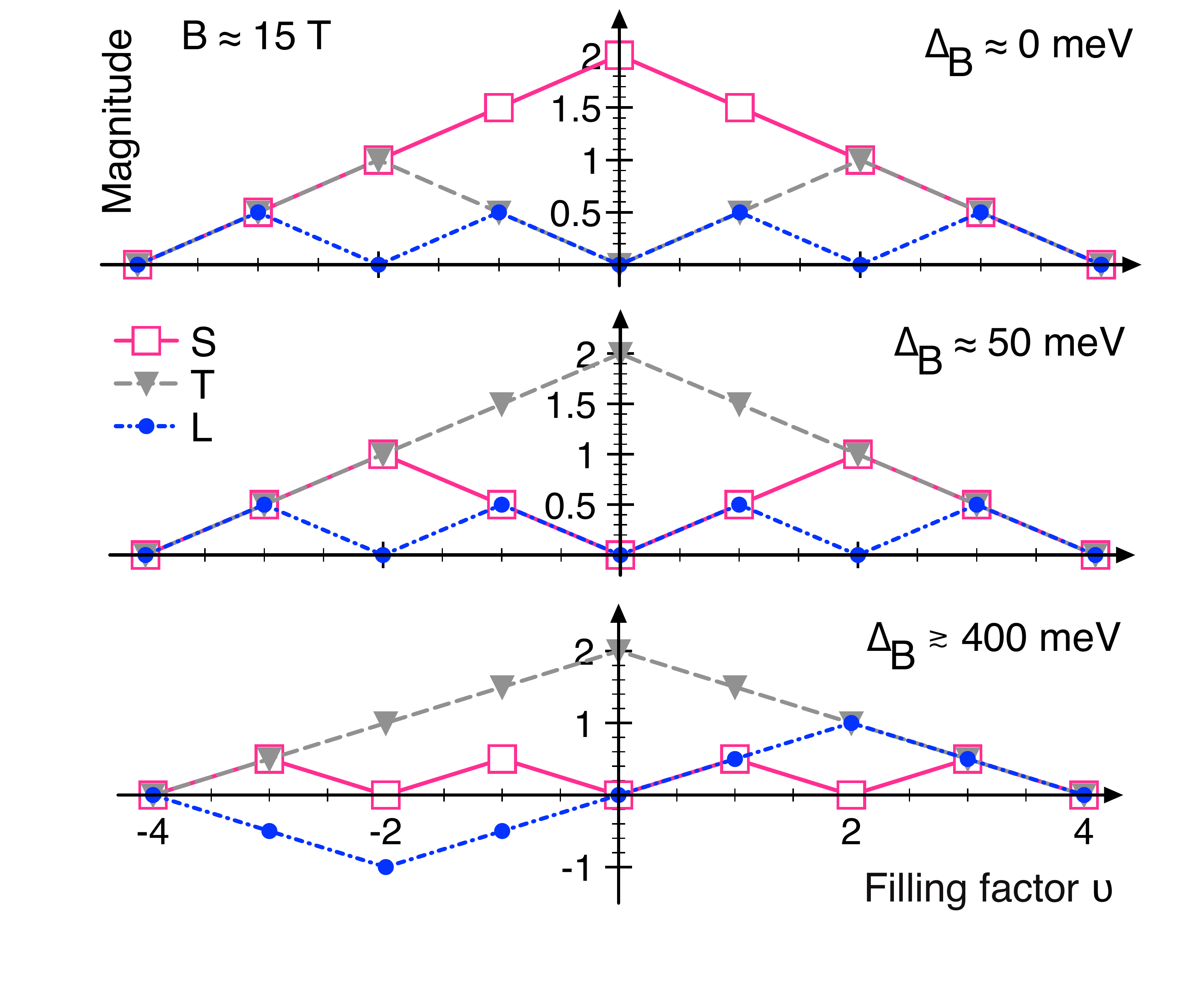}
  \caption{Octet polarization properties: At magnetic field $B=15$ T and for 
different values of the bias potential $\Delta_B$, we plot the magnitude of the 
spin vector (pink, solid line, empty squares), the valley isospin vector (gray, 
dashed line, filled triangles) and the orbital isospin vector (blue, 
dashed-dotted line, filled circles) as a function of the filling factor $\nu$. 
The values of $B$ and $\Delta_B$ are chosen as representative examples, similar 
behavior occurs over a broad parameter range in throughout the phase diagrams. The \textit{magnitude} of an isospin vector is to be understood as magnitude[$\mathbf{e}_z$]=magnitude[$\mathbf{e}_x$]=1.}
\label{fig:Hund}
 \end{figure}

We first analyze the spin and isospin polarization properties within the octet. 
For the unbiased case $\Delta_B\equiv0$, dependence on the system's polarization 
on the filling factor has been studied previously\cite{Barlas2008}, 
establishing Hund's rules for the SP level occupation when the states of the 
octet are gradually filled with electrons. In 
Fig.~\ref{fig:Hund}, we show the pseudospin 
polarizations for $B=15$ T and three bias values $\Delta_B=0$ meV, 
$\Delta_B=50$ meV, and $\Delta_B=400$ meV. 
For the unbiased case, $\Delta_B\equiv0$, we recover the results of Barlas et 
al.\cite{Barlas2008}. First the real 
spin degree of freedom is polarized. Second, under the restrictions imposed by 
the spin 
configuration, the polarization of the valley isospin and third the polarization 
of the orbital isospin is maximized to the greatest possible extent. This 
behavior is shown in the upper plot of Fig.~\ref{fig:Hund}.
The examples at non-zero values of the bias, $\Delta_B>0$, (central and 
lower plot of Fig.~\ref{fig:Hund}) demonstrate that this picture may
change if the system is biased. In the case of intermediate bias, 
$\Delta_B=50$ meV, the role of real spin and valley isospin are reversed: here, 
the valley degree of freedom is maximized first. In the case of stronger 
bias, here for $\Delta_B=400$ meV, the properties of the orbital 
isospin polarization can be altered: we observe states which are 
antiferromagnetically polarized in the orbital degree of freedom. 

A remark about the generality of these statements is in order: the 
examples we show in Fig.~\ref{fig:Hund} represent cuts through the broadest 
phases of the phase diagrams we show in Fig.~\ref{fig:GSPDs} for all the 
different $\nu$. Due to the rich structure apparent from Fig.~\ref{fig:GSPDs} 
exhibiting a variety of different phases, many cuts through the phase 
diagrams are possible which yield octet polarization diagrams different from the 
ones shown in Fig.~\ref{fig:Hund}. 

\begin{figure}[!htb]
  \centering
   \begin{subfigure}[!htb]{0.45\textwidth}
\includegraphics[width=\textwidth]{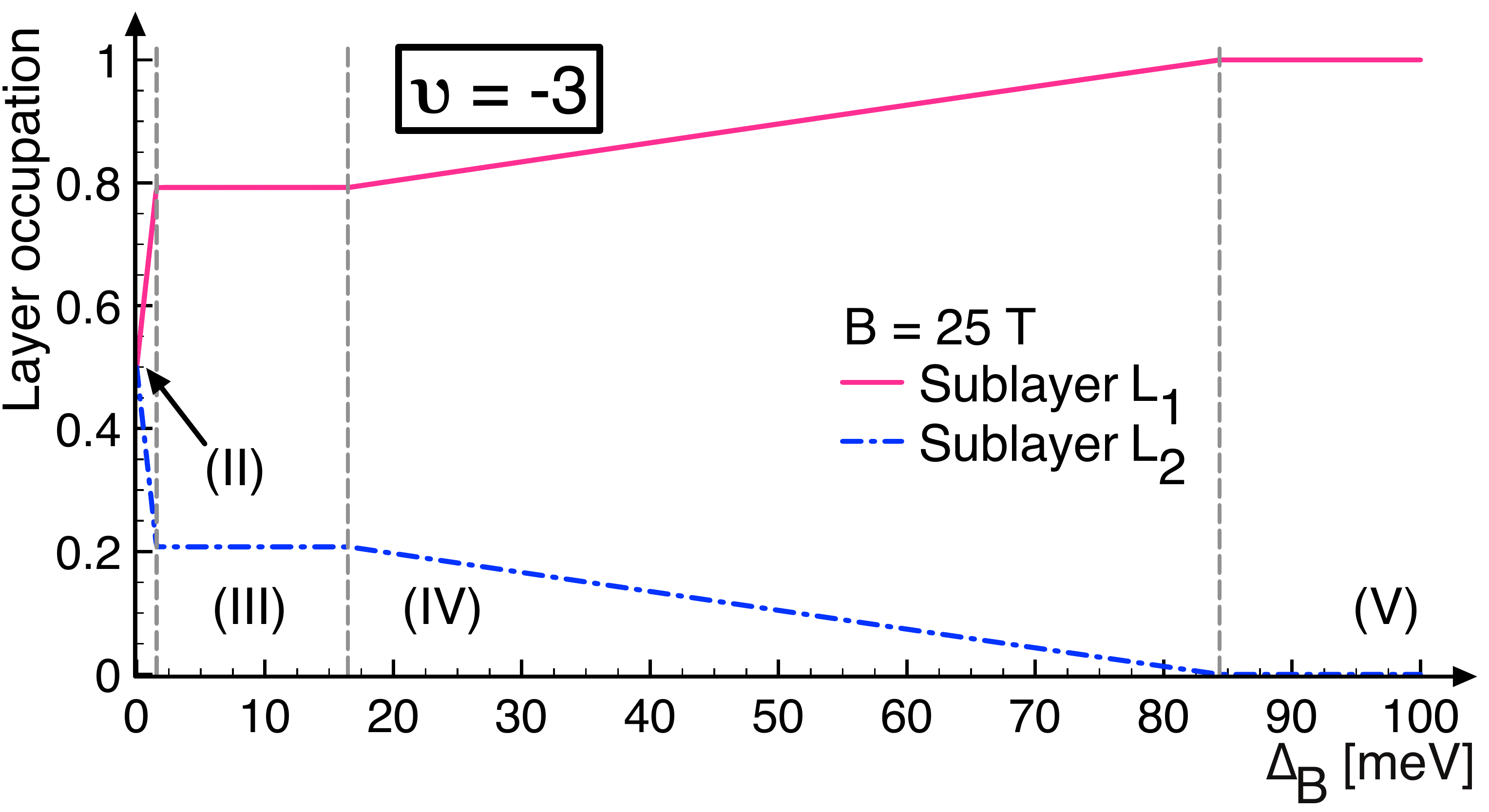}
  \caption{ .}
\label{sfig:b1iofM}
 \end{subfigure}
   \begin{subfigure}[!htb]{0.45\textwidth}
\includegraphics[width=\textwidth]{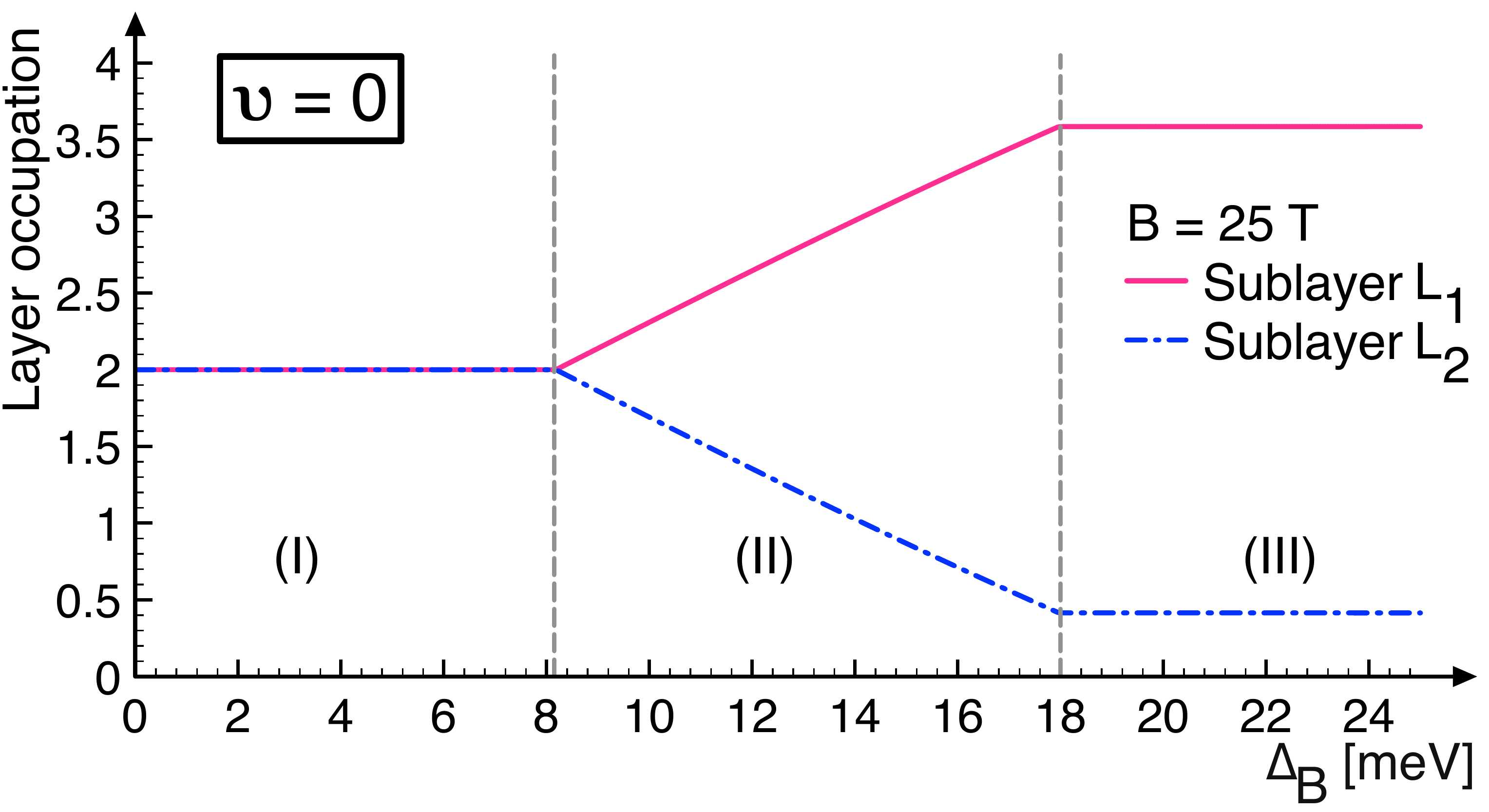}
  \caption{ .}
\label{sfig:b1iofM}
 \end{subfigure}
   \begin{subfigure}[!htb]{0.45\textwidth}
\includegraphics[width=\textwidth]{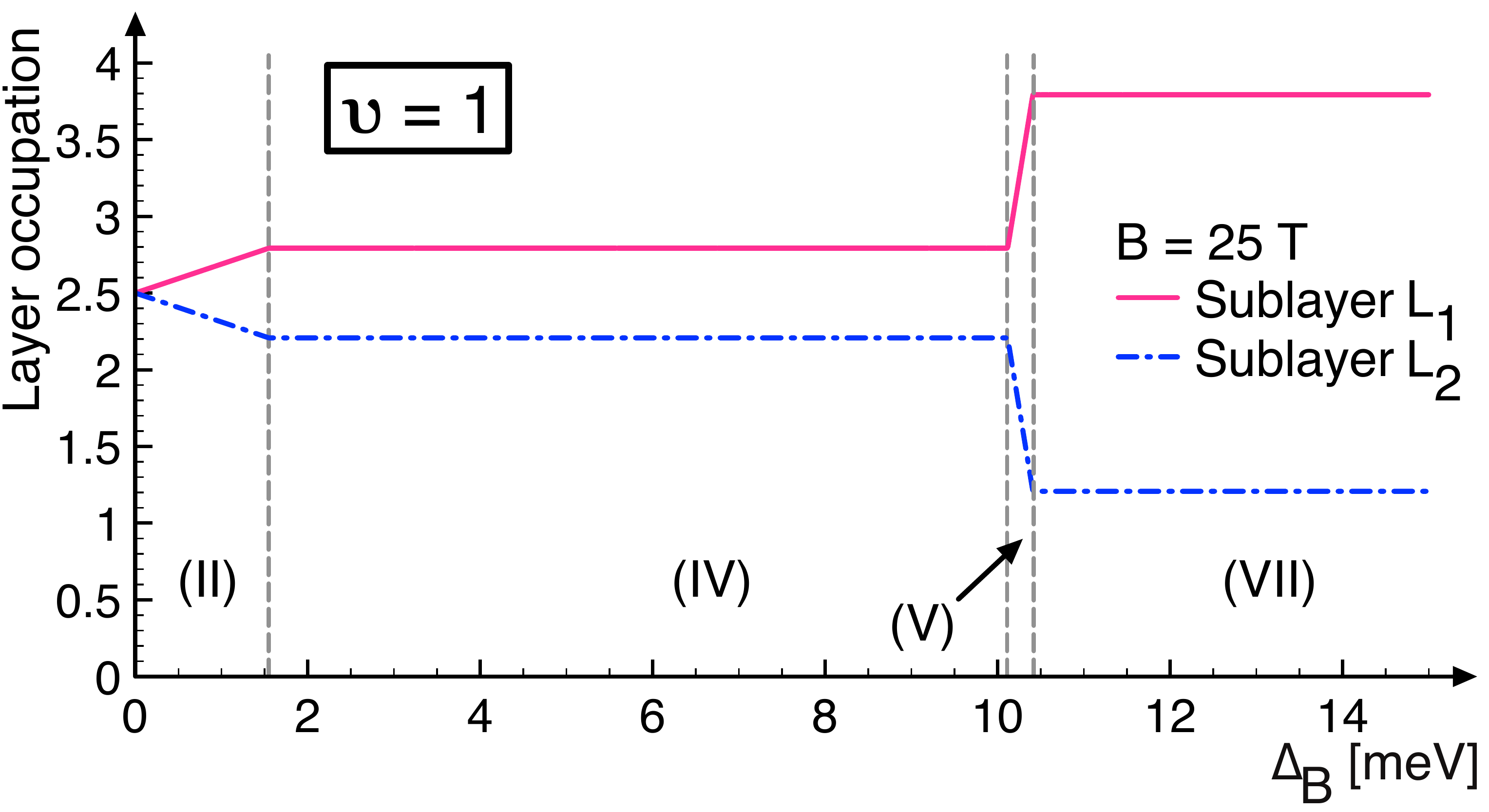}
  \caption{ .}
\label{sfig:b1iofM}
 \end{subfigure}
   \begin{subfigure}[!htb]{0.45\textwidth}
\includegraphics[width=\textwidth]{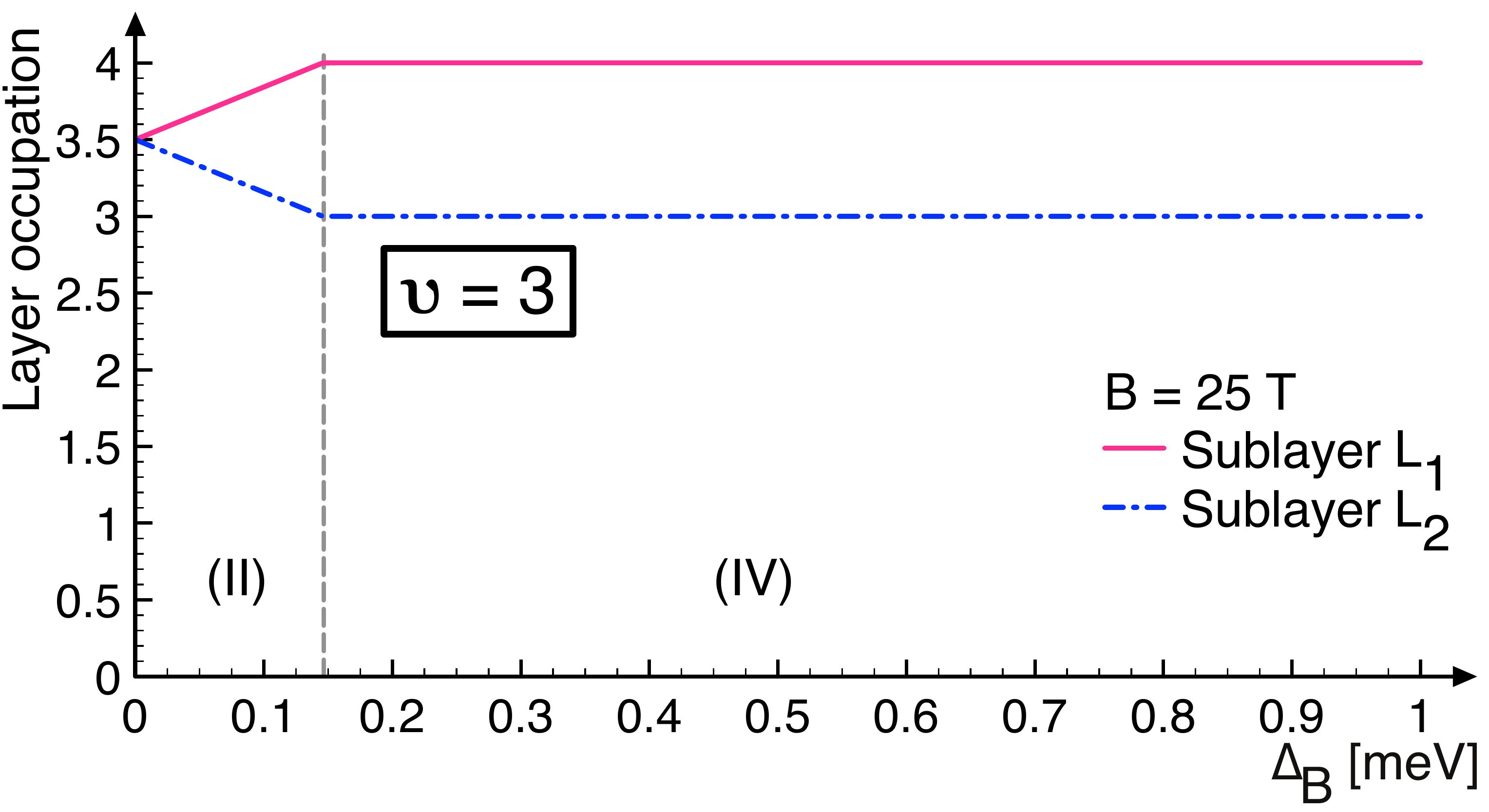}
  \caption{ .}
\label{sfig:b1iofM}
 \end{subfigure}
   \caption{Electronic distribution between the two sublayers, labeled as layer
1 and layer 2, in different  phases at filling factors $\nu=-3, \nu=0, \nu=1$, 
and $\nu=3$. }
 \label{fig:RealSpaceEx}
\end{figure}

 The electronic distribution between the two graphene layers has 
frequently been discussed in previous works\cite{Cote2010, Zhang2012, Min2008}, 
in relation with the formation of  states exhibiting either interlayer 
coherence or being fully layer polarized. This is related to 
the formation of electronic dipoles\cite{Cote2010} or the anomalous condensation 
of excitons\cite{Barlas2010}. These studies, however, have been carried out 
within the effective two-band model of BLG\cite{McCann2006}. In 
this approximate description, there is a direct correspondence between the value 
of the valley index assigned to the electrons and the graphene layer. 
Therefore, this model automatically predicts a state which 
is valley-polarized also to be \textit{layer-polarized}. This is not the case in the 
 four-band model. As pointed out  in Sec.~\ref{ssec:HF_Int}, it is clear from the form of the four-spinor states, 
Eq.~\ref{eqn:H_psi_01}, together with the behavior of the coefficients for the 
respective entries given in Eq.~\ref{eqn:psi_01_coeff} that the one-to-one 
correspondence between valley index and layer occupation is not exact in the description using all four bands.
While for electrons occupying 
the $n=0$ orbital the identification  valley $\leftrightarrow$ layer can still 
be made, for electrons in the $n=1$ orbital also for a well-defined valley 
index $+$ or $-$, occupation of both layers is enforced as soon as the bias 
$\Delta_B$ takes non-zero values. 
This implies important consequences for the 
properties of the  phases we identified in the phase diagrams of 
Fig.~\ref{fig:GSPDs}. In general, valley polarized phases can not be 
automatically identified with fully layer polarized states. In fact, as evident 
from the form of the state in Eq.~\ref{eqn:H_psi_01}, full layer polarization 
can only be achieved if two conditions are met simultaneously: the electrons 
must form a state polarized in the valley degree of freedom and at the same time 
all of them exclusively occupy the $n=0$ orbital. We observe phases fulfilling 
 these two requirements in the large bias regime of the two smallest filling 
factor: in phase (V) at filling factor $\nu=-3$ and in phase (III) for filling 
$\nu=-2$. The other states at negative filling factors $\nu=-1$ and $\nu=0$, 
respectively, tend towards partially polarized states in the limit of large 
$\Delta_B$. Although the overall orbital isospin is partially negatively 
polarized along the $z$-axis, in these cases not only $n=0$, but also $n=1$ 
orbitals are partially occupied. Therefore, the layer occupation does not tend 
towards exact layer polarization. Nevertheless, in this regime we do find states 
 in which the occupation of one of the two layers largely dominates over the 
occupation of the other layer. This, however, is not the case for the positive 
filling factors $\nu=1,2,3$. In these cases the states at large bias exhibit 
overall positive orbital polarization, hence occupation of the $n=1$ orbital 
dominates over occupation of the $n=0$ state. As a consequence, no such thing as 
full
layer polarization can be seen. Even in the limit of large bias, the electrons 
will be distributed between both layers. Furthermore, for the unbiased system at 
$\Delta_B\equiv0$, we observe the electrons to be equally distributed between 
both graphene layers for all values of the filling factor $\nu\in[-3,3]$. We 
illustrate these different types of behavior for the examples $\nu=-3, \nu=0, 
\nu=1$, and $\nu=3$ in Fig.~\ref{fig:RealSpaceEx}.

\subsection{Extrapolation to zero magnetic Field}

\begin{figure}[!htb]
  \centering
\includegraphics[width=0.25\textwidth]{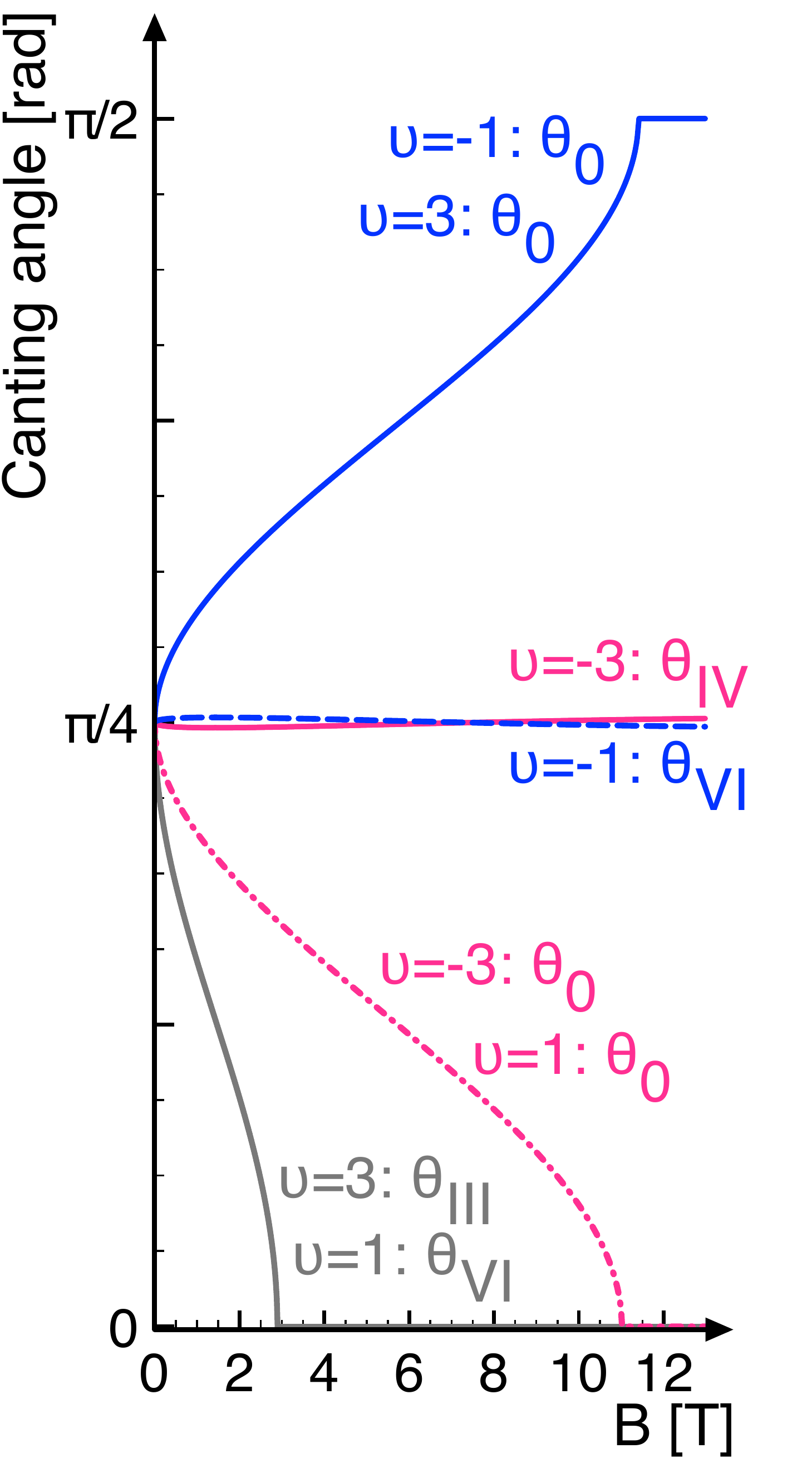}
  \caption{Canting angles in orbital space at different filling factors at low magnetic field: Blue curves for $\nu=-1$ or $\nu=3$: Unbiased case $\theta_0$ according to Eqs.~\ref{eqn:NuM1_t0}, \ref{eqn:Nu3_t0} (solid line) and phase (VI) $\theta_{VI}$ according to Eq.~\ref{eqn:NuM1_tVI}  (dashed line). Pink curves at $\nu=-3$ or $\nu=1$: Phase (IV) $\theta_{IV}$ according to Eq.~\ref{eqn:NuM3_tIV}  (solid line) and unbiased case $\theta_{0}$ according to Eqs.~\ref{eqn:NuM3_t0} , \ref{eqn:Nu1_t0}  (dashed line). Gray curve at $\nu=3$ or $\nu=1$: Phase (III) $\theta_{III}$ according to Eq.~\ref{eqn:Nu3_tIII} and Phase (VI) $\theta_{VI}$ according to Eq.~\ref{eqn:Nu1_tVI}. For the corresponding angles at filling factors $\nu=2$ and $\nu=-2$ we find $\theta_0\equiv\frac{\pi}{4}$ for all $B$.  }
\label{fig:AngleB20}
 \end{figure}

Experiments have studied in detail the limit of vanishing magnetic field. 
Indeed it has been argued that in the absence of 
any magnetic field, there is magnetic ordering of the spin and isospin degrees 
of freedom that
spontaneously breaks underlying symmetries\cite{Min2008, 
Zhang2010,Nandkishore2010,Jung2011} and this may lead to spontaneous QH 
states\cite{Zhang2011, Zhang2012,Zhang2012a}. Furthermore, it has been discussed 
how these spontaneous  QH states might be related to the QH states at nonzero 
magnetic field\cite{Nandkishore2010a, Kharitonov2012}. Recent experimental 
investigation draws the following picture: for charge neutral BLG, the 
existence 
of a gapped phase at zero magnetic field in sufficiently clean samples at 
sufficiently low temperatures is generally established\cite{Freitag2012, 
Bao2012}. This phase evolves continuously in the gapped $\nu=0$ QH state 
as the magnetic field increases\cite{Veligura2012a, Jr2012, Weitz2010}. For 
filling factor $\nu=2$, the observations of Ref.~\onlinecite{VelascoJr2014} 
suggest that the behavior with $B \rightarrow 0$ depends on the bias potential 
applied: while for small $\Delta_B$ the system extrapolates to vanishing gap, 
for 
sufficiently large bias, when the system presumably has entered a phase 
different from the low bias phase, the gap remains finite as $B$ goes to 
zero. Reference \onlinecite{Shi2016} reports for $\nu=1$ a vanishing gap with 
vanishing magnetic field independently of the bias potential, e.g., for 
the two different phases observed in this study.
We now analyze the limit $B\rightarrow0$ in 
our approach. We summarize for each filling factor the properties of the 
unbiased 
case $\Delta_B\equiv0$, as well as the phases that extend to the low magnetic 
field regime of the phase diagrams in Fig.~\ref{fig:GSPDs}. For the odd filling 
factors these phases go along with canting of the orbital degree of freedom 
(phase (IV) at $\nu=-3$, phase (VI) at $\nu=-1$, phase (VI) at $\nu=1$). To 
understand the behavior at low magnetic field, we show the evolution as a 
function of $B$ of the different canting angles in these respective phases as 
well as the orbital canting angles of the zero bias phases in Fig.~\ref{fig:AngleB20}. The states which follow from the 
naive extrapolation$B\rightarrow0$ are summarized in Table \ref{tab:StatesB20}. 
While for  even filling factors $\nu=-2, 0,2$ the GS configurations 
 decompose into simple product states in the orbital 
degree of freedom, at odd fillings $\nu=-3, -1, +1, +3$ we find states with 
non-trivial orbital coherence in the limit $B\rightarrow0$. These orbitally
coherent states explicitly rely on the quantization of the LL modes by the 
external magnetic field and thus do not have an obvious counterpart in the 
zero-field
case. This means that the states at odd filling factors 
 behave differently from the even filling factor states when the field is 
 decreased to zero. While at even fillings the GS might be connected 
smoothly to gapped spontaneous QH states at $B=0$, such extrapolation is not  
obvious for odd filling factors. Here, the zero magnetic field GS 
might be gapless. Indications for such behavior have been seen 
experimentally, e.g., in Refs.~\onlinecite{Veligura2012a, Jr2012, Weitz2010}, 
and \onlinecite{Shi2016}. We note, however, that the description of 
BLG in our model is valid really only in the limit of high magnetic 
fields since LL mixing will be important at low fields.
The states we extract for $B\rightarrow0$ in Table \ref{tab:StatesB20} 
 can serve only as hints to connect the high 
magnetic field region and the case $B=0$ where 
spontaneous QH states have been predicted. We can not exclude the existence of  
additional phases in the regime of small but nonzero magnetic field, as 
 conjectured, e.g., in Refs.~\onlinecite{ Weitz2010} or 
\onlinecite{Nandkishore2010a}.

\begin{table}
\centering
\begin{tabular}{| l | c |c| c|}
\hline
$\nu$ & Phase & GS in the limit $B\rightarrow0$\\
\hline
 $ -3$ & $\Delta_B\equiv0$ &$ |v_1\RD=\frac{1}{{2}} \Big[\,|1,\uparrow,+\RD+|1,\uparrow,-\RD + |0,\uparrow,+\RD+|0,\uparrow,-\RD\,\Big] $ \\
   &    (IV) & $ |v_1\RD=\frac{1}{\sqrt{2}} \Big[|0,\uparrow,+\RD+ |1,\uparrow,+\RD \Big] $\\
 \hline
 $ -2$&  $\Delta_B\equiv0$   & $  |v_1\RD=\frac{1}{\sqrt{2}}  \Big[|1,\uparrow,+\RD+ |1,\uparrow,-\RD \Big],
 |v_2\RD=\frac{1}{\sqrt{2}} \Big[ |0,\uparrow,+\RD+   |0,\uparrow,-\RD \Big] $ \\
  &  (II) &  $ |v_1\RD= |1,\uparrow,+\RD ,  |v_2\RD=|0,\uparrow,+\RD $\\
    \hline
  $ -1$ & $\Delta_B\equiv0$ &   $    |v_1\RD= -\frac{1}{\sqrt{2}} \Big[\,|1,\uparrow,+\RD+|1,\uparrow,-\RD\,\Big], 
  |v_2\RD= \frac{1}{\sqrt{2}} \Big[\,|0,\uparrow,+\RD+|0,\uparrow,-\RD\,\Big]$,\\
&  &$|v_3\RD=-\frac{1}{{2}} \Big[\,|1,\uparrow,+\RD-|1,\uparrow,-\RD +\,|0,\uparrow,+\RD-|0,\uparrow,-\RD\,\Big]$  \\
   & (VI) &  $ |v_1\RD=  |1,\uparrow,+\RD  , |v_2\RD=  |0,\uparrow,+\RD , |v_3\RD= \frac{1}{\sqrt{2}} \Big[ |1,\downarrow,+\RD + |0,\downarrow,+\RD \Big]$\\
\hline
  $ 0$ & $\Delta_B\equiv0$  &  $ |v_1\RD=  |1,\uparrow,+\RD ,
 |v_2\RD=  |1,\uparrow,-\RD ,
 |v_3\RD=  |0,\uparrow,+\RD ,
  |v_4\RD=  |0,\uparrow,-\RD $  \\
   & (III) & $|v_1\RD=  |1,\uparrow,+\RD ,
 |v_2\RD=  |1,\downarrow,+\RD ,
 |v_3\RD=  |0,\uparrow,+\RD ,
 |v_4\RD=  |0,\downarrow,+\RD $ \\
\hline
  $ 1$ &  $\Delta_B\equiv0$ & $ |v_1\RD=  |1,\uparrow,+\RD, 
  |v_2\RD=  |1,\uparrow,-\RD , 
  |v_3\RD= |0,\uparrow,+\RD,
  |v_4\RD= |0,\uparrow,-\RD$,\\
& &$|v_5\RD=\frac{1}{{2}} \Big[\,|1,\downarrow,+\RD+|1,\downarrow,-\RD - |0,\downarrow,+\RD-|0,\downarrow,-\RD\,\Big]$ \\
   &  (VI)  & $ |v_1\RD=  |1,\uparrow,+\RD , 
  |v_2\RD=  |1,\downarrow,+\RD , 
  |v_3\RD= |0,\uparrow,+\RD,  |v_4\RD= |0,\downarrow,+\RD,  |v_5\RD=   \frac{1}{\sqrt{2}} \Big[\, |1,\uparrow,-\RD +  |0,\uparrow,-\RD\, \Big]$  \\
\hline
  $ 2$ & $\Delta_B\equiv0$  & $ |v_1\RD=  |1,\uparrow,+\RD , 
  |v_2\RD=  |1,\uparrow,-\RD , 
  |v_3\RD= |0,\uparrow,+\RD,
  |v_4\RD= |0,\uparrow,-\RD$ ,\\
 & &$|v_5\RD= \frac{1}{\sqrt{2}}  \Big[\,|1,\downarrow,+\RD+  |1,\downarrow,-\RD\,\Big] ,
|v_6\RD= \frac{1}{\sqrt{2}}\Big[\,|0,\downarrow,+\RD+  |0,\downarrow,-\RD\,\Big]$  \\
   &  (II)&$  |v_1\RD=  |1,\uparrow,+\RD , 
  |v_2\RD=  |1,\uparrow,-\RD , 
  |v_3\RD= |0,\uparrow,+\RD,
  |v_4\RD= |0,\uparrow,-\RD ,
 |v_5\RD= |1,\downarrow,+\RD,\
|v_6\RD=|0,\downarrow,+\RD $   \\
\hline
  $ 3$ & $\Delta_B\equiv0$   & $|v_1\RD=  |1,\uparrow,+\RD , 
 |v_2\RD=  |1,\uparrow,-\RD , 
 |v_3\RD= |0,\uparrow,+\RD,
 |v_4\RD= |0,\uparrow,-\RD , |v_5\RD=\frac{1}{{2}} \Big[\,|1,\downarrow,+\RD+|1,\downarrow,-\RD+|0,\downarrow,+\RD+|0,\downarrow,-\RD\,\Big],$ \\
& &$|v_6\RD=\frac{1}{{2}} \Big[\,|1,\downarrow,+\RD+|1,\downarrow,-\RD-|0,\downarrow,+\RD-|0,\downarrow,-\RD\,\Big] , |v_7\RD=\frac{1}{{2}} \Big[\,|1,\downarrow,+\RD-|1,\downarrow,-\RD+|0,\downarrow,+\RD-|0,\downarrow,-\RD\,\Big]  $ \\
   &  (III) &  $|v_1\RD=  |1,\uparrow,+\RD , 
  |v_2\RD=  |1,\uparrow,-\RD , 
 |v_3\RD= |0,\uparrow,+\RD,
 |v_4\RD= |0,\uparrow,-\RD$,\\
&  & $|v_5\RD= |1,\downarrow,+\RD ,
 |v_6\RD= |0,\downarrow,+\RD ,
|v_7\RD=\frac{1}{\sqrt{2}} \Big[\, |1,\downarrow,-\RD+ |0,\downarrow,-\RD\,\Big]$\\
\hline
\end{tabular}\caption{States in the limit $B\rightarrow0$}\label{tab:StatesB20}.\end{table}

\section{Relation to Experiment and to  theoretical Studies}
\label{sec:Relation}

 The effect of external magnetic and electric fields on
graphene mono- and multilayers has been under intense experimental  
investigation\cite{Weitz2010,  Kim2011, Bao2012, 
Jr2012, Maher2013,  VelascoJr2014, Lee2014, Maher2014, Shi2016, Hunt2016}.
We first compare our work 
with experimental findings, before discussing similarities and differences with 
 theoretical approaches\cite{Castro2010, Cote2010, Lambert2013, 
Shizuya2012}.
The fact that external fields  influence the 
ordering of spin, valley, and orbital degrees of freedom, and that 
transitions between states of different spin and isospin order can be induced by 
tuning externally applied fields has  been realized several years 
ago\cite{Weitz2010,Kim2011}. 
Recently, there has been tremendous improvement 
in the quality of the samples, and data became available in a
much wider parameter range. This has lead to detailed insights about the nature 
of 
the different phases at different filling factors. By carefully 
monitoring sudden changes in the  conduction properties, one infers the 
number of phase transitions  upon varying the bias potential at 
fixed magnetic field $B$.  At $\nu=\pm3$, a single phase 
transition has been seen\cite{Weitz2010, Maher2013, Maher2014} at zero 
bias $\Delta_B\equiv0$. For $\nu=\pm2$, Refs.~\onlinecite{Weitz2010, 
VelascoJr2014, Lee2014, Hunt2016, Maher2014} report transitions at nonzero bias 
 while there is no sign of  phase transition  at zero 
bias. Both types of transitions, at $\Delta_B\equiv0$ as well as at 
$|\Delta_B|\neq 0$, have been observed\cite{Weitz2010, Shi2016, Hunt2016, 
Maher2014} 
at $\nu=\pm1$. The properties at charge neutrality $\nu=0$ have been 
investigated in Refs.~\onlinecite{Weitz2010, Kim2011, Bao2012, Jr2012, 
Lee2014, Hunt2016, Maher2013}. While early investigations reported one 
transition at nonzero bias\cite{Weitz2010,  Kim2011}, 
more recent studies report signatures of transitions at two different values 
of the bias potential implying at least three different phases. Common belief is 
that for large bias potential the system will 
be in a spin and isospin configuration that maximizes layer polarization. 
Accordingly, in the opposite limit of very small or vanishing bias, the spin and 
isospin ordering is assumed to be different from maximally possible 
layer polarization. 

We compare these experimental observations to the predictions of our 
calculations. In 
parameter ranges comparable to those of the respective experiments, we examine 
the different phases and the number of phase transitions at fixed magnetic 
and increasing bias:

* For filling factor $\nu=-3$, we obtain the following picture: for 
 $B<11$ T we see the sequence of transitions 
(I)$\rightarrow$ (IV) $\rightarrow$ (V), whereas for higher magnetic fields 
$B>11$ T the series of transitions (II)$\rightarrow$ (III) $\rightarrow$ (IV) 
$\rightarrow$ (V) is observed.

Maher et al.~\onlinecite{Maher2014} as well as 
Hunt et al.~\onlinecite{Hunt2016} have studied  the BLG system at  $\nu=\pm3$ 
h in the range of the bias  $|\Delta_B|\approx0-34$ meV for magnetic 
fields  $B=9$ T and $B=31$ T, respectively. We may attribute the single 
transition close to zero bias observed in both references  to the 
transitions (I)$\rightarrow$ (IV) at  lower magnetic field or  
(II)$\rightarrow$ (III) at  higher magnetic field value, respectively. The 
values of the bias potential at which these transitions occur in our model  are 
both small compared to the energy scales 
of the other phases of the phase diagram: $\Delta_B\approx0.185$ meV and  
$\Delta_B\approx2.5$ meV, respectively. The fact that no second phase transition 
is observed by Hunt et al.~\onlinecite{Hunt2016} 
may imply that phase (IV) has not yet been reached at these values of the bias.
If the zero-bias phases we find in the HF treatment are destroyed by 
fluctuations beyond HF then this may explain a zero-bias transition between
oppositely polarized states.

* For $\nu=\pm 2$ we see for all values of the 
magnetic field the sequence of transitions (I)$\rightarrow$ (II) $\rightarrow$ 
(III) as a function of increasing bias. The second transition  (II) 
$\rightarrow$ (III), however, occurs at much higher values of the bias potential 
than those shown in experimental data: $\Delta_B\gtrsim300$ meV 
in 
Fig.~\ref{fig:GSPDs}. Our predictions  are consistent with the 
observations at  $\nu=\pm2$ of Velasco et al.~\onlinecite{VelascoJr2014}, 
Maher et al.~\onlinecite{Maher2014}, Hunt et al.~\onlinecite{Hunt2016}, Lee et 
al.~\onlinecite{Lee2014} identifying one
phase transition at nonzero bias $\Delta_B>0$. So the low-bias phase has valley 
coherence and this coherence is destroyed beyond a critical bias.
The slope of the I/II transition line in Velasco et al. is 0.72mV nm$^{-1}$ 
T$^{-1}$ while the HF value is 0.55mV nm$^{-1}$ 
T$^{-1}$.

* At filling $\nu=-1$ for magnetic fields $B<11.3$ T, we go through the 
sequence (I)$\rightarrow$ (III) $\rightarrow$ (VI) $\rightarrow$ (VII), in the 
opposite case $B>11.3$ T we find (II)$\rightarrow$ (IV) 
$\rightarrow$ (V) $\rightarrow$ (VI) $\rightarrow$ (VII) when increasing 
$\Delta_B$. In the case $\nu=+1$, at  small magnetic field
$B<11.3$ 
T, the sequence is  (I)$\rightarrow$ 
(III) $\rightarrow$ (VI) / (VII), whereas for larger field  $B>11.3$ T it is
 (II)$\rightarrow$ (IV) $\rightarrow$ (V) $\rightarrow$ (VI).

This may be compared to the experimental results of Shi et al.~\onlinecite{Shi2016},  Hunt et al.~\onlinecite{Hunt2016}, and Maher et al.~\onlinecite{Maher2014}, where the 
states $\nu=\pm1$ 
are probed for $B=28$ T in the range $|\Delta_B|\approx 0 - 17$ meV, and in the 
range $|\Delta_B|\approx 0 - 34$ meV at magnetic fields $B=31$ T and $B=9$ T, 
respectively. The observed transition near zero bias can be attributed to the 
phase transitions (I)$\rightarrow$ (III) or (II)$\rightarrow$ (IV), 
respectively, which occur in our model at relatively small values of $\Delta_B$ 
compared to the range of the broadest phases of the phase diagram and to the 
overall 
range of the bias. 
The phase II has valley coherence as proposed in Shi et al.
A second transition observed in experiment 
at nonzero value of the bias might be identified with the transitions  (III) 
$\rightarrow$ (VI) or (IV) $\rightarrow$ (V)/(VI) at $\nu=-1$ and  (III) 
$\rightarrow$ (VI) / (VII) or  (IV) $\rightarrow$ (V)/(VII) at $\nu=1$, 
respectively. In fact Maher et al. have proposed that the finite bias 
transition they see at $\nu =+1$ is the IV/VII transition.

* In the $\nu=0$ case, for any value of the magnetic field, we observe two 
successive phase transitions (I)$\rightarrow$ (II) $\rightarrow$ (III) upon 
increasing bias potential. The phase II is a spin-valley coherent phase 
flanked by simpler incoherent phases I, III. This is consistent with  recent 
experimental 
studies of the  $\nu=0$ state by Lee et al.~\onlinecite{Lee2014} and 
Hunt et al.~\onlinecite{Hunt2016}, where two transitions at two distinct 
nonzero values of 
the bias potential have been observed. The $\nu=0$ phase diagram shown in 
Ref.~\onlinecite{Hunt2016} Fig.~2D also agrees well with the corresponding 
phase 
diagram predicted by our calculations. Furthermore, Maher et al.~\onlinecite{Maher2013} 
have also observed a critical bias 
 increasing as a function of  the magnetic field.

From this discussion we see that our calculation reproduces
several features observed experimentally in BLG at different filling factors. 
Notably, for 
every $\nu$, we are able to identify phase transitions detected in experiment 
with transitions predicted by our model.  The range 
of the 
bias achieved in the various different experimental studies only covers part of 
the phase diagrams presented in Fig.~\ref{fig:GSPDs}. In particular the phase 
of maximal orbital polarization, corresponding the phase 
with the highest number in each case, presumably has not been reached in 
experiments  
for the filling factors $\nu=-3, \nu=-2, \nu=-1$, and $\nu=2$. As a 
consequence, according to the properties of the 
four-band model as discussed in Sec.~\ref{sec:HFRESII}, maximally possible 
layer polarization has not 
been achieved experimentally. Furthermore, from the phase diagrams of 
Fig.~\ref{fig:GSPDs}, we conjecture that for example at fillings $\nu=-3$ or 
$\nu=1$ a richer picture of different phases and phase transitions may
emerge for an extended range of $B$ and $\Delta_B$.

It should be noted, however, that the various experiments often differ in 
the way the sample is prepared, e.g., Refs.~\onlinecite{Weitz2010,Jr2012, VelascoJr2014, Shi2016} 
investigate the properties of suspended BLG, Ref.~\onlinecite{Lee2014} uses 
double BLG heterostructures separated by a hexagonal boron nitride dielectric 
while in Refs.~\onlinecite{Hunt2016, Maher2013, Maher2014} the BLG samples 
are encapsulated by hexagonal boron nitrate. We have not tried to take into 
account the 
additional effects due to these different substrates, gatings, dielectrics, or 
encapsulations. These differences may change the physics of the phase 
competition.

\begin{figure}[!htb]
  \centering
   \begin{subfigure}[!htb]{0.32\textwidth}
\includegraphics[width=\textwidth]{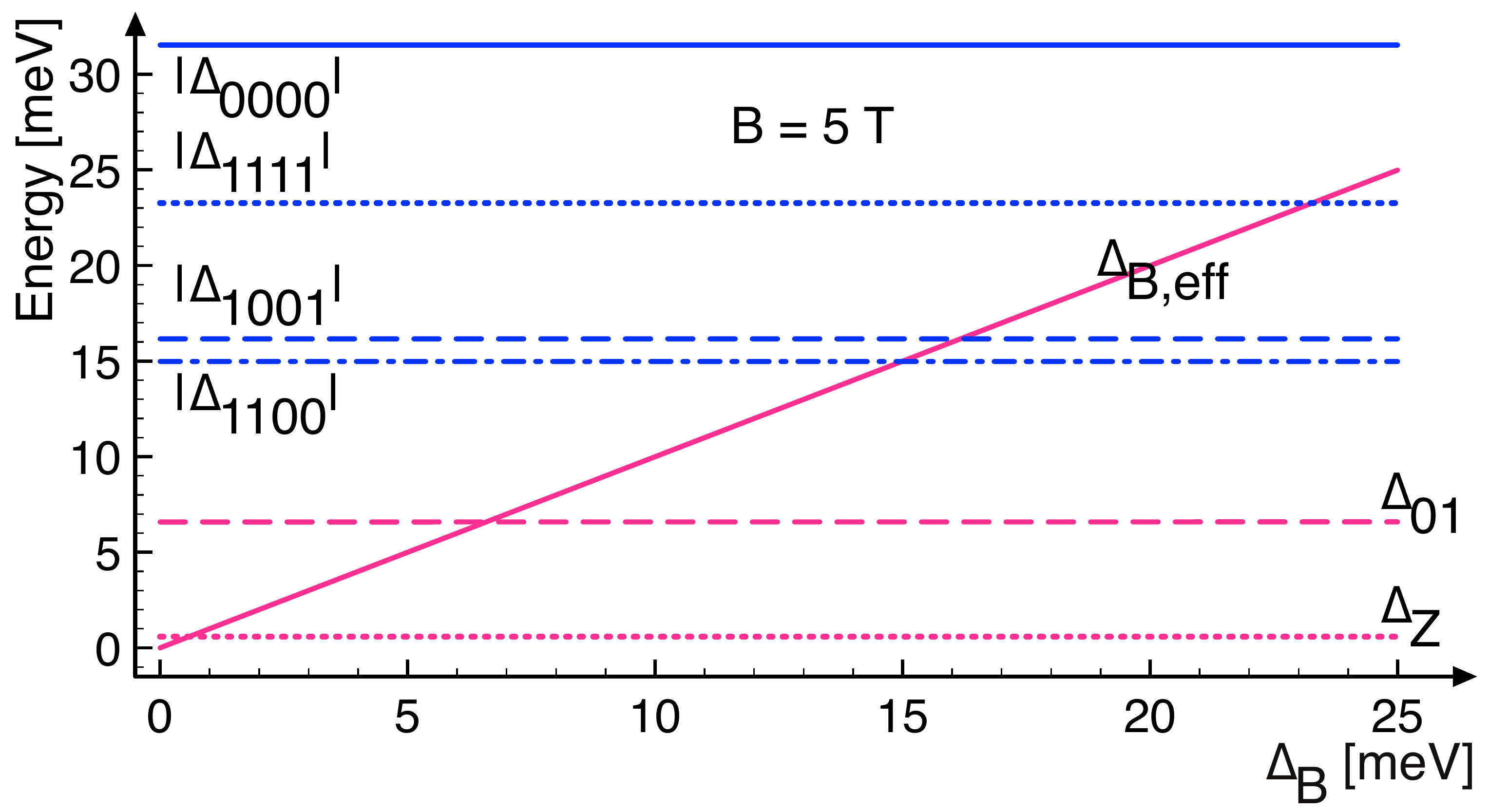}
\label{sfig:SplitElmts_B5}
 \end{subfigure}
   \begin{subfigure}[!htb]{0.32\textwidth}
\includegraphics[width=\textwidth]{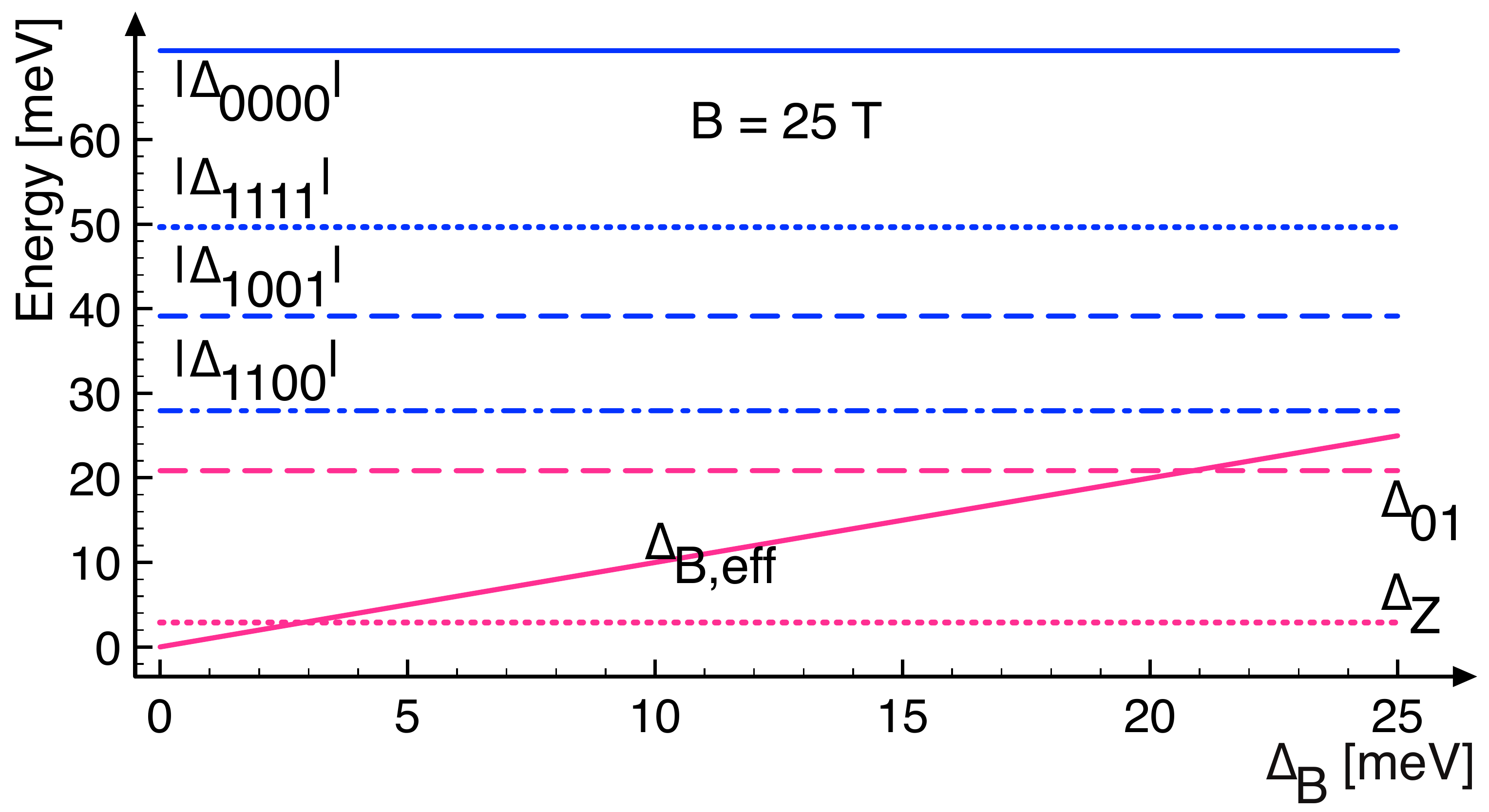}
\label{sfig:SplitElmts_B25}
 \end{subfigure}
   \begin{subfigure}[!htb]{0.32\textwidth}
\includegraphics[width=\textwidth]{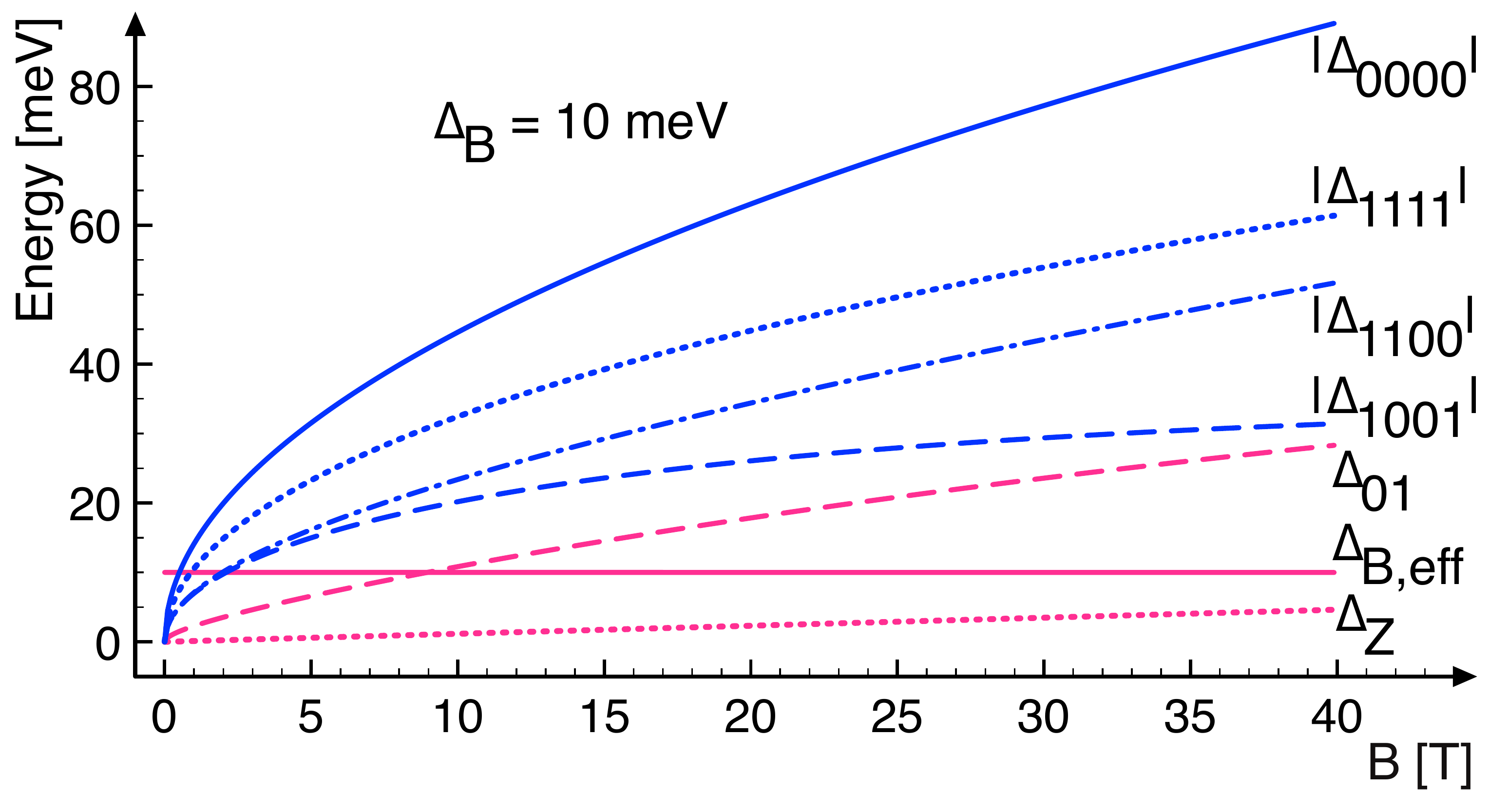}
\label{sfig:SplitElmts_DB10}
 \end{subfigure}
   \caption{Dependence on the magnetic field and on the bias potential of the 
energy splittings in valley space, orbital space, and spin space, 
$\Delta_{B,eff}, \Delta_{01}$, and $\Delta_{Z}$, respectively, as well of the
valley conserving exchange matrix elements of the Coulomb interaction, 
$\Delta_{1111}, \Delta_{0000}, \Delta_{1001}$, and $\Delta_{1100}$, as computed from 
Eq.~\ref{eqn:X}. The curves for the valley breaking matrix elements, $X_{1111}, X_{0000}, 
X_{1001}$, and $X_{1100}$ are slightly offset with respect to the valley 
conserving terms but comparable in their overall behavior and are therefore not 
shown for the sake of visibility.}
 \label{fig:SplitElmts}
\end{figure}

Let us now  compare our results to previous 
theoretical investigations.  Ref.~
\onlinecite{Lambert2013} presents a detailed HF study of BLG zero energy octet 
with an effective two-band model. 
They obtained the phase diagram of their model for all different filling 
factors 
$\nu\in[-3,3]$ for $B=10$ T as a function of the  bias. The vast majority 
of states they 
deduce from their model is orbitally incoherent. 
Phases exhibiting orbital coherence emerge only at very large values of 
the bias.
These authors do not take into 
account the presence of the Dirac sea. It has become 
clear, however, that these electrons of the Dirac sea do play a non-silent 
role: 
As we discuss in Sec.~\ref{ssec:HF_Int}, Shizuya shows in Ref.~\onlinecite{Shizuya2012} in a four-band model the importance of this effect.
The GS 
configurations identified in this treatment e.g.~at zero bias can be 
coherent superpositions of the $n=0$ and $n=1$ states. Moreover, in this 
analysis, the $n=1$ state lies lower in energy than the $n=0$ orbital while in 
Ref.~\onlinecite{Lambert2013} generally the $n=0$ state is populated 
first. These results, however, were obtained from
 a somewhat simplified model with respect to Lambert 
and Cot\'e in Ref.~\onlinecite{Lambert2013}.

Our treatment contains the ingredient of a realistic band structure, i.e., four 
bands with all the $\gamma_i$ couplings and we have included the Dirac sea 
exchange.
\section{ Conclusion}
\label{sec:DiscConcl}

We have derived the phase diagram of the Bernal-stacked bilayer graphene as a 
function of the applied magnetic field and potential bias between the layers. 
We have focused on the octet of levels near neutrality for which the filling 
factor is in the range $\left[-3,+3\right]$. We have used a HF method 
which is 
known to capture the main features of quantum Hall ferromagnetism. Our
 tight-binding model includes hoppings $\gamma_0,\gamma_1,\gamma_3,\gamma_4$
that breaks weakly particle-hole symmetry and we have retained the four bands. 
In the 
HF calculation we have included the exchange within the occupied Dirac sea 
which restores the particle-hole symmetry in the absence of $\gamma_4$. 
The splitting between $n=0$ and $n=1$ orbitals is thus governed by the 
competition between band structure effects and Lamb-shift-like exchange 
interactions. 
The spin and isospin 
configuration hence is governed by a careful balance between all these 
different 
symmetry breaking terms. This is illustrated in Fig.~\ref{fig:SplitElmts}, 
where 
we show the evolution of the energy splittings in spin space, valley isospin 
space, and orbital isospin space, $\Delta_Z, \Delta_{B,eff}$, and $\Delta_{01}$, 
 as well as the matrix elements of the Coulomb interaction as 
computed in Sec.~\ref{ssec:HF_Int} as functions of the external magnetic 
and 
electric fields for different parameters. 
In the regime of 
small bias and large magnetic field, $\Delta_{01}$ plays a pivotal role before 
being washed out at sufficiently strong bias by $\Delta_{B,eff}$ acting as a 
"Zeeman-like" splitting in valley space. 

For even filling factors $\nu=0,\pm 2$ our results are the same as the HF 
treatment of Lambert and C\^ot\'e\cite{Lambert2013}. However for
odd fillings $\nu=\pm 1, \pm 3$ we find phases with non-trivial 
 orbital coherence: see
Fig.~\ref{fig:GSPDs}. These 
phases are thus of fundamentally different nature than those predicted in 
Ref.~\onlinecite{Lambert2013}. As these orbital coherent phases 
appear at experimentally accessible values of the bias potential, it is 
plausible that they are among the phases actually observed  in 
experiment. For fillings $\nu=-3,-1$ they extend to all values of the magnetic 
field but require a specific range of bias. For $\nu=+3,+1$ the orbital phases 
are restricted to the small-field regime which may be out of range of our 
approach due to Landau level mixing.

For  odd filling 
factors we observe at  small bias a transition from an orbital coherent 
phase to orbitally incoherent phases as a function of the magnetic field 
strength:  the vector of  orbital isospin 
rotates from a canted position at small magnetic field to a partially polarized 
configuration above a critical field strength $B_{crit}$. Such transitions with 
$B$ have not been reported previously in the literature as e.g.~Ref.~
\onlinecite{Lambert2013} restricts its investigations of the GS phases to the 
phase diagram at a single fixed value of the magnetic field. We conclude that 
varying the magnetic field can trigger the emergence of  phase 
transitions for all odd $\nu$. We thus conjecture the existence of more phases 
and even richer phase 
diagrams when the BLG is studied over a a sufficiently large range of 
$B$ values. 


\begin{acknowledgments}
We acknowledge discussions with A. H. MacDonald, Feng Cheng Wu, and Ren\'e 
C\^ot\'e. AK would like to thank R. A. R\"omer for discussions about numerical HF calculations and for pointing out References \onlinecite{Sohrmann2007} and  \onlinecite{Romer2008}. AK gratefully acknowledges support by the German Academic Scholarship Foundation and by the German Academic Exchange Service.
\end{acknowledgments}


\bibliography{BilayerGraphene_HFGSPaper.bbl}

\end{document}